\documentclass[prd,twocolumn,nofootinbib,superscriptaddress,amsmath,amssymb,eqsecnum,longbibliography]{revtex4-1}
\usepackage[utf8]{inputenc}
\usepackage[T1]{fontenc}
\usepackage{mathtools}
\usepackage{graphics}
\usepackage{graphicx}
\usepackage{dcolumn}
\usepackage{bm}
\usepackage{dsfont} 
\usepackage{amsmath,amssymb}
\usepackage{hyperref}
\usepackage{tabularx}
\usepackage{epstopdf}
\usepackage{tensor}
\usepackage{mathrsfs}
\usepackage[normalem]{ulem}
\usepackage[usenames]{color}
\hypersetup{
    colorlinks=true,
    linkcolor=blue,
    filecolor=magenta,      
    urlcolor=blue,
    citecolor=blue
}
\usepackage[dvipsnames]{xcolor}
\usepackage{mathrsfs}
\usepackage{microtype}

\newcommand{\shammi}[1]{\textcolor{blue}{\it{\textbf{(st: #1)}}} }

\newcommand\be{\begin{equation}}
\newcommand\ba{\begin{eqnarray}}
\newcommand\ee{\end{equation}}
\newcommand\ea{\end{eqnarray}}
\newcommand\bw{\begin{widetext}}
\newcommand\ew{\end{widetext}}
\newcommand{\lb}{\left(}
\newcommand{\rb}{\right)}
\newcommand{\ETH}{\text{\DH}}
\newcommand\m{\mathcal}
\newcommand\td{\tilde}

\newcommand{\BD}{{\mbox{\tiny BD}}}
\newcommand\dxi{\delta_{\xi}}

\newcommand{\DN}[1]{\textcolor{ForestGreen}{\textbf{[DN: #1]}}}

\newcommand{\UVA}{Department of Physics, University of Virginia, P.O.~Box 400714, 382 McCormick Road, Charlottesville, Virginia 22904-4714, USA}

\begin{document}
\allowdisplaybreaks

\title{Gravitational-wave memory effects in the Damour-Esposito-Far\`ese extension of Brans-Dicke theory}

\author{Shammi Tahura}
\email{st.tahura@gmail.com}
\affiliation{Department of Physics and Astronomy, University of Iowa, Iowa City, IA 52242, USA}

\author{David A.\ Nichols}
\email{david.nichols@virginia.edu}
\affiliation{\UVA}

\author{Kent Yagi}
\email{ky5t@virginia.edu}
\affiliation{\UVA}

\date{\today}

\begin{abstract} 
Gravitational-wave memory effects are lasting changes in the strain and its time integrals.
They can be computed in asymptotically flat spacetimes using the conservation and evolution equations in the Bondi-Sachs framework. 
Modified theories of gravity have additional degrees of freedom with their own asymptotic evolution equations; these additional fields can produce differences in the memory effects in these theories from those in general relativity. 
In this work, we study a scalar-tensor theory of gravity known as the Damour-Esposito-Far\`ese extension of Brans-Dicke theory.
We use the Bondi-Sachs framework to compute the field equations in Bondi-Sachs form, the asymptotically flat solutions, and the leading gravitational-wave memory effects.
Although Damour-Esposito-Far\`ese theory has additional nonlinearities not present in Brans-Dicke theory, these nonlinearities are subleading effects in an expansion in inverse luminosity distance; thus, the two theories share many similarities in the leading (and some subleading) solutions to hypersurface equations, asymptotic symmetries, and types of memory effects. 
The conservation equations for the mass and angular momentum aspects differ between the two theories, primarily because of the differences in the evolution equation for the scalar field.
When used to compute the time dependence of the gravitational-wave memory signals from quasicircular inspirals of binary neutron stars and black-hole--neutron-star binaries, they give rise to new features that are of second-order in a small coupling parameter in this theory.
The observational bound that the coupling is small suggests that it would be challenging to use memory effects during the inspiral to distinguish between Brans-Dicke and Damour-Esposito-Far\`ese theories.
Nevertheless, these results can be used to analyze and interpret memory effects from numerical-relativity simulations of binaries in this theory.
\end{abstract}

\maketitle

\section{Introduction}

Gravitational wave (GW) observations opened a new window to test general relativity (GR) in the strong-field regime of GR~\cite{LIGOScientific:2019fpa,LIGOScientific:2020tif,LIGOScientific:2021sio}.
They are also allowing the nature of compact objects to be investigated (e.g.,~\cite{Isi:2019aib,Bhagwat:2019dtm,Isi:2020tac,CalderonBustillo:2020rmh}), enabling the properties of the astrophysical population of merging black holes to be inferred~\cite{LIGOScientific:2018jsj,LIGOScientific:2020kqk,KAGRA:2021duu} and constraining the Hubble constant~\cite{LIGOScientific:2017adf}. 
These results were obtained from the first three observing runs of the LIGO-Virgo-KAGRA (LVK) collaboration, during which almost one-hundred compact-binary coalescences have been observed~\cite{LIGOScientific:2018mvr,LIGOScientific:2020ibl,KAGRA:2021vkt}. 

During the fourth observing run, as the LVK detectors improve, the number of detections is expected to more than double~\cite{KAGRA:2013rdx}. 
In the next decade, several new ground-based and space-based detectors are anticipated to operate (e.g.,~\cite{LISA:2017pwj,Punturo:2010zz,Yagi:2013du,TianQin:2015yph,Gong:2021gvw,Reitze:2019iox}), which will both allow new sources to be detected in new frequency ranges and permit GWs from compact binaries to be detected at much larger rates and with significantly higher precision.
Particularly the high-precision measurements will permit GR to be tested more precisely and to probe GW phenomena that have not yet been detected.
The currently unobserved phenomena that will be the focus of this paper are GW memory effects.

\subsection{Background on memory effects in GR}

The GW memory effect was first predicted in the 1970s~\cite{Zeldovich:1974gvh} (see also~\cite{Smarr1977,1978Natur.274..565T,1978ApJ...223.1037E,1979ApJ} and earlier related discussion in~\cite{Newman:1966ub}) and was computed in the context of linearized gravity.
It has subsequently been called ``linear GW memory,'' because Christodoulou~\cite{Christodoulou:1991cr} predicted a nonlinear memory effect, and Blanchet and Damour~\cite{Blanchet:1992br} independently computed the effect in the context of multipolar post-Minkowski expansion (see, e.g.,~\cite{Blanchet:2013haa} for a review of the post-Minkowski formalism).
This latter effect has been called ``nonlinear GW memory.''
The study of GW memory effects has expanded since these first computations several decades ago largely because of developments in their theoretical study and  observational prospects, which we review next.

The measurement prospects for detecting the displacement memory had been studied prior to the observation of GWs~\cite{Braginsky:1986ia,1987Natur.327..123B,Kennefick:1994nw,Favata:2009ii,Pollney:2010hs}, but these studies focused on individual sources of GWs, including binary black hole (BBH), black-hole neutron-star (BH-NS) and binary neutron star (BNS) mergers.
After the LVK's detection of many BBH mergers, it was realized that detecting evidence for the memory effect in the entire population of BBH mergers would be possible~\cite{Lasky:2016knh}.
Subsequent forecasts based on the BBH population inferred from the LVK detections have indicated that evidence for the memory effect in the BBH population is likely to be measured during the fifth LVK observing run~\cite{Hubner:2019sly,Boersma:2020gxx,Grant:2022bla}.
There are active searches for the memory effect in BBH populations using LVK data~\cite{Hubner:2019sly,Hubner:2021amk,Cheung:2024zow} and from individual events from subsolar mass BBHs~\cite{Ebersold:2020zah} or with pulsar timing arrays~\cite{Wang:2014zls,NANOGrav:2023vfo}.
It will also be possible to detect the memory with future GW detectors on the ground~\cite{Johnson:2018xly,Grant:2022bla,Goncharov:2023woe} and in space~\cite{Favata:2009ii,Islo:2019qht,Goncharov:2023woe,Gasparotto:2023fcg,Inchauspe:2024ibs} from individual BBH systems.
As more BH-NS mergers are detected by the LVK in upcoming observing runs and by next-generation detectors, the memory effect can be detected from these systems and used to distinguish BBH from BH-NS mergers~\cite{Tiwari:2021gfl}. 

There have been developments in the study of GW memory effects from a theoretical perspective, too.
First, GW memory effects were found to be part of an infrared triangle (see, e.g.,~\cite{Strominger:2014pwa,Strominger:2017zoo}) which relates them to the asymptotic symmetries of the spacetime and to soft theorems of gravitational scattering~\cite{Weinberg:1965nx}.
Second, several additional GW memory observables were predicted that generalize the linear and nonlinear effects of~\cite{Zeldovich:1974gvh,Christodoulou:1991cr}, respectively.
To describe these many new effects and distinguish them from the memory effects of~\cite{Zeldovich:1974gvh,Christodoulou:1991cr}, this led to new nomenclature for the effects, which we now review.

First, the linear and nonlinear effects of~\cite{Zeldovich:1974gvh,Christodoulou:1991cr} are collectively referred to as ``displacement memory''~\cite{Nichols:2017rqr}.
It was given this name because it can be measured by the change in separation of two comoving freely falling observers after the passage of a burst of GWs (i.e., the solution of the geodesic deviation that depends on initial relative separation)~\cite{Flanagan:2018yzh}.
However, if the observers have a relative velocity, then there will be a residual change in their displacement after the burst of waves that depends on their initial velocity and the time integral of the shear~\cite{Flanagan:2018yzh}; this effect is independent of the displacement memory~\cite{Grant:2023ged}.
This subleading GW memory~\cite{Flanagan:2018yzh,Grant:2021hga} (also called ``drift memory''~\cite{Grant:2023ged}) is closely connected with the spin~\cite{Pasterski:2015tva,Flanagan:2015pxa,Nichols:2017rqr} and center-of-mass (CM)~\cite{Nichols:2018qac} GW memory effects, which have also been related to infrared triangles~\cite{Strominger:2017zoo} (and quadrilaterals~\cite{Compere:2018ylh}).
Observers with a relative acceleration can measure a hierarchy of ``higher GW memory effects,'' which depend on the initial time derivatives of the relative acceleration and two or more integrals of the GW strain~\cite{Flanagan:2018yzh,Grant:2021hga,Grant:2023ged,Grant:2023jhd,Siddhant:2024nft}.

Analogously to the linear and nonlinear contributions to the displacement memory, all of these higher GW memory effects can be expressed in terms of (i) integrals of ``fluxes'' (or ``pseudofluxes''), which are nonlinear in the metric variables and vanish in the absence of radiation, and (ii) changes in ``charges,'' which can be constructed to be constant in the absence of radiation (and which include some terms that are linear in the metric variables)~\cite{Grant:2021hga}.
In some cases, the charges and fluxes are associated with underlying symmetries of asymptotically flat spacetimes.
This has been shown most comprehensively for the displacement memory effect, which is closely connected to the supertranslation symmetries of asymptotically flat spacetime in the Bondi-Metzner-Sachs (BMS) group~\cite{Bondi:1962px,Sachs:1962wk} (and the conjugate supermomentum charges) via the infrared triangle discussed above.
The spin and CM memory effects are related to the extended~\cite{Barnich:2009se,Barnich:2010eb,Barnich:2011mi} or generalized~\cite{Campiglia:2014yka,Campiglia:2015yka} BMS algebras through the charges and fluxes, though not quite as straightforwardly as with the standard BMS group~\cite{Compere:2018ylh}.
The charges from which the higher GW memory effects arise also may be related to a (loop) $w_{1+\infty}$ algebraic structure, though this is still being investigated (see, e.g.,~\cite{Strominger:2021mtt,Freidel:2021ytz,Freidel:2022skz,Compere:2022zdz,Geiller:2024bgf}).

There are several methods to compute GW memory effects, including using different perturbative techniques, the charge and flux relationships described above, and certain numerical-relativity (NR) simulations.
Memory effects have been computed during the inspiral phase of a binary system using the (nonlinear) flux terms and post-Newtonian (PN)~\cite{Favata:2008yd,Favata:2011qi,Ebersold:2019kdc} or self-force~\cite{Elhashash:2024thm,Cunningham:2024dog} waveforms.
These calculations are consistent with full PN calculations at 4PN order which do not make use of the balance laws~\cite{Trestini:2023wwg,Blanchet:2023bwj,Blanchet:2023sbv} for the common PN orders in each calculation method.
The (nonlinear) fluxes in the Bondi-Sachs framework also can be used to compute memory effects approximately using NR waveforms~\cite{Nichols:2017rqr,Mitman:2020bjf,Mitman:2020pbt,Khera:2020mcz,Grant:2023jhd}.
This is necessary, because NR simulations that extract waveforms from the computational domain using extrapolation do not capture all the GW memory effects (see, e.g.,~\cite{Boyle:2019kee}).

Although NR simulations using extraction do not resolve the displacement memory effect, there are two closely related numerical frameworks that do.
The first is Cauchy-characteristic evolution (CCE), which evolves NR data at a finite outer boundary to future null infinity (after the finite-domain NR simulation has been run).
The second is Cauchy-characteristic matching (CCM), which evolves on a domain of finite extent and a compactified infinite domain simultaneously (see, e.g., the review~\cite{Winicour:2008vpn}). 
CCE and CCM compute gravitational waveforms at future null infinity that include memory effects in a particular BMS frame (see the review~\cite{Mitman:2024uss} for more details about the choice of the frame). 
The NR code SpECTRE~\cite{Moxon:2020gha,Moxon:2021gbv,Ma:2023qjn,spectrecode}, for example, is under development and is implementing both the CCE and CCM algorithms.

\subsection{Memory effects in gravitational theories beyond GR} \label{subsec:introBeyondGR}

Computing memory effects in beyond-GR theories is a newer enterprise.
The oscillatory GWs from compact-binary sources have been studied extensively in a number of beyond-GR theories (see, e.g.,~\cite{Yunes:2013dva,Yunes:2024lzm}), and they have been constrained by the LVK observations~\cite{LIGOScientific:2020tif,LIGOScientific:2021sio}.
Memory effects in beyond-GR theories, however, have been studied in less detail.
There have been a handful of calculations performed in the past few years, including in Brans-Dicke (BD) theory~\cite{Tahura:2020vsa,Tahura:2021hbk,Hou:2020tnd,Hou:2020wbo} and dynamical Chern-Simons gravity~\cite{Hou:2021oxe,Hou:2021bxz}, which were studied using the Bondi-Sachs framework. 
Post-Newtonian calculations of the displacement and spin memory have been performed in BD theory~\cite{Tahura:2021hbk} for compact binary inspirals.\footnote{Note that black holes in many scalar-tensor theories satisfy no-hair theorems (see, e.g.,~\cite{Capuano:2023yyh}), so BBH systems are not expected to be sources of scalar waves or scalar memory effects. Thus, the calculations in~\cite{Tahura:2021hbk} were most relevant for BH-NS and BNS systems.}
The displacement GW memory effect also was computed in a general scalar-vector-tensor theory~\cite{Heisenberg:2023prj,Heisenberg:2024cjk} (which encompasses a wide class of specific beyond-GR theories) using the Isaacson effective stress-energy tensor~\cite{Isaacson:1968hbi,Isaacson:1968zza} in beyond-GR theories (see also, e.g.,~\cite{Stein:2010pn}). 
There have been linearized-theory calculations of asymptotic symmetries and their connections to GW memory effects in the Lorentz-violating Einstein-\AE{}ther theory~\cite{Hou:2023pfz} and in diffeomorphism-invariant theories containing tensor degrees of freedom that closely resemble those in GR~\cite{Hou:2024exz}.

One obstacle to computing memory effects in beyond GR theories has been the limited number of systematic analyses of the asymptotic field equations, which identify an appropriate space of asymptotically flat solutions and conservation or evolution equations that can be used to compute memory effects (as in GR).
These analytical analyses are also helpful for identifying the asymptotic symmetries, conserved quantities, and other elements of the infrared triangles (e.g.,~\cite{Seraj:2021qja}).
A second limitation is the restricted number of NR simulations of compact-binary mergers in beyond GR theories (a few examples of such simulations in scalar-tensor theories include~\cite{Barausse:2012da,Shibata:2013pra,Palenzuela:2013hsa}).
To use the asymptotic field equations to compute memory effects, NR waveforms that lack the memory can be input into these equations to obtain the unresolved memory signal that is missing from the NR simulations.
Finally, it would be beneficial to have beyond-GR NR simulations that use CCE or CCM to analyze and compare with the conservation-law calculations.

Progress is being made on the NR simulations with and without CCE or CCM.
Specifically, NR simulations of BH-NS binaries in Damour-Esposito-Far\`ese (DEF) theory with a single scalar field~\cite{Damour:1992we} were performed recently.
These simulations started early enough in the inspiral that they could be compared with PN calculations (see~\cite{Ma:2023sok}).
In addition, while most of the development of CCE and CCM has been performed in the context of GR, there has been recent work in~\cite{Ma:2024bed}, which solved the Einstein-Klein-Gordon system by incorporating a massless scalar field into the framework.
This will help with computations of GW memory effects from compact binary coalescences in scalar-tensor (ST) theories of the Bergmann–Wagoner types~\cite{Wagoner1970,Bergmann:1968ve} or in DEF theory, in which the field equations take the form of the Einstein-Klein-Gordon system in a particular conformal frame.

\subsection{Objectives and results of this paper}

To further enable the calculation and interpretation of GW memory effects in beyond-GR theories, it will be important to continue developing the systematic analyses of the asymptotic field equations and to perform PN calculations of the GW memory effects to match with the inspiral portion of the NR simulations.
Our focus in this paper will be to derive the conservation laws and compute the PN memory waveforms in DEF theory with a single scalar field in Bondi-Sachs coordinates.
We focus on DEF theory specifically because of the recent high-accuracy simulations in DEF theory~\cite{Ma:2023sok} and because it is a representative of general ST theories with one scalar field.
We use Bondi-Sachs coordinates, because they are well adapted to the structure of asymptotically flat spacetimes and to the coordinates used in the CCE and CCM frameworks.

In addition, DEF theory contains a rich phenomenology, because of nonlinear interactions of the scalar field with itself, which produce nontrivial strong-field dynamics (such as spontaneous scalarization for neutron stars~\cite{Damour:1993hw}).
These nonlinearities, in principle, could allow for stronger GW memory effects in all polarizations of the waves.
The Bondi-Sachs framework is well adapted for studying nonlinear dynamics of massless fields in the wave zone, and it will allow us to understand what strong-field features of DEF theory also have an imprint on the generation and propagation of GWs far from the source.
Another feature of DEF theory that will prove useful is that it reduces to BD theory when the coupling function in the action between gravity and the scalar field becomes a constant.
This will allow us to more easily compare with our prior work~\cite{Tahura:2020vsa,Tahura:2021hbk}, which we henceforth refer to as Papers I and II, respectively.
In particular, we can more easily determine which features of the memory waveforms are associated with having an additional, noninteracting scalar degree of freedom as in BD theory, and which are associated with the additional nonlinearities in the dynamics of the scalar field as in DEF theory.

Thus, in this paper, we will focus on the similarities and differences of aspects of asymptotically flat spacetimes and memory effects in DEF and BD theories, and we will write the results in a way that highlights the similarities and differences with the results of Papers I and II.
For the memory effects, we will focus on the displacement and spin memory effects, because they are, respectively, the largest amplitude electric and magnetic-parity memory effects from compact binary sources that have been computed in GR (see, e.g.,~\cite{Nichols:2017rqr,Mitman:2020bjf,Grant:2023jhd,Siddhant:2024nft}) as well as in BD theory (Papers I and II). 
We find that most differences in the field equations and memory effects arise from the evolution of the scalar field, especially the subleading (in an expansion in the inverse Bondi radial coordinate) dynamical part of the scalar field.
We also compute how these differences affect the PN inspiral memory waveforms that are computed from compact-binary sources (with at least one NS) on quasicircular orbits.

We now summarize our main findings.
In our calculations of the DEF field equations in Bondi-Sachs coordinates, we found that the evolution and conservation equations, when expanded in inverse Bondi-coordinate radius, can be written in a form identical to those in BD theory; however, to write them as such, the metric functions in the Bondi expansion and the subleading part of the scalar field must be modified by nonlinear terms in the leading dynamical part of the scalar field.
This result has several implications.
For computing the flux (nonlinear) contributions to the displacement and spin memory effects, the formal expressions used will be the same as those in BD theory (see Papers I and II).
However, because the radiative data that enters into these calculations for a given source in the two theories will not be the same, the time dependence of the memory signals will generically differ.

As an example of this different time dependence, we discuss the calculation of the PN memory signal during the quasicircular inspiral of a compact binary, which is most relevant for BH-NS or BNS systems.
There we find that there would be difference in the Newtonian-order memory signals in BD and DEF theories.
However, the differences arise at quadratic order in a coupling parameter that is of order $10^{-5}$.
Thus, we expect that differences in the inspiral memory signals between BD and DEF theory (at least during the inspiral) would be too small for current or future GW detectors to measure.
We do not perform a detection study of the memory in DEF theory using the PN waveforms because, as in GR, the amount of signal-to-noise ratio (SNR) accumulated during the inspiral is a small fraction of the total SNR, which comes about primarily during the merger and ringdown or disruption (see, e.g.,~\cite{Lasky:2016knh,Tiwari:2021gfl}).
Given that there are not enough full NR waveforms in DEF theory available, we do not perform such a detection study either.
However, our results in this paper will allow us to perform such studies once the NR waveforms are available.

\subsection{Organization and conventions of this paper}

The organization of this paper is as follows:
In Sec.~\ref{sec:BSframework}, we review aspects of ST theories in the Einstein frame in Bondi-Sachs coordinates.
We use the Einstein-frame results to determine appropriate asymptotically flat boundary conditions and the corresponding form of the metric in the Jordan frame.
We then compute the field equations of DEF theory in the Jordan frame and derive asymptotically flat solutions. 
In Sec.~\ref{sec:memory-effects}, we discuss the leading and subleading tensor and scalar memory effects in DEF theory.
We then describe how the tensor effects can be computed using the conservation equations for the mass and angular momentum aspects. 
The scalar effects are computed from charge contributions, and the subleading scalar has a new nonlinear contribution in DEF theory that we discuss.
We also describe how the PN inspiral memory signals from compact binaries differ between DEF and BD theories and how our results could be used to study the detection prospects of the memory effects, once NR waveforms in DEF theory are available.
We present our conclusions in Sec.~\ref{sec:Conclusions}. 
Finally, we discuss BMS symmetries and the transformation of the Bondi metric functions and scalar fields under these symmetries in Appendix~\ref{sec:Symmetries}.

Throughout this paper, we use units in which the speed of light is $c=1$ and in which the gravitational constant in the Einstein frame satisfies $\tilde G_E=1$. 
We use the conventions for the metric and curvature tensors given in~\cite{Misner:1974qy}. 
Greek indices ($\mu,\nu,\alpha, \dots$) represent four-dimensional spacetime indices, and uppercase Latin indices $(A, B, C, \dots)$ represent indices on the 2-sphere. 
We used the \textsc{xAct} suite~\cite{xAct,Martin-Garcia:2008ysv} to perform some of the calculations.
A \textsc{Mathematica} notebook with these calculations is available in a GitHub repository~\cite{def_jordan}.

\section{Bondi-Sachs framework for DEF theory}\label{sec:BSframework}

In this section, we review the formulation of DEF theory and give the forms of the field equations in Bondi-Sachs coordinates. 
We will typically refer to the theory as DEF theory, though we may also refer to it as ST theory, as is done in~\cite{Ma:2023sok}. 
First, we give a brief overview of DEF theory in the Einstein frame in Sec.~\ref{sec:EinsteinFrame}; the form of the field equations and the assumptions that we make for the boundary conditions for asymptotic flatness are identical to those in BD theory (Paper I and~\cite{Hou:2020tnd}).
Next, we transform to the Jordan frame to obtain the corresponding fall-off conditions of the Bondi metric functions and the scalar field; the transformation is of precisely the same form as Paper I in BD theory.
Finally, we compute the field equations in the Jordan frame and derive asymptotically flat solutions in Sec.~\ref{sec:JordanFrame}. 

As discussed in more detail in Paper I, we follow this approach, because it is simplest to determine asymptotically flat fall-off conditions in the Einstein frame, where the field equations resemble the Einstein-Klein-Gordon system.
However, matter fields follow geodesics of the Jordan frame metric, so it is more straightforward to interpret a GW detector's response in the Jordan frame.
Thus, we found it most straightforward to transform the fall-off conditions on the fields to the Jordan frame and directly solve the field equations in the Jordan frame.

\subsection{Einstein-frame field equations and fall-off conditions} \label{sec:EinsteinFrame}

In the Einstein frame, the action of a ST theory in the absence of additional matter fields takes the same form as in BD theory~\cite{Wagoner1970,Berti:2015itd}:
\be\label{Eq:EinAction}
S=\int d^{4} x \sqrt{-\tilde{g}}\left[\frac{\tilde{R}}{16 \pi}-\frac{1}{2} \tilde{g}^{\rho \sigma}\left(\tilde{\nabla}_{\rho} \Phi\right)\left(\tilde{\nabla}_{\sigma} \Phi\right)\right] .
\ee
We use the notation $\tilde g_{\mu\nu}$ for the metric in the Einstein frame, $\tilde R$ for the Ricci scalar, and $\tilde{\nabla}_\mu$ for the covariant derivative compatible with $\tilde g_{\mu\nu}$. 
The field $\Phi$ is a massless real scalar, and we set the ``bare'' gravitational constant\footnote{The phrase ``bare'' gravitational constant refers to the constant that appears in the action. 
The locally measured value of Newton's constant can differ from the bare one in modified theories of gravity.} $\tilde G_E$ in the Einstein frame equal to one. 
Varying the action with respect to the metric and the scalar field gives the field equations of DEF theory:
\begin{subequations}\label{Eq:Field_Eq_Ein}
\be
\tilde G_{\mu\nu} = 8\pi \tilde{T}_{\mu \nu}^{(\Phi)} , \qquad 
\tilde \nabla_{\mu}\tilde \nabla^{\mu} \Phi = 0 .
\ee
Here $\tilde G_{\mu\nu}=\tilde R_{\mu\nu}-(1/2)\tilde R \, \tilde g_{\mu\nu}$ is the Einstein tensor, and the stress-energy tensor $\tilde T_{\mu\nu}^{(\Phi)}$ has the usual form for that of a massless, real scalar field $\Phi$:
\be
\tilde T_{\mu \nu}^{(\Phi)} = \tilde{\nabla}_{\mu} \Phi \tilde{\nabla}_{\nu} \Phi-\tilde{g}_{\mu \nu}\left[\frac{1}{2} \tilde{g}^{\alpha \beta} \tilde{\nabla}_{\alpha} \Phi \tilde{\nabla}_{\beta} \Phi\right] .
\ee
\end{subequations}

We now introduce Bondi coordinates $(\tilde u,\,\tilde r,\,\td x^A)$, which are well adapted for studying outgoing radiation from an isolated source.
The variable $\td u$ is the retarded time, $\td r$ is a radial coordinate, and $\td x^A$ are coordinates on two-sphere surfaces of constant $\tilde u$ and $\td r$ ($A$ can be either $1$ or $2$). 
The explicit form of the metric is (see~\cite{Bondi:1962px,Sachs:1962wk,Madler:2016xju})
\begin{align} \label{eq:Metric_Ein}
ds^2 = & -\frac{\td V}{\td r} e^{2 \td \beta} d \td u^{2}-2 e^{2 \td \beta} d \td u d \td r \nonumber \\
& + \td r^{2} \td h_{A B}\left(d \td x^{A}-\td U^{A} d \td u\right)\left(d \tilde x^{B}-\tilde U^{B} d \tilde u\right) .
\end{align}
The four Bondi metric functions---$\tilde V$, $\tilde \beta$, $\tilde U^A$ and $\td h_{AB}$, which depend on all four Bondi coordinates---contain six degrees of freedom, because the symmetric tensor $\td h_{AB}$ contains only two degrees of freedom when the determinant condition of Bondi gauge is imposed: namely,
\begin{equation}
    \det\left[\td h_{A B}\right] = q\left(\td x^{C}\right).
\end{equation}
The function $q\left(\td x^{C}\right)$ is commonly selected to be the determinant of the round two-sphere metric $q_{AB}$, which gives $\tilde r$ the interpretation of being an areal radius.
The form of the metric in Eq.~\eqref{eq:Metric_Ein} already has the remaining three independent Bondi gauge conditions imposed:
\be 
\td g_{\td r A}=\td g_{\td r \td r}=0 .
\ee
These conditions ensure that the hypersurfaces of constant $\tilde u$ are null and that the angular coordinates $\tilde x^A$ are constant along outgoing null rays. 

Note that the field equations in Eq.~\eqref{Eq:Field_Eq_Ein} have the same form as the Einstein-Klein-Gordon system in GR with a massless real scalar field, precisely as in BD theory. 
We thus assume the same fall-off for $\Phi$ as a power series expansion in $1/\td{r}$, with the leading-order piece being constant, as in Paper I:
\ba\label{eq:Phi-Ein-off}
\Phi\left(\td u, \td r, \td x^{A}\right) = \Phi_{0} + \frac{\Phi_{1}\left(\td u, \td x^{A}\right)}{\td r} + O(\td r^{-2}) .
\ea
We also impose the same fall-off conditions on metric functions as in Paper I:
\begin{subequations}\label{eq:Met_Ein_falloff}
\begin{align}
\td h_{AB} = {} & q_{AB}(\td x^C)+\frac{\td c_{AB}(\td u,\, \td x^C)}{\td r}+\mathcal{O}\lb \frac{1}{\td r^2}\rb, \\
\td\beta = {} & \frac{\tilde\beta_1(\td u,\, \td x^C)}{\td r}+\mathcal{O}\lb \frac{1}{\td r^2}\rb\,,\\
\label{eq:UEinstein}
\td U^A = {} & \frac{\td U_2^A(\td u,\, \td x^C)}{\td r^2}+\m O \lb\frac{\log \td r}{\td r^3}\rb,\,\\
\td V = {} & \td r+\td V_0(\td u,\, \td x^C)+\mathcal{O}\lb \frac{1}{\td r}\rb\,.
\end{align}
\end{subequations}
The coefficient of the $1/\td{r}$ piece of conformal 2-metric, $\td c_{AB}$, is the shear tensor, which is traceless with respect to $q_{AB}$ (i.e., $q^{AB}\td c_{AB}=0$); this follows from the determinant condition of Bondi gauge. 
The shear tensor encodes the two transverse-traceless degrees of freedom of tensor GWs. 

Because the fall-off conditions and the evolution equations have precisely the same form as in BD theory, the solutions to the field equations will also have the same form.
For example, they give that $\tilde \beta_1 = 0$ and $\td U_2^A=-(1/2)\eth_B \td c^{AB}$, where $\eth_B$ is covariant derivative compatible with $q_{AB}$.
A more detailed discussion is given in Paper I.
As in Paper I, the detailed form of the equations is not essential for the remaining calculations, because we use the Einstein-frame results to determine the fall-off conditions in the Jordan frame.
We will solve the field equations in the Jordan frame rather than transforming the full solution in the Einstein frame to the Jordan frame.

\subsection{Conformal transformation to the Jordan frame and Jordan-frame fall-off conditions} \label{subsec:JordanBCs}

The spacetime metric in the Einstein and Jordan frames are related by a conformal transformation~\cite{Damour:1992we,Carroll:2004st}:
\be\label{Eq:Conf_Transf}
\td g_{\mu\nu}=\lambda g_{\mu\nu} .
\ee
The function $\lambda(x^\mu)$ is the scalar field in the Jordan frame, and in generic scalar-tensor theories, it is related to $\Phi$ through a differential equation involving a ``coupling'' function $\omega(\lambda)$:
\ba\label{eq:omega-to-lambda-Phi}
\frac{d\Phi}{d\log\lambda} = \sqrt{\frac{3+2\omega(\lambda)}{16\pi}} .
\ea
By postulating the relationships in Eqs.~\eqref{Eq:Conf_Transf} and~\eqref{eq:omega-to-lambda-Phi}, the Einstein-frame action in Eq.~\eqref{Eq:EinAction} can be transformed to the action in the Jordan frame,
\be
\label{eq:JordanAction}
S = \frac 1{16\pi} \int d^{4} x \sqrt{-g}\left[ \lambda  R- \frac{\omega(\lambda)}{\lambda} g^{\mu \nu} \nabla_{\mu} \lambda  \nabla_{\nu} \lambda  \right] 
\ee
(see, e.g.,~\cite{Damour:1992we}).
Here $\nabla_\mu$ is the covariant derivative compatible with the Jordan frame metric $g_{\mu\nu}$ and $R$ is the corresponding Ricci scalar. 
Given that we chose $\tilde G_E=1$ in the Einstein frame, the scalar field is dimensionless (rather than having units of the inverse of Newton's constant, as in BD theory~\cite{Brans:1961sx}).

In BD theory, $\omega(\lambda)$ is assumed to be a constant, which allows Eq.~\eqref{eq:omega-to-lambda-Phi} to be integrated and $\log\lambda$ is proportional to $\Phi$ minus its asymptotic value.
In DEF theory, a similar relationship between $\log\lambda$ and $\Phi$ is postulated, but instead of assuming that $\log \lambda$ is linear in $\Phi$, the Einstein-frame scalar $\Phi$ is Taylor expanded to quadratic order around its asymptotic value $\Phi_0$~\cite{Damour:1993hw}.
We use the notation of~\cite{Ma:2023sok} to write the expression as
\be\label{eq:lambda-to-Phi}
\log \lambda= -4\sqrt{\pi}\alpha_0(\Phi-\Phi_0)-4\pi \m B_0(\Phi-\Phi_0)^2 ,
\ee
where $\alpha_0$ and $\m B_0$ are the two coupling constants of DEF theory. 
By computing $d\log\lambda/d\Phi$ from Eq.~\eqref{eq:lambda-to-Phi}, the inverse can be substituted into the differential equation~\eqref{eq:omega-to-lambda-Phi} and solved for $\omega(\lambda)$, with $\Phi(\lambda)$ being given implicitly (see~\cite{Ma:2023sok}): 
\be\label{eq:omega-to-Phi}
\omega(\lambda) = -\frac{3}{2}+\frac{1}{2 \left[\alpha_0+2 \sqrt{\pi } \m B_0 (\Phi -\Phi_0)\right]^2}\,.
\ee
Note that DEF theory reduces to BD theory when $\m B_0$ goes to zero and $\alpha_0=-(3+\omega_{\text{BD}})^{-1/2}$, and that it reduces to GR when $\m B_0=\alpha_0=0$.

We now use the expansion of the Einstein-frame scalar field and metric functions, the conformal transformation of the metric, and the relationship between $\lambda$ and $\Phi$ in Eq.~\eqref{eq:lambda-to-Phi} to determine the asymptotic boundary conditions on the fields  $\lambda$ and $g_{\mu\nu}$ in the Jordan frame. 
Substituting Eq.~\eqref{eq:Phi-Ein-off} into \eqref{eq:lambda-to-Phi} and exponentiating gives
\be \label{eq:lambda-Phi1}
\lambda=1-\frac{4 \sqrt{\pi } \alpha_0 \Phi_1}{\td r}+\m O \lb \td r^{-2}\rb.
\ee
Given that the differences between DEF and BD theories arise at quadratic order in the scalar field, it is reasonable that the scalar field has the same asymptotic behavior as in BD theory: namely, it goes as a constant plus a function of the coordinates $\td u$ and $\td x^A$ normalized by $\td r$.
Note, however, that the leading order piece of $\lambda$ has been chosen to be one here, whereas in Paper I it was allowed to be an arbitrary constant $\lambda_0$. 
Thus, to compare results here with those in Paper I, one should set $\lambda_0=1$ in the results of Paper I.\footnote{In ST theories, the value of $\lambda_0$ is related to the asymptotic value of the gravitational constant, $G$ by $G=(4+2\omega_0)/[\lambda_0(3+2\omega_0)]$. 
To set $G=1$ as $r\rightarrow\infty$, then $\lambda_0$ would be different from one, in general (see, e.g., the discussion after Eq.~(4.6) of Ref.~\cite{Mirshekari:2013vb} and also footnote 5 of Paper I). 
Note, however, that the GR limit is $\omega_0\rightarrow\infty$, which leads to $G=1$ with $\lambda_0=1$. In this paper, we set $\lambda_0 = 1$ because we do not require the asymptotic value of $G$ to be 1 at infinity. \label{fn:G-value}}
Thus, the asymptotic boundary conditions on the fields in the Jordan frame in DEF theory will be the same as those in BD theory, which are described in more detail in Paper I.

We now remind the reader that when performing the conformal transformation in Eq.~\eqref{Eq:Conf_Transf} to obtain the Jordan-frame metric, the Bondi gauge conditions $g_{\td r\td r}=g_{\td r A}=0$ are satisfied, but the determinant condition is not satisfied: namely the determinant $\det[g_{AB}] = \tilde r^4 \lambda^{-2} q(x^B)$ is no longer independent of $\td u$ and its $\td r$ dependence is no longer an overall multiplicative factor of $\tilde r^4$. 
To restore the determinant condition, we use the same transformation used in BD theory in Paper I, where we define new coordinates by
\be\label{eq:Coord_Transf}
u=\td u,\quad r=\frac{\td r}{\sqrt{\lambda}},\quad x^A=\td x^A .
\ee
In the new set of coordinates, the Jordan frame metric can be written in Bondi-Sachs form as
\begin{align}\label{eq:Met_Jordan}
g_{\mu \nu} d x^{\mu} d x^{\nu} = & - \frac{V}{r} e^{2 \beta} d u^{2}-2 e^{2 \beta} d u d r \nonumber \\
& + r^{2} h_{A B}\left(d x^{A}-U^{A} d u\right)\left(d x^{B}-U^{B} d u\right) ,
\end{align}
where $\det[h_{AB}] = q(x^C)$.
Using the conformal transformation in Eq.~\eqref{Eq:Conf_Transf}, the coordinate transformation in Eq.~\eqref{eq:Coord_Transf}, and the boundary conditions on the metric and scalar field in the Einstein frame [given in Eqs.~\eqref{eq:Met_Ein_falloff} and~\eqref{eq:Phi-Ein-off}], we find the Jordan frame metric functions and scalar field should have the following scaling with $r$:
\begin{subequations} \label{eq:Jordan-fall-off}
\begin{align}
\lambda = {} &1+\frac{\phi}{r}+\m O \lb\frac{1}{r^2}\rb , \\
h_{AB} = {} & q_{AB}+\frac{c_{AB}}{r}+\m O \lb\frac{1}{r^2}\rb , \\
\label{eq:Beta_Ein_To_Jordan}
\beta = {} & -\frac{1}{2r}\phi+\m O \lb\frac{1}{r^2}\rb\,,\\
U^{A} = {} & \frac{1}{2r^2}(\eth^A \phi-\eth_B c^{AB})+\m O \lb\frac{\log r}{r^3}\rb,\\
V = {} & (1+\partial_u\phi)r+\m O (r^0)\,.
\label{eq:V_Ein_To_Jordan}
\end{align}
\end{subequations}
Two-sphere indices were raised with $q^{AB}$ in the expressions above.
The two-sphere tensor $c_{AB}$ is trace-free with respect to $q_{AB}$ (i.e., $q^{AB}c_{AB}=0$), and it has the same form as in the Einstein frame [i.e., $c_{AB}(u,x^A)=\td c_{AB}(\td u, \td x^A)$].
We have also defined $\phi=-4 \sqrt{\pi } \alpha_0\Phi_1$.

Because we will solve the Jordan-frame field equations in the next section, the emphasis here is not on determining the detailed relationship between the Einstein-frame and Jordan-frame metric quantities and scalar fields (the leading-order results are the same as in BD theory; see Paper I).
Rather, we are just determining what are the expected fall off rates in $r$ of the metric functions in the Jordan frame.
For example, by enforcing the determinant condition of the Bondi gauge in the coordinates ($u,\,r,\,x^A$), the leading order part of the metric component $g_{uu}$ becomes a function of $u$ and $x^A$ (this would not occur if we used a gauge that did not enforce this condition; see~\cite{Hou:2020tnd}).
Because scalar-tensor theories admit a scalar ``breathing-mode'' polarization of GWs, the time dependence of the leading part of $g_{uu}$ is where this radiative degree of freedom enters the metric if the Bondi determinant condition is enforced; the curvature tensors and other properties of the solution are consistent with our notions of asymptotically flat boundary conditions defined in the earlier parts of this section.

\subsection{Jordan-frame field equations}\label{sec:JordanFrame}

In this section, we solve the field equations in Jordan frame using the Bondi coordinates and the boundary conditions on the fields that were discussed in Sec.~\ref{subsec:JordanBCs}. 
We first give the field equations in the Jordan frame, and then we determine solutions to these equations in which the fields are expanded as a series in $1/r$. 

\subsubsection{Field equations in Bondi gauge}
 
The modified Einstein equations are obtained from varying the Jordan-frame action in Eq.~\eqref{eq:JordanAction} with respect to the inverse metric.
The scalar field equation is obtained from varying the action with respect to $\lambda$ (up to boundary terms) and combining the result with the trace of the modified Einstein equations.
The resulting field equations are of the form
\begin{subequations} 
\begin{align} \label{eq:Mod_Ein_Jordan_1}
G_{\mu\nu}
&=\frac{1}{\lambda}\lb 8\pi T_{\mu\nu}^{(\lambda)}+\nabla_{\mu}\nabla_{\nu}\lambda - g_{\mu\nu}\Box \lambda\rb , \\
\label{eq:scalar-field-eq}
\Box \lambda&=-\frac{1}{3+2\omega(\lambda)} \frac{d\omega}{d\lambda} \nabla_{\mu}\lambda \nabla^{\mu}\lambda .
\end{align}
\end{subequations}
Here $G_{\mu\nu}=R_{\mu\nu}-(1/2)R \, g_{\mu\nu}$ is the Einstein tensor, and $ \Box= g^{\mu\nu} \nabla_\mu \nabla_\nu$ is the wave operator constructed using the Jordan frame metric. 
The stress-energy tensor of $\lambda$ is given by 
\be\label{eq:StressEnergyJordan}
T_{\mu\nu}^{(\lambda)}=\frac{\omega(\lambda)}{8 \pi\lambda} \left(\nabla_{\mu}\lambda\nabla_{\nu}\lambda-\frac{1}{2}g_{\mu\nu}\nabla^{\alpha}\lambda\nabla_{\alpha}\lambda\right) .
\ee

The modified Einstein equations~\eqref{eq:Mod_Ein_Jordan_1} have the same form as the ones in BD theory, except the factor of $\omega$ is now $\lambda$-dependent. 
The scalar field equation in BD theory is  $\Box\lambda=0$, which is consistent with the fact that $\omega$ is independent of $\lambda$ in BD theory.
In DEF theory, there is a nonzero term on the right side, which depends on $\lambda$. 
It will be at least quadratic in $\lambda$ and its derivatives.

To help in referring to specific components of the modified Einstein equations, we introduce the notation
\be\label{eq:Mod_Ein_Jordan}
\m G_{\mu\nu} \equiv G_{\mu\nu}-\frac{1}{\lambda}\lb 8\pi T_{\mu\nu}^{(\lambda)} -\nabla_{\mu}\nabla_{\nu}\lambda - g_{\mu\nu}\Box \lambda\rb = 0 .
\ee
For the subsequent calculations, it will also be useful to write~\eqref{eq:omega-to-Phi} explicitly as a function of $\log\lambda$ rather than implicitly, as it is currently written.
To do so, we solve the quadratic equation for $\Phi - \Phi_0$ given in Eq.~\eqref{eq:lambda-to-Phi}.
The solution consistent with our assumptions for $\Phi$ and $\lambda$ in Eqs.~\eqref{eq:Phi-Ein-off} and~\eqref{eq:lambda-to-Phi} is given by
\begin{equation} \label{eq:Phi-Phi0-sol}
    \Phi - \Phi_0 = \frac{\alpha_0}{2\sqrt\pi \mathcal B_0} \left(-1 + \sqrt{1-\frac{\mathcal B_0}{\alpha_0{}^2}\log\lambda}\right).
\end{equation}
Substituting this expression into Eq.~\eqref{eq:omega-to-Phi} gives the result for $\omega$ as a function of $\lambda$.
Given that $\lambda$ has an expansion of the form in Eq.~\eqref{eq:lambda-Phi1}, then $\log\lambda$ will be a small parameter.
We can then Taylor expand the result in Eq.~\eqref{eq:Phi-Phi0-sol} in $\log\lambda$; at the order in the expansion in $1/r$ in which we work in the remainder of the paper, it will be sufficient to work to linear order in $\log\lambda$.
Thus, we have that 
\be\label{eq:omega-to-lambda}
\omega(\lambda) =  -\frac{3}{2} + \frac{1}{2\alpha_{0}^2} + \frac{\mathcal B_0}{2\alpha_0{}^4} \log\lambda + \mathcal O[(\log\lambda)^2].
\ee

Because the modified Einstein equations~\eqref{eq:Mod_Ein_Jordan_1} in DEF theory have the same form as in BD theory when the constant $\omega$ of BD theory is replaced by the function $\omega(\lambda)$, the field equations formally have the same form under this replacement.
The explicit form of the field equations in the Jordan frame were given in the Appendix of Paper I. 
Because there are some subtleties related to integrating the ``hypersurface equations'' that were not discussed there, we describe this in more detail below.

First, the $\m G_{rr}=0$ components of the field equations give rise to a hypersurface-type equation
\begin{align} \label{eq:Err} 
    & 2\left(\frac{2}{r}+\partial_r\log\lambda \right) \partial_r\beta = \nonumber \\
    & \frac{1}{\lambda}\partial^2_r\lambda+\frac{\omega}{\lambda^2}(\partial_r\lambda)^2
+\frac{1}{4}h^{AB}h^{CD}\partial_r h_{AC}\partial_r h_{BD} .
\end{align}
For general solutions (those that are not necessarily asymptotically flat), the term multiplying $\partial_r \beta$ could be zero, which prevents one from generically dividing by this term and integrating $\partial_r\beta$ to obtain $\beta$ as a first integral of $\lambda$, $h_{AB}$ and their radial derivatives on a surface of fixed retarded time $u$.
However, for the asymptotically flat solutions that we consider in this paper, the term $\partial_r \log\lambda$ will scale as $1/r^2$, so the multiplicative factor will be nonzero for finite $r>0$, and Eq.~\eqref{eq:Err} can be integrated as a hypersurface equation as in GR or in DEF theory in the Einstein frame.

Next, the $\m G_{rA}=0$ components of the field equations are those that produce a hypersurface equation for $U^A$.
The form of this equation is given by
\begin{align} \label{eq:ErA}
    & \frac{1}{2r^2}\left(\partial_r + \partial_r \log \lambda \right)(r^4 e^{-2\beta} h_{AB}\partial_r U^B) = \nonumber \\
    & r^{2} \partial_r\left(\frac{1}{r^{2}} D_{A} \beta\right)
- \frac{1}{2} h^{B C} D_{B}\left(\partial_r h_{A C}\right) -\frac{D_A\lambda}{\lambda r} \nonumber \\
& +\frac{\omega}{\lambda^2}\partial_r\lambda D_A \lambda - \frac{1}{\lambda}D_A\beta\partial_r\lambda
- \frac{1}{2\lambda} h^{BC} D_B\lambda\partial_r h_{AC} \nonumber \\
& +\frac{1}{\lambda}\partial_rD_A\lambda .
\end{align}
Here $D_A$ is the covariant derivative compatible with metric $h_{AB}$.
Note that unlike in GR or in DEF theory in the Einstein frame in which the only term in the differential operator acting on $r^4 e^{-2\beta} h_{AB}\partial_r U^B$ is the $\partial_r$ term, there is now a second term involving $\partial_r \log\lambda$ on the left-hand side.
Thus, in general, to solve for $\partial_r U^B$ (and subsequently $U^B$) one would need to invert the more complicated differential operator that acts on $r^4 e^{-2\beta} h_{AB}\partial_r U^B$ (instead of just integrating with respect to $r$ as in GR or DEF theory in the Einstein frame).
However, for the asymptotically flat boundary conditions that we consider in this paper, the second term involving $\partial_r \log\lambda$ will be subleading to the first term; thus, the hierarchy for solving for a given order in $1/r$ of the expansion of $U^B$ will be determined by the expansions of $h_{AB}$, $\lambda$, $\beta$, and lower order parts of $U^B$.
This is a slight modification of the typical description of the hierarchy of the hypersurface equations.

The last of the hypersurface equations is the one for $V$, which can be obtained from the trace of the components of the field equations $\m G_{AB}=0$.
As with the previous hypersurface equations, the differential operator acting on $V$ will now be $\lambda$-dependent (unlike in GR or in DEF theory in the Einstein frame, but similarly to BD theory in the Jordan frame).
In addition, the wave operator acting on the scalar field $\Box \lambda$ enters into the hypersurface equation for $V$.
In BD theory, assuming the scalar-field equation of motion is satisfied in vacuum ($\Box \lambda = 0$) then this term vanishes; in DEF theory, one can instead replace $\Box \lambda$ with the right-hand side of Eq.~\eqref{eq:scalar-field-eq} when the scalar-field equation is satisfied. 
Thus, the hypersurface equation for $V$ can be written in the form
\bw
\begin{align} \label{eq:EABtrace}
    2\left[\partial_r + \partial_r\log\lambda \left(1 - \frac 14 \frac{\partial \log(2\omega+3)}{\partial \log r} \right)\right] V = {} & e^{2\beta}[\mathscr R -2h^{AB}\lb D_A D_B\beta+D_A\beta D_B\beta\rb ] + \frac{1}{r^2} D_A\partial_r(r^4U^A) \nonumber \\
    &-\frac{1}{2}r^4e^{-2\beta}h_{AB}\partial_rU^A\partial_rU^B -\frac{e^{2\beta} h^{AB}}{\lambda^2} (\omega D_{A}\lambda D_{B}\lambda + \lambda D_{B}D_{A}\lambda) \nonumber \\
    & + \frac r\lambda \left[2 (\partial_u \lambda +  U^A D_A \lambda) + r(D_A U^A) \partial_r \lambda \right]  \nonumber \\
    & - \frac{d\omega/d\lambda}{\lambda(2\omega+3)} [ 2 r^2 \partial_u \lambda \partial_r\lambda + 2 r^2 \partial_r \lambda U^A D_A \lambda - e^{2\beta} h^{AB} D_A \lambda D_B \lambda] .
\end{align}
\ew

We introduced $\mathscr R$ to be the Ricci scalar of the metric $h_{AB}$. We also used the relation $\partial_r\omega (\lambda)=(\partial_r\lambda)d\omega/d\lambda$.
In addition to the more complicated radial differential operator that needs to be inverted, note that the retarded time derivative, $\partial_u \lambda$, also appears in the hypersurface equation for $V$.
This occurs both in BD theory and DEF theory in the Jordan frame (see the first term in the third line).
The hypersurface equations in GR, or in the Einstein frame for DEF theory, can be integrated using the metric functions $\beta$, $U^A$ and $h_{AB}$ (and the scalar field $\lambda$ for DEF theory) and spatial derivatives of these fields.
In both BD and DEF theories information about how the scalar field changes off of a hypersurface of constant $u$ is required to integrate the hypersurface equations.
In the Jordan frame, one would generically need to simultaneously solve the hypersurface equation with the evolution equation for the scalar field.
Thus, we give this evolution equation next.

To write the scalar field equation, it is helpful to note that in~\cite{Grant:2021hga}, it was observed that the evolution equation for the trace-free part of $h_{AB}$, rather than being interpreted as an evolution equation, could also be considered to be a hypersurface equation for the trace-free part of $\partial_u h_{AB}$.
In this spirit, and given that $\partial_u \lambda$ appears in the hypersurface equations, it is convenient to write the scalar-field evolution equation in this form, as follows:
\bw
\begin{align} \label{eq:scalar-full}
2 r^2 \left[\partial_r + \frac 1r\left(1 + \frac 12 \frac{\partial \log(2\omega+3)}{\partial \log r} \right) \right] \partial_u\lambda = & -2 r U^{A} (D_{A}\lambda + r D_{A}\partial_r\lambda) + e^{2 \beta} h^{AB} (2 D_{A}\beta D_{B}\lambda + D_{B}D_{A}\lambda) \nonumber \\
& -  r^2 D_{A}\lambda (\partial_r U^{A}) - r^2 D_{A}U^{A} \partial_r\lambda + (V + r \partial_rV) \partial_r\lambda  + r V \partial_r^2\lambda \nonumber \\
& -\frac{d\omega /d\lambda }{(3 + 2 \omega)}\left[- r (\partial_r\lambda)^2 V -  e^{2 \beta} h^{AB} D_{A}\lambda D_{B}\lambda + 2 r^2 (U^{A} D_{A}\lambda) \partial_r\lambda \right] .
\end{align}
\ew
Writing the scalar field equation in this form highlights the similar structure it has with the hypersurface equation for $V$ in Eq.~\eqref{eq:EABtrace}.
Without the assumption of asymptotic flatness, Eqs.~\eqref{eq:EABtrace} and~\eqref{eq:scalar-full} would need to be solved as two coupled hypersurface equations for $V$ and $\partial_u \lambda$.

With the asymptotically flat boundary conditions assumed in Sec.~\ref{subsec:JordanBCs}, it will be possible to solve for the leading-order in $1/r$ evolution of the scalar field without needing $V$; then $V$ can be solved in terms of the leading part of $\partial_u \lambda$.
This process can be iterated to obtain higher orders in $1/r$ of $\partial_u\lambda$ and then higher orders in $1/r$ of $V$.
We describe the asymptotic solution of these hypersurface equations in more detail in the next part.

We do not give the evolution equations (the trace-free part of $\mathcal G_{AB}=0$) or the supplementary equations ($\mathcal G_{uu}=0$ and $\mathcal G_{uA}=0$) in full generality here.
The expressions are lengthy, and in the solution with asymptotically flat boundary conditions, we will work at an order in $1/r$ in which there are no differences from the results in BD theory given in Paper I.

\subsubsection{Asymptotically flat solutions in the Jordan frame}

Given the structure of the hypersurface equations described above, if we postulate a form for the expansion of the 2-metric $h_{AB}$ and $\lambda$, then we can solve the hypersurface equations to determine corresponding expansions of $\beta$, $U^A$, $\partial_u\lambda$ and $V$.
The evolution equations and supplementary equations give information about the dynamics of the metric functions and the scalar field.

Based on the fall-off conditions posed in the Einstein frame, which we transformed to the Jordan frame, we write, as in BD theory, the expansion of the 2-metric $h_{AB}$ and scalar field $\lambda$ as
\begin{subequations}\label{Eq:Jordan_FallOff}
\ba
&&h_{AB}=q_{AB}+\frac{c_{AB}(u,x^A)}{r}+\frac{d_{AB}(u,x^A)}{r^2}+ \m O (r^{-3}),\nonumber\\ \\
&&\lambda(u,r,x^A) =1 + \frac{\lambda_1\left(u,x^A\right)}{r} + \frac{\lambda_2\left(u,x^A\right)}{r^2}+ \m O (r^{-3})\,.\nonumber\\
\label{Eq:lambda_Exp}
\ea
\end{subequations}
The determinant condition of Bondi gauge requires that $q^{AB}c_{AB}=0$.  
It also constrains the $d_{AB}$ trace portion (with respect to $q_{AB}$), but it does not constrain the trace-free part.
We thus write $d_{AB}$ as
\be
d_{AB}=D_{AB}+\frac{1}{4}c_{CD}c^{CD}q_{AB} ,
\ee
with $ q^{AB}D_{AB}=0$.

Because $\omega$ is a function of $\lambda$, then the expansion in Eq.~\eqref{Eq:lambda_Exp} allows us to write $\omega$ as a series in $\lambda$.
Substituting Eq.~\eqref{Eq:lambda_Exp} into Eq.~\eqref{eq:omega-to-lambda} gives
\be \label{eq:omega-expand}
\omega=\omega_{0}{} + \frac{\omega_{1}{}}{r} + \m O(r^{-2}) ,
\ee
with the coefficients in the expansion being
\begin{equation}
\label{eq:omega-to-alpha}
\omega_0 = \frac{1}{2} \left(-3 + \frac{1}{\alpha_{0}^2}\right) , \qquad
\omega_1 = \frac{\mathcal{B}_{0}}{2\alpha_{0}{}^4}  \lambda_{1}.
\end{equation}
The fact that $\omega$ in Eq.~\eqref{eq:omega-expand} at leading order is a constant and differs from a constant at order $1/r$ allows us to estimate the orders in $1/r$ at which asymptotically flat solutions of DEF theory will first differ from those of BD theory.

Specifically, the right-hand side of Eq.~\eqref{eq:scalar-field-eq} for the evolution of the scalar field will be of order $1/r^3$.
This implies that the evolution equation for $\lambda_1$ will be the same in BD and DEF theories, but the evolution of $\lambda_2$ will differ.\footnote{In more detail, $\partial_u \lambda_1$ will be unconstrained ``radiative data'' in both BD and DEF theories. 
Having $\lambda_1$ being the free radiative data in both theories, however, does not imply that the value of $\lambda_1$ produced by ``equivalent'' sources would be the same. 
As we will discuss in Sec.~\ref{subsec:PNbinries}, for example, the tensor and scalar gravitational waveforms from compact binaries with the same numerical values of the masses and sensitivities in BD and DEF theories will differ.\label{fn:unaffected}}
The fact that $\lambda_2$ differs from BD theory then allows us to determine the orders in $1/r$ for which there are differences in the stress-energy tensor $T_{\mu\nu}^{(\lambda)}$ and in the remaining terms on the right-hand side of the modified Einstein equation~\eqref{eq:Mod_Ein_Jordan_1}.
The leading-order terms will come from those linear in $\lambda$ in Eq.~\eqref{eq:Mod_Ein_Jordan_1}, specifically from the $\lambda_2$ terms which evolve differently in BD theory from in DEF theory.
Given its scaling with $1/r$, we anticipate differences to occur at order $1/r^2$ in $\beta$, $1/r^3$ in $U^A$, and $1/r^2$ in $h_{AB}$; they would also be expected to affect the conservation equations for the mass aspect and angular momentum aspect, which will be introduced later.\footnote{These are just expectations based on scaling arguments, but there could be cancellations that arise that make the differences first enter at higher orders. 
For example, there will be no additional changes to the $1/r^3$ part of $U^A$ from solving the hypersurface equation in Eq.~\eqref{eq:ErA}, because several terms proportional to the angular derivatives of $\lambda_2$ cancel; however $U^A$ will differ at order $1/r^3$ because the evolution of the angular momentum aspect does not have such a cancellation.}

Given that we will need the evolution equation for $h_{AB}$ at leading and next-to-leading order in $1/r$, where we expect no difference from BD theory, it will be simplest to review the solutions of this equation first (although, as discussed in Paper I, for example, the hypersurface equations generally should be solved before the evolution equations are).
As in BD theory, the leading-order in $1/r$ part of the evolution equation for $h_{AB}$ gives that $\partial_u c_{AB} = N_{AB}$, where $N_{AB}$ is the news tensor.
The news (for short) is the free tensor radiative data in this formalism (i.e., the evolution equations do not constrain the evolution of the shear).
The next order in $1/r$ of the evolution equation for $h_{AB}$ gives 
\be \label{eq:DAB-evolve}
\partial_u \left(D_{AB} + \frac 12  \lambda_1 c_{AB} \right) = 0 .
\ee
This implies that $D_{AB} + (\lambda_1/2) c_{AB}$ is a constant of motion (i.e., it is determined by the initial values of these fields).
As we discuss below, we will chose the initial data so as to eliminate an order $r^{-3}\log r$ term in $U^A$.

In the discussion of the hypersurface equations, it was observed that, in general, solving the scalar-field evolution equation requires solving the $\beta$ and $U^A$ hypersurface equations before, and solving the hypersurface equation for $V$ simultaneously. 
However, given the expected fall-offs of the Bondi metric functions and the scalar field in Eq.~\eqref{eq:Jordan-fall-off}, then as in BD theory, the leading-order part of the scalar-field equation shows that $\partial_u \lambda_1$ is unconstrained by the evolution equation (making it the free radiative data, which was referred to as the ``scalar news'' in Paper I).
Thus, $\lambda_1$ does not depend on the solutions to the hypersurface equations.
The evolution of $\lambda_2$ will, so we postpone the discussion of its solution until after we solve the hypersurface equations.

Solving the hypersurface equations proceeds similarly to the procedure in BD theory.
We can obtain the solution for $\beta$ expanded in a series in $1/r$ by 
substituting Eq.~\eqref{Eq:Jordan_FallOff} into Eq.~\eqref{eq:Err} and integrating $\partial_r\beta$ with respect to $r$ to obtain $\beta$.
The solution for $U^A$ is obtained by using the result for $\beta$ and the expressions in Eq.~\eqref{Eq:Jordan_FallOff} in the hypersurface equation~\eqref{eq:ErA}.
The resulting equation can be integrated twice with respect to $r$ to obtain a series expansion for $U^A$.
This double integration permits the introduction of a function of integration at order $1/r^3$, which is denoted $L_A(u,x^B)$ and called the angular momentum aspect.
It also allows for a logarithm term at order $r^{-3} \log r$.
However, as in BD theory, the coefficient of this term turns out to be non-dynamical.
Specifically, it is the two-sphere divergence of the constant of motion in Eq.~\eqref{eq:DAB-evolve}.
For simplicity in Paper I, we chose initial data that admits expansions of the metric functions in $1/r$ (as in~\cite{Bondi:1962px}) rather than the ``polyhomogeneous'' expansion with terms of the from $(\log r)^m / r^n$ (for $n > m$; see, e.g.,~\cite{1998JGP....24...83F}).
We will also make this assumption of a smooth expansion in $1/r$ here.
Finally, using $U^A$, $\beta$, $\partial_u \lambda_1$ and Eq.~\eqref{Eq:Jordan_FallOff}, we can substitute these expressions into the hypersurface equation for $V$ in Eq.~\eqref{eq:EABtrace} to obtain the leading and next-to-leading parts of $V$.
Because only $\partial_u \lambda_1$ is required at these orders, the hypersurface equation has effectively decoupled from the scalar field equation.
It is now possible to integrate $V$, which has a new function of integration, the mass aspect $M(u,x^A)$.
In summary, we find that the hypersurface equations have solutions
\begin{subequations} \label{eq:beta-UA-V-expand}
\begin{align} 
    \beta = {} & (\beta)_{\omega_0} + \m O(r^{-3}) , \\
    U^{A} = {} & \lb U^A\rb_{\omega_0} + \m O(r^{-4}) , \\
    V = {} & (V)_{\omega_0} + \m O(r^{-1}) ,
\end{align} 
where
\begin{align} \label{eq:beta_sol}
(\beta)_{\omega_0} = &-  \frac{\lambda_{1}{}}{2 r }- \frac 1{r^2} \left( \frac{1}{32} c_{AB} c^{AB}  + \frac{\omega_0-1}{8} \lambda_{1}{}^2 + \frac{3}{4} \lambda_{2}\right) \nonumber \\
& + \m O(r^{-3}) , \\
\label{eq:U_sol}
\lb U^A\rb_{\omega_0} = & -\frac{1}{2r^2}\lb\eth_B c^{AB}-\eth^A\lambda_1\rb \nonumber \\
 &+\frac{1}{3 r^3}\left[ c^{AD}\eth^B c_{DB} -c^{AB}\eth_{B}\lambda_1 +\lambda_1\eth_B c^{AB}\right. \nonumber \\
&\left. - \lambda_1\eth^A\lambda_1 +6L^A (u,x^A)\right] + \m O\lb r^{-4}\rb , \\
\label{eq:Vsol}
(V)_{\omega_0} = {} & (1 +\partial_{u}\lambda_{1})r - 2M(u,x^A)+\m O\lb r^{-1} \rb .
\end{align}
\end{subequations}
We introduced the notation with a subscript $\omega_0$ [e.g., $(\beta)_{\omega_0}$] to indicate that the corresponding expressions in BD and DEF theories have the same form when written in terms of $M$, $L_A$, and the coefficients of the expansions of $h_{AB}$ and $\lambda$ in $1/r$ (and when $\omega_0\rightarrow \omega_{\BD}$).
Thus, the integration of the hypersurface equations produces a solution that has the same mathematical form as the solution in BD theory at the given orders in $1/r$.

The evolution of subleading scalar field $\lambda_2$ can now be obtained by substituting the expansion of $\lambda$ and $h_{AB}$ in~\eqref{Eq:Jordan_FallOff} into  Eq.~\eqref{eq:scalar-full}, and using the results for the Bondi metric functions in Eqs.~\eqref{eq:beta-UA-V-expand}.
It is given by
\be\label{eq:lambda2_Ev}
\partial_u \lambda_2 = -\frac{1}{2} \ETH^2\lambda_{1}{} -  \frac{\mathcal{B}_{0}}{2\alpha_0{}^2} \lambda_{1} \partial_{u}\lambda_{1} .
\ee
We introduced the notation $\ETH^2=\eth^A\eth_A$ for the Laplacian operator above.
Because the parameter $\m B_0$ is zero in BD theory, this is the first expression that we have written down that has a manifest difference from that in BD theory.
It arises from the term proportional to $d\omega/d\lambda$ on the right-hand side of scalar-field equation. 
Note, however, that the expression for $\beta$ in Eq.~\eqref{eq:beta_sol} depends on $\lambda_2$.
Thus, the expression for $\beta$ in BD and DEF theories will generically differ because of the different evolution of $\lambda_2$ in the two theories..

Because the quadratic term in Eq.~\eqref{eq:lambda2_Ev} is a total derivative, we can also define
\begin{equation}
    \ell_2 \equiv \lambda_2 + \frac{\mathcal B_0}{4\alpha_0{}^2} (\lambda_1)^2 . 
\end{equation}
This new variable $\ell_2$ satisfies an evolution equation of the form
\begin{equation}
    \partial_u \ell_2 = -\frac{1}{2} \ETH^2\lambda_{1} ,
\end{equation}
which is of the same form as the evolution equation for $\lambda_2$ in BD theory.

This approach of redefining the expansion coefficients of the fields in $1/r$ will also prove useful in analyzing the ``conservation equations'' that describe the evolution of the Bondi mass and angular-momentum aspects.
The conservation equation for the mass aspect is determined by the component $\m G_{uu}=0$ of the modified Einstein equations at order $1/r^2$.
By making the following redefinition of the mass aspect,
\be \label{eq:mass-aspect-redef}
\m M = M - \frac{1}{8}\left( 1 + \frac{\mathcal{B}_{0}}{\alpha_0{}^2} \right) \partial_u (\lambda_1)^2 ,
\ee
then this modified mass aspect satisfies an evolution equation with the same form as that in BD theory.
Namely, one can write
\begin{equation}
    \partial_u \m M = (\partial_u \m M)_{\omega_0} ,
\end{equation}
where
\begin{align}\label{eq:MassEv}
(\partial_u \m M)_{\omega_0} = & - \frac{1}{8} N_{AB} N^{AB}+ \tfrac{1}{4} \eth_{B}\eth_{A}N^{AB} \nonumber\\
& -\frac{1}{4}(3+2\omega_0)(\partial_u\lambda_{1}{})^2+ \tfrac{1}{4} \partial_u\ETH^2\lambda_{1} .
\end{align}
Note that we did not use the relationship that $3+2\omega_0 = 1/\alpha_0{}^2$ here, so as to make the $\omega_0$ dependence more manifest.

As discussed in Paper I, the $\m B_0$-independent term in the modified mass aspect $\mathcal M$ was introduced so that the right-hand side of Eq.~\eqref{eq:MassEv} is strictly decreasing when averaged over the 2-sphere.
This allows the terms quadratic in the news and scalar news to be interpreted as an energy flux carried by tensor and scalar waves, analogously to the right-hand side in the Bondi mass loss formula~\cite{Bondi:1962px} in the context of GR or in BD theory (as noted in Paper I).\footnote{We have also used a similar normalization of the mass aspect here as was used in Paper I in BD theory.
From Eq.~(4.5) of~\cite{Mirshekari:2013vb}, the Newtonian-order metric of a point mass, $m$, has a potential $-G m/r$, where $G = (4+2\omega_0)/(3+2\omega_0)$ is the gravitational constant in units with $\lambda_0=1$, as discussed further in Footnote~\ref{fn:G-value}.
By relating the Bondi and harmonic-gauge metrics, it was shown in Appendix A of Paper II (in the context of BD theory) that the mass that appears in the metric in Bondi coordinates generally differs from the mass in the harmonic-gauge metric (which was obtained from a computation using the matter stress-energy tensor) by $\omega_0$-dependent terms.
Thus, the mass-aspect parameter $M$ used in this paper is generally not the same as a mass parameter associated with any matter stress-energy sources; to relate the two notions of mass in specific cases, further detailed calculations would be needed.
Similar statements also apply for the angular momentum aspect $L_A$ and the angular momentum associated with the matter stress-energy tensor.}
In DEF theory, a similar argument holds for redefining the mass aspect by including the $\m B_0$-dependent term when $\m B_0 > 0$, so as to ensure that energy fluxes decrease the mass of the system.
While this reason is less compelling when $\m B_0 < 0$, it seems like a reasonable prescription to also absorb this term in the mass aspect in this case, so that the definition of $\mathcal M$ is not dependent on the sign of $\mathcal B_0$.

The presence of this term proportional to $\mathcal B_0$ arises because the evolution equation for the mass aspect $M$ has a term proportional to $\partial_u^2 \lambda_2$; thus it is the difference in the evolution of $\lambda_2$ that is the reason why we make a $\mathcal B_0$-dependent definition of $\mathcal M$.

The conservation equation for the angular-momentum aspect comes from the order $1/r^2$ part of the expression $\m G_{uA}=0$.
If we make a modified definition of the angular momentum aspect
\begin{equation}
    \mathscr L_A = L_A + \frac{\m B_0}{16\alpha_0{}^2}\eth_A (\lambda_1)^2 \, ,
\end{equation}
then we find that
\begin{equation} \label{eq:LEv}
    \partial_u \mathscr L_A = (\partial_u L_A)_{\omega_0} ,
\end{equation}
where for completeness, we give an equivalent lengthy expression for $(\partial_u L_A)_{\omega_0}$ as that given in Paper I
\bw
\begin{align}
(\partial_u L_A)_{\omega_0} = {} & \frac{1}{12} \eth_C(\eth^C\eth_{B}c_{A}{}^{B} -\eth_{A}\eth_{B}c^{BC}) - \frac 1{48} \left[ \eth_A(c_{BC} N^{BC}) - 8\eth_B(c^{BC} N_{AC}) + 4 c^{BC} \eth_A N_{BC} \right] -\frac{1}{3} \eth_{A}\mathcal{M} \nonumber \\
&  - \frac{1}{24} \eth_{A}\ETH^2\lambda_{1} -  \frac{1}{12} (\omega_0 + 4)\lambda_{1}{} \eth_{A}\partial_u\lambda_{1} + \frac{1}{12} (2+3\omega_0) \eth_{A}\lambda_{1}{} \partial_u\lambda_{1}  + \frac{1}{12} \partial_u(\lambda_{1}{} \eth^{B}c_{AB} - c_{AB} \eth^{B}\lambda_{1}) .
\end{align}
\ew

The presence of a term proportional to $\mathcal B_0$ in $\mathscr L_A$ arises because the evolution equation for the angular-momentum aspect $L_A$ has a term proportional to $\partial_u \eth_A \lambda_2$.
As with the mass aspect, it is the difference in the evolution of $\lambda_2$ that is the reason why we chose to define $\mathscr L_A$ with the $\mathcal B_0$-dependent term.

We give a brief summary of the differences in the form of the asymptotically flat solutions in BD and DEF theories.
Explicit differences in the form of the Bondi metric functions arise at a relative order of $1/r^2$ from the leading-order term in each metric function.
The evolution of scalar field also differs at an order of $1/r^2$ from the leading, constant $\lambda_0 = 1$, because the evolution equation for $\partial_u \lambda_2$ contains the parameter $\m B_0$ in DEF theory times a nonlinear term in the scalar field $\lambda_1$. 
This change in the evolution of $\lambda_2$ has a leading-order affect on the evolution of mass and angular momentum aspects; however, these metric functions can be redefined in a $\m B_0$-dependent (and nonlinear in $\lambda_1$) manner to make the evolution of the redefined variables have the same form as in BD theory.
However, the differences in the mass and angular-moments aspects as well as the evolution of $\lambda_2$ imply that the expansion of the Bondi metric functions $\beta$, $U^A$ and $V$ will differ at a relative order of $1/r$ from the leading-order part of the solution.

The remaining evolution equations in the Bondi-Sachs formalism are those for the order $1/r^3$ and higher-order expansion functions for $h_{AB}$.
These equations are necessary for computing the higher memory effects discussed, for example, in~\cite{Grant:2021hga,Grant:2023jhd,Siddhant:2024nft}.
However, because we will focus on the displacement and spin memory effects in the remainder of this paper, we will not need to compute these higher-order evolution equations.

\section{Gravitational-wave memory effects} \label{sec:memory-effects}

Because the metrics of asymptotically flat solutions in DEF theory have the same form as those in BD theory, then the leading-order, $1/r$ part of the Riemann tensor will also have a similar form.
Thus, the leading-order solutions to the equation of geodesic deviation will have the same form as given in Paper I.
This implies that there will be the same types of displacement memory effects and subleading effects (drift memory) in DEF theory as in BD theory.

The first is the ``tensor'' displacement memory effect, which also occurs in GR.
It is produced by a change in the shear tensor $c_{AB}$ from before to after a burst of GWs.
For simplicity, we will consider the burst of waves to start at a retarded time $u_1$ and end at $u_2$ (though one can also permit $u_1$ to go to minus infinity and $u_2$ to plus infinity).
We introduce the $\Delta$ notation for this purpose:
\begin{equation}
\Delta c_{AB} \equiv c_{AB}(u_2) - c_{AB}(u_1) = \int_{u_1}^{u_2} N_{AB}.
\end{equation}
In the language of~\cite{Grant:2021hga}, the leading tensor displacement memory effect is the ``zeroth (temporal) moment'' of the news tensor.
The second is the ``scalar'' displacement memory effect, which does not arise in GR, because it is sourced by the scalar waves.
It is given by 
\begin{equation}
\Delta \lambda_1 \equiv \lambda_1(u_2) - \lambda_1(u_1) = \int_{u_1}^{u_2} \partial_u \lambda_1 .
\end{equation}
In analogy with the terminology of~\cite{Grant:2021hga}, it is natural to refer to this scalar memory effect as the zeroth moment of the scalar news.

As we describe in more detail below, the shear tensor can be decomposed into an ``electric'' and a ``magnetic'' parity part.
The evolution equation for the Bondi mass aspect can be used to constrain the electric part of the displacement memory effect. 
The magnetic-parity part is not constrained by the Bondi mass aspect.
Rather it was shown in~\cite{Satishchandran:2019pyc} that certain forms of stress-energy are responsible for producing a magnetic-parity displacement memory.
We will focus on the electric-parity tensor memory effects in this paper.
The scalar waves (being scalar, as opposed to pseudoscalar) are only electric parity.

The next class of memory effects are those related to the first moment of the news tensor and scalar news.
These effects go by several different names: subleading displacement memory, drift memory, or spin and center-of-mass memory effects, in the tensor case.
In scalar-tensor theories, one can also compute the first moment of the scalar news, which is closely related to the scalar analogue of the CM memory that was discussed in Paper I.
Two different notions of moments of the news were used for the tensor memory effects in the context of GR.
The ``Mellin moments'' were used in~\cite{Grant:2021hga}, which are related to the Mellin transform of the news, and the ``Cauchy moments'' were used in~\cite{Siddhant:2024nft}; the two types of moments are simply related (see~\cite{Siddhant:2024nft}).
The Mellin moments were more natural for describing the leading-order behavior of the curve-deviation observable~\cite{Flanagan:2018yzh} in asymptotically flat spacetimes~\cite{Grant:2021hga}, whereas the Cauchy moments were better adapted for computing higher memory signals associated with a given charge, flux or pseudoflux term in the moment of the news or the corresponding GW strain~\cite{Siddhant:2024nft}.

We will follow the Cauchy prescription for computing moments of the news here, though we use a notation more similar to that used in Paper I than in~\cite{Grant:2021hga,Siddhant:2024nft}.
Specifically, we define the first moment of the news tensor to be
\begin{equation}
    \Delta \mathcal C_{AB} = \int^{u_2}_{u_1} c_{AB} du .
\end{equation}
Similarly, we will denote the first moment of the scalar news as 
\begin{equation}
    \Delta \Lambda_1 = \int^{u_2}_{u_1} \lambda_1 du .
\end{equation}
As with the zeroth moment, the first moment of the scalar news will be just electric parity.
The first moment of the news tensor will have both electric- and magentic-parity parts which can be constrained by flux terms.
We discuss this decomposition and the fluxes in more detail next.

\subsection{Computing GW memory effects from fluxes} \label{sec:ConstrainMemory}

The conservation equations for the Bondi mass aspect~\eqref{eq:MassEv} and angular-momentum aspect~\eqref{eq:LEv}, and their equivalents in GR, can be interpreted as consistency relationships that the moments of the news tensor must satisfy.
For the displacement memory effect in GR, for example, the nonlinear energy flux of GWs acts like a source of secondary GWs, and the memory can be computed from integrating this nonlinear term.
This can be done in a nonlinearity expansion, such as the multipolar post-Minkowski approach (see, e.g.,~\cite{PhysRevD.45.520,Favata:2008yd,Nichols:2017rqr})
or also in numerical-relativity simulations of black-hole binaries without CCE, in which the memory effect is not captured by the waveform extraction methods used (see, e.g.,~\cite{Favata:2009ii,Mitman:2020bjf}).
The remaining terms in the conservation equations for the mass and angular-momentum aspects that are neither a flux or linear in the news or shear are the ``charge'' terms.
These terms are responsible for the ``linear'' memory effects.

In BD theory, there were potential ambiguities that arose when splitting the conservation equations for the mass and angular-momentum aspects into nonlinear flux terms, charge terms, and memory terms.
For the mass aspect, for example, there was a nonlinear term involving the scalar field $\lambda_1$, which entered into the redefinition of the mass aspect, as in Eq.~\eqref{eq:mass-aspect-redef}, which, in principle could have been considered as part of the flux.
There also was a linear term in the scalar field proportional to $\ETH^2 \partial_u \lambda_1$, which could have been interpreted either as a scalar memory term or as part of the charge term after integration with respect to $u$.

While there is not a unique prescription for making the split into charge, flux, and memory terms, the choice made in Paper I was to use the Wald-Zoupas prescription~\cite{Wald:1999wa} to define the flux.
In addition, it was assumed that the conservation equations constrain just the tensor memory effects; any remaining terms in the conservation equations were considered to be parts of the charge terms.
We will follow this prescription in this paper as well.
Because the Wald-Zoupas flux is conformally invariant, and because the leading and subleading parts of the metric and the scalar field have the same form in DEF theory as in BD theory, this prescription will be applicable in DEF theory.

To quantitatively describe the computation of the memory effects below, it will be useful to decompose the shear tensor into its electric- and magnetic-parity parts.
These are parameterized by two scalar functions on the 2-sphere, $\Theta$ and $\Psi$:
\be \label{Eq:cdecomp}
c_{AB} = \left(\eth_A \eth_B - \frac{1}{2} q_{AB} \ETH^2\right) \Theta + \epsilon_{C(A} \eth_{B)} \eth^C \Psi .
\ee
We used $\epsilon_{AB}$ for the antisymmetric tensor on 2-spheres of constant $u$ and $r$.
The leading, tensor displacement memory effect can be derived from the change $\Delta\Theta$ using the conservation equation for the mass aspect.

\subsubsection{Tensor displacement memory effect}

To compute the displacement memory, we first substitute the decomposition of shear in Eq.~\eqref{Eq:cdecomp} into the term linear in the news in the conservation equation for the mass aspect in Eq.~\eqref{eq:MassEv}.
Next, we integrate the equation with respect to $u$ and commute angular derivatives to write the result in the form (as in~\cite{Flanagan:2015pxa,Nichols:2017rqr})
\begin{subequations}
\be\label{eq:Memory_Supertrans}
    \ETH^2 (\ETH^2+2) \Delta \Theta = \Delta \m E + 8 \Delta Q_{\m M} .
\ee
We made the definitions 
\begin{align}
    \label{eq:mem-nonlinear}
    \Delta \m E = {} & \int du \left[ N_{AB} N^{AB} +(6+4\omega_0) (\partial_u \lambda_1)^2 \right] , \\
    \label{eq:Memory_Supertrans_Ord}
    Q_{\m M} = {} & \m M - \frac 1{4} \ETH^2 \lambda_1 .
\end{align}
\end{subequations}
When $\Delta \Theta$ is decomposed into scalar spherical harmonics, the left-hand side of Eq.~\eqref{eq:Memory_Supertrans} vanishes for the $l=0$ and $l=1$ harmonics; thus, the right-hand side of the equation encodes information about the losses of the Bondi 4-momentum.
The first term on the right-hand side, $\Delta \m E$, is the change in the flux radiated in scalar and gravitational waves, and the second is the change in the charge.

For the $l\geq 2$ moments of $\Delta \Theta$, the term $\Delta \m E$ is the nonlinear (or ``null''~\cite{Bieri:2013ada} or ``flux''~\cite{Grant:2021hga}) contribution to the memory effect, and $\Delta Q_{\m M}$ is the linear (or ``ordinary''~\cite{Bieri:2013ada} or ``charge''~\cite{Grant:2021hga}) contribution.
Note that only $\Delta\Theta$ appears in this equation, which is why the displacement memory is an electric-parity effect (the magnetic-parity part is not constrained by the mass-aspect conservation equation).
In practical computations, it is most convenient to decompose $\Delta \Theta$, the mass aspect $\m M$, the news tensor and the scalar news into appropriate spherical harmonics, which makes inverting the differential operator $\ETH^2(\ETH^2+2)$ acting on $\Delta\Theta$ simpler.

Although our notation differs slightly from that in Paper I, the procedure of computing the displacement memory effect in DEF theory is equivalent to that in BD theory.
The nonlinear term has the same form in terms of the news tensor and the scalar news (though note that this does not imply that the solutions for the scalar news or news tensor would be the same for similar sources in BD and DEF theories; thus, there could still be residual differences in the nonlinear memory for such similar sources in these theories). 
The charge term also has the same form in terms of $\mathcal M$ and $\lambda_1$, but when there is nonzero scalar news, $\partial_u \lambda_1$, the definitions of the mass aspects differ in DEF and BD theories by a $\m B_0$-dependent term.

We also note that our result for the nonlinear displacement memory is also consistent with a similar calculation in scalar-vector-tensor theories that was performed in Ref.~\cite{Heisenberg:2023prj} using the Isaacson effective stress-energy tensor.

\subsubsection{Tensor spin memory effect}

To compute the spin memory effect it is useful to first take the ``curl'' (namely $\epsilon^{BA}\eth_B$) of the evolution equation for the angular-momentum aspect~\eqref{eq:LEv}.
This removes terms that are gradients on the two-sphere. 
Integrating the equation gives a constraint on the time-integral of the magnetic part of the shear, which we denote by 
\be
    \Delta \Sigma = \int_{u_1}^{u_2} \Psi du\,.
\ee
We then use the Wald-Zoupas prescription for defining the flux, as in Paper I, and we determine the change in $\Delta\Sigma$ via
\begin{subequations}
\begin{equation} \label{eq:spin-memory}
    \ETH^4 (\ETH^2 + 2) \Delta\Sigma = \Delta \mathcal J + 8 \Delta Q_{\mathscr L} , 
\end{equation}
where we have defined
\begin{align}
    \Delta \mathcal J = {} & 2 \int_{u_1}^{u_2} du \, \epsilon^{DA} \eth_D [2 \eth_B( c_{AC} N^{BC}) - N^{BC} \eth_A c_{BC} \nonumber \\
    & \qquad \qquad  -(3+2\omega_0)(\partial_u \lambda_1 \eth_A \lambda_1 - \lambda_1 \eth_A \partial_u \lambda_1)] , \\
    Q_{\mathscr L} = {} & \epsilon^{CA} \eth_C \left[3 \mathscr L_A + \frac 14 (c_{AB} \eth^B \lambda_1 - \lambda_1 \eth^B c_{AB}) \right] .
\end{align}
\end{subequations}
As with the displacement memory, the differential operator acting on $\Delta\Sigma$ annihilates $l=1$ spherical harmonics; thus, the $l=1$ multipoles of the right-hand side give the result that the change in the angular-momentum charge $\Delta Q_{\mathscr L}$ equals the flux of angular momentum $\Delta \mathcal J$.
For the $l \geq 2$ moments, the change in the spin memory $\Delta \Sigma$ has a (nonlinear) flux contribution from $\Delta \mathcal J$ and a charge contribution from $\Delta Q_{\mathscr L}$.

The flux term in DEF theory has the same form as in BD theory.
As with the displacement memory, however, this does not necessarily imply that the scalar field $\lambda_1$ and shear tensor $c_{AB}$ will have the same values for similar sources in BD and DEF theories.
Thus, the memory computed from these fluxes in the two theories could differ.
Although the charge involves $\mathscr L_A$ in DEF theory rather than in $L_A$ in BD theory, the $\mathcal B_0$-dependent term in the redefinition of $\mathscr L_A$ is a divergence and does not affect the spin memory, which is computed from taking a curl of $\mathscr L_A$.
This implies that the form of the ordinary or charge contribution to the spin memory effect has the same form in BD and DEF theories, too.

By taking a divergence rather than a curl of the evolution equation for the angular momentum aspect, one could also obtain the equation for the CM memory effect; however, we consider just the leading electric-parity displacement effect and the leading magnetic-parity spin memory effect in this paper.

\subsection{Scalar memory effects}

The scalar memory effect, $\Delta \lambda_1$ could not be computed through conservation equations as with the tensor memory effects.
In the Bondi framework, $\partial_u \lambda_1$, the scalar news, is unconstrained radiative data; thus, to determine $\Delta \lambda_1$; one would need to perform a detailed calculation of $\lambda_1$.
This has been done in both the PN context~\cite{Mirshekari:2013vb,Lang:2013fna,Lang:2014osa} and in numerical relativity~\cite{Ma:2023sok}, for example.
As in BD theory, we will not compute the scalar displacement memory effect in DEF theory either.

There is also a subleading scalar memory effect, $\Delta \Lambda_1$, which is the scalar analogue of the CM memory effect.
In BD theory, this could be computed, in principle, by integrating the evolution equation for $\lambda_2$ once in time.
Following such a procedure would demonstrate that the change in $\lambda_2$ is the source of this higher scalar memory effect.
Thus, if the initial and final values of $\lambda_2$ are known, it would be straightforward to compute this memory, which has only an ordinary (or charge) contribution.

However, in DEF theory, integrating Eq.~\eqref{eq:lambda2_Ev} gives
\begin{equation}
    \ETH^2 \Delta \Lambda_1 = -2 \Delta \lambda_2 - \frac{\mathcal B_0}{2\alpha_0{}^2} \Delta (\lambda_1)^2 .
\end{equation}
While the scalar analogue of the CM memory has a linear contribution to the charge term $\Delta \lambda_2$, there is also a nonlinear contribution to the charge term proportional to $\mathcal B_0$ and the square of $\lambda_1$.
Thus, in DEF theory, when a system has a nonzero scalar displacement memory, there will also generally be a higher memory in $\Delta \Lambda_1$ that is sourced by the leading scalar memory. 
This could be used to compute a contribution to the scalar analogue of the CM memory effect.

\subsection{Memory effects from post-Newtonian compact binaries} \label{subsec:PNbinries}

In Paper II, we computed the nonlinear displacement and spin memory GW signals in the PN approximation, within the context of BD theory.  
We computed the waveforms to Newtonian (0PN) order and to zeroth and first order in the small BD parameter $\xi=1/(2+\omega_{\mathrm{BD}})$.
Both the displacement and spin memory effects had negative PN terms proportional to $\xi$, which arose from dipole radiation in the scalar field.
We worked in the approximation that the energy losses from scalar dipole radiation were subleading to the tensor quadrupole radiation.
We now comment on the anticipated PN orders and orders in $\xi$ at which we would anticipate differences between the displacement and spin memory GW signals in BD and DEF theories.
[Note that in DEF theory, it is natural to define $\xi$ as $\xi=1/(2+\omega_0)$ with $\omega_0$ defined in Eq.~\eqref{eq:omega-to-alpha}; our parameter $\xi$ is twice the parameter $\zeta$ used in~\cite{Mirshekari:2013vb}.]

As we described in the previous section, in DEF theory, it was natural to define the nonlinear contributions to the displacement and spin memory effects to be of the same form (in terms of $c_{AB}$ and $\lambda_1$) as in BD theory.
When evaluating the integrals required to compute the memory waveforms, differences could arise if the amplitude or GW frequency of the tensor or scalar waveforms differ for similar sources; in addition, if the binary's binding energy or its evolution differs for these sources in BD and DEF theories, then this would also affect the memory computation, because we have assumed energy balance holds when evaluating the memory integrals (see Paper II for further details). 

The PN expression for the binary dynamics, the tensor GWs, and the scalar GWs were computed in a sequence of papers (Refs.~\cite{Mirshekari:2013vb,Lang:2013fna,Lang:2014osa}; see also~\cite{Sennett:2016klh,Bernard:2018hta,Bernard:2018ivi,Bernard:2022noq,Trestini:2024zpi,Trestini:2024mfs}); we can make use of these existing results to determine the orders in $\xi$ and the PN orders at which we anticipate differences between memory signals in BD and DEF theories.
For this purpose, it will be useful to make use of the scalar-tensor and equation-of-motion parameters defined in Table I of Ref.~\cite{Mirshekari:2013vb}.
The terms related to $d\omega/d\lambda$ (or $\mathcal B_0$ in DEF theory) appear in the Newtonian (0PN) expression for the scalar waveform and at 1PN in the tensor waveform.
The binding energy differs at 1PN order, but the GW luminosity differs at Newtonian order.
Thus, we find from PN counting that the PN waveforms in BD and DEF theories will differ at Newtonian order for sources with the same masses and sensitivities.

However, the terms proportional to $d\omega/d\lambda$ in the waveforms, binding energy and luminosity are proportional to $\xi^2$ (or in other cases $\xi^3$), and they are being multiplied by coefficients that are of order unity (for BH-NS or BNS systems where the sensitivities are order unity).
Because $\xi$ is constrained to be small ($\xi\leq 10^{-5}$) from solar system experiments~\cite{Will:2014kxa,Bertotti:2003rm}, then the differences in the PN signals in DEF and BD theories will generally be of order $\xi^2$ at Newtonian order.

\subsection{Detection prospects of memory effects in DEF theory}

We conclude this section with a few comments on the detection prospects for the different memory effects.
Similar points were discussed in the conclusions of Paper II in the context of BD theory.
As was argued above, the difference between the inspiral memory signals in BD and DEF theories arises at Newtonian order in a PN expansion, but at second order in the small coupling parameter $\xi$.
Given that the inspiral portion of the memory signal has a small SNR compared with the merger and ringdown (or disruption) portions of the signals~\cite{Lasky:2016knh,Tiwari:2021gfl}, simply measuring the order $\xi$ PN corrections in BD theory was argued to be likely beyond the capabilities of the LVK or next-generation interferometers.
It follows that distinguishing BD and DEF theories using the memory is less likely.

However, as was discussed in Paper II (for BD theory), most of the SNR for the memory comes from the times of the coalescence near the merger, which PN theory cannot accurately describe.
To have more robust forecasts of the detection prospects of memory effects in DEF theory (and whether they could be distinguished from those in BD theory), we would need to use the full scalar and tensor polarization waveforms from NR simulations.
The number of such simulations that have been performed is limited, and the waveforms are not available in public catalogs, such as those for GR (e.g.,~\cite{Boyle:2019kee}).
Once such waveforms are available, however, we can use the conservation equations for the mass and angular-momentum aspects that we derived in Eqs.~\eqref{eq:mass-aspect-redef}--\eqref{eq:LEv} (and the multipolar expansion of the nonlinear flux terms in these equations from Paper II) to compute the memory effects.
With NR waveforms in DEF theories, studies focused on the detection and distinguishing memory effects in DEF and BD theories will be possible and more realistic than those that use PN waveforms.
We thus save such precise studies for future work.

Subject to the several caveats and uncertainties discussed above, we nevertheless estimate some aspects related to the detection prospects of the GW memory signal in GR, BD and DEF theories.
Forecasts of these prospects in the context of GR were performed in~\cite{Grant:2022bla} (see also~\cite{Goncharov:2023woe}) which showed that with Cosmic Explorer an individual BBH merger will be measured with an accompanying memory signal with a SNR${}>5$ at a rate of roughly $1/{}$yr. 
It has also been shown that the memory signal can help distinguish BH-NS and BBH mergers, though we are not aware of forecasts for the detection rate of the memory signal from BH-NS as was done in~\cite{Grant:2022bla}.
Given the comparable local rates of BH-NS and BBH mergers in the most recent GW catalog~\cite{LIGOScientific:2025pvj} and the smaller amplitude memory signal for BH-NS systems, the rate of SNR${}>5$ observations of the memory signal from BH-NS systems with Cosmic Explorer will likely be a factor of several lower than that of BBH systems (i.e., a few per decade).

In BD and DEF theory, the differences in the memory waveform from the GR signal first enter at linear order in the $\xi$ parameter in the inspiral (which was noted above to be constrained to be of order $10^{-5}$).
Assuming that there still is a well-defined perturbative expansion in the $\xi$ parameter during the merger and ringdown (namely, that the parts of the solution associated with the beyond-GR modifications to the field equations remain small compared with $1/\xi$), one would expect that there would be negligible difference in the forecasts for the detection of memory signals in GR, BD or DEF theories, because the differences in the SNR between GR and BD or DEF theories would be of order $\xi$.
Because of dynamical scalarization (see, e.g.,~\cite{Barausse:2012da,Palenzuela:2013hsa}) where scalar charges and sensitivities can become large during the late inspiral (see also~\cite{Kuntz:2024jxo}), the difference could, conceivably, be larger than $\xi \sim 10^{-5}$ (though likely not at a level that it could change signal enough to affect the detection forecasts for the memory given the existing GW constraints from a LVK BH-NS merger~\cite{Takeda:2023wqn}).

There is also a separate question of whether GW measurements of the memory signal could be used to distinguish the signals in GR and BD or DEF, as well as whether the signals in BD and DEF theories could be distinguished.
Again, if the differences between the memory signals are of order $\xi \sim 10^{-5}$, it is difficult to see how the GR memory signal could be distinguished from that in BD or DEF theories (or how BD and DEF could be distinguished) from a single BH-NS observation.
For ``stacking'' the evidence from the entire population of BH-NS observations measured by Cosmic Explorer (which could include of order $10^5$ detections, but which for the majority, the SNR of the memory signal is below the threshold of detection), the prospects also look pessimistic, because the evidence grows with the square root of the number of events.
Thus, the contributions from the beyond-GR modifications to the memory signal would need to be enhanced during the merger and ringdown by several orders of magnitude from the order $\xi$ effects during the inspiral.

Given that differences between DEF and BD theories enter at order $\xi^2$ in the inspiral memory signals, it is unlikely that these theories could be distinguished through a direct measurement of the memory signal.
However, given that spontaneous and dynamical scalarization have been shown to occur in DEF not BD theory, then if there are sufficiently large scalar effects to produce a measurable difference between the GR and beyond-GR theories, then the presence of such large scalar charges and sensitivities could be interpreted as an indirect indicator of the underlying theory being that of DEF rather than of BD.

\section{Conclusions and Discussion}\label{sec:Conclusions}

In this paper, we investigated Damour-Esposito-Far\`ese scalar-tensor theory using the Bondi-Sachs formalism, which generalized prior work (Papers I and II~\cite{Tahura:2020vsa,Tahura:2021hbk}) that had been performed for Brans-Dicke theory.
DEF theory contains a scalar field that is nonminimaly coupled to gravity through a coupling function $\omega(\lambda)$. 
There are two coupling parameters $(\alpha_0, \m B_0)$ in the theory which are related to coefficients in a Taylor expansion of $\omega (\lambda)$ about the asymptotic value of the scalar field $\lambda_0$.
The parameter $\alpha_0$ appears at leading order in this Taylor expansion, which implies that $\omega_0=\omega(\lambda_0)$ is a function of $\alpha_0$ only.
Thus, $\alpha_0$ is related to the gravitational constant in the theory~\cite{Berti:2015itd}.
The parameter $\m B_0$ determines the leading behavior of how the scalar field acts as a source of scalar waves.

Our focus in this paper was on the general equations of motion in the Einstein and Jordan frames, the boundary conditions associated with asymptotic flatness, the asymptotically flat solutions consistent with these boundary conditions, and the leading GW memory effects that can be computed using the conservation laws of the asymptotic field equations. 

The asymptotically flat solutions in DEF theory are similar to those in BD theory; the field equations reduce to those of Paper I in BD theory when $\omega (\lambda)\rightarrow\omega_{\BD}$. 
The most manifest difference arises from the evolution equation for the scalar field. 
In particular, the order $1/r^2$ function in the expansion of the scalar, $\lambda_2$, has a nonlinear term that does not arise in BD theory (which is proportional to the coefficient $\mathcal B_0$).
The hypersurface equations (which allow the Bondi metric functions to be solved in terms of the expansion of the scalar field, the shear tensor and other Bondi-metric functions) have a similar form as those in BD theory at the order at which we computed the results, in the sense that they did not contain any explicit $\mathcal B_0$ dependence.
The same was also true of the evolution equations for the coefficients of the expansion of the Bondi 2-metric.
Our scaling arguments implied that any explicit dependence on the DEF parameter $\m B_0$ enters at the first order in $1/r$ beyond what we computed.

This does not imply that the metric functions in the BD and DEF theories are the same.
The scalar field $\lambda_2$ explicitly appears in some of the metric functions.
In addition, the conservation equations for the Bondi mass and angular-momentum aspects depend on the parameter $\m B_0$. 
For the mass and angular momentum aspects, however, it was possible to make a $\mathcal B_0$-dependent redefinition so that its evolution in DEF theory followed an equation of the same form as that in BD theory.

As in BD theory, there will be memory effects related to the tensor GWs and the scalar waves.
For the tensor memory effects, there will be a hierarchy of different higher memory effects related to the moments of the news tensor, as in GR.
There also will be a set of scalar memory effects related to the moments of the scalar news.
In this paper, we focused on the leading displacement effects and the subleading drift effects for both the scalar and tensor GW memory effects.
The tensor effects are constrained by the conservation equations for the mass and angular-momentum aspects.
The leading scalar memory effect cannot be computed through the evolution equation for the scalar field in Bondi-Sachs coordinates, because the scalar news $\partial_u \lambda_1$ is part of the unconstrained radiative data in the Bondi-Sachs formalism.
The evolution equation for the subleading part of the scalar field $\lambda_2$ can be used to constrain the subleading scalar memory in terms of the leading scalar field $\lambda_1$.

For the tensor memory effects, the conservation equations can be split into flux and charge pieces, as is done in GR or in BD theory. 
We used the same conventions as in BD theory to make this split, which was based on the flux terms computed from the Wald-Zoupas procedure.
Thus, what we defined to be the nonlinear contributions to the (electric-parity) tensor displacement memory effect and the (magnetic-parity) spin memory effect had the same form as in BD theory. 
The solutions for scalar field $\lambda_1$ and the Bondi shear would generically differ for solutions in BD and DEF theories, so one would not expect that the precise time dependence of the memory signals in these theories would be the same (despite the same definitions of the flux).
Because the charge terms used the redefined, $\mathcal B_0$-dependent mass and angular-momentum aspects, their contributions to the displacement and spin memory effects would likely differ between BD and DEF theories.

As a specific example, we commented on the GW memory effects generated by PN compact-binary sources on quasicircular orbits in DEF theory. 
The scalar and tensor waveforms, the binding energy, and the GW luminosity are used to compute the memory effects via the fluxes, and it had been previously found that these quantities all differ between BD theory and DEF theory.
However, the differences in the tensor and scalar multipole moments of the waveforms and in the binary's equations of motion depend on the DEF parameter $\m B_0$ at a higher order in the small parameter $\xi$ in scalar-tensor theories. 
Thus, while the memory signals generically differ between BD and DEF theory, when working to linear order in the parameter $\xi$, the PN memory signals in BD theory that were computed in Paper II will also apply to DEF theory (with an appropriate mapping of the coupling parameters in the two theories). 
The flux laws and the PN waveforms could prove helpful in comparisons of GW memory effects computed from numerical relativity simulations with Cauchy characteristic extraction or matching~\cite{Ma:2023sok,Ma:2024bed}.

\acknowledgments

We thank Sizheng Ma, Leo Stein, and Vijay Varma for discussions about memory effects in scalar-tensor theories in the early stages of this work.
D.A.N.\ acknowledges support from the NSF grant No.\ PHY-2309021 and the NSF CAREER Award PHY-2439893. 
K.Y. acknowledges support from the NSF grant No. PHY-2309066, the NSF CAREER Award PHYS-2339969, and the
Owens Family Foundation.

\appendix

\section{Asymptotic symmetries}\label{sec:Symmetries}

In this appendix, we discuss the asymptotic symmetries of the Bondi-Sachs metric in the Jordan frame for DEF theory.
First, recall that the metric functions expanded in $1/r$ in Eq.~\eqref{eq:beta-UA-V-expand} have the same form in BD and DEF theories at the order at which we are working.
Next, note that the asymptotic symmetries are generated by a vector field $\xi^\mu$ that preserves the boundary conditions imposed and that maintains the Bondi gauge conditions when taking the Lie derivative of the Bondi-gauge metric.
Because the metrics have the same form and the field equations are not used in computing the Lie derivative, then the vector field $\xi^\mu$ will have the same form as in Eq.~(3.2) of Paper I.
This vector field can be expressed in terms of the metric functions $c_{AB}$ and $D_{AB}$; the scalar field functions $\lambda_1$ and $\lambda_2$; and a conformal Killing vector on the 2-sphere $Y^A$ and scalar function constructed from $Y^A$ and $\alpha(x^A)$ denoted by $f$:
\begin{subequations}
\begin{align}
& \eth_C Y_D+\eth_D Y_C=\psi q_{AB},\quad \psi=\eth_A Y^A ,\\
& f(u,x^B)=\frac{1}{2}u\psi+\alpha (x^B) .
\end{align}
\end{subequations}
Thus, it follows that the globally defined asymptotic symmetries in DEF theory are again the BMS group.
The vector field $Y^A$ parametrizes Lorentz transformations and the scalar function $\alpha$ generates supertranslations.
We reproduce the components below for convenience:
\begin{subequations} \label{eq:xi-vector}
\begin{align}
\xi^{u} = {} & f\lb u,x^A\rb\,, \\
\xi^{r} = & -\frac{1}{2}r\eth_A Y^A+\frac{1}{2}\ETH^2 f-\frac{1}{4r}\bigg( c^{AB}\eth_B\eth_A f \nonumber \\
& + 2 \eth_A f \eth_B c^{AB} +\lambda_1 \ETH^2 f\bigg) + O\lb r^{-2}\rb\,, \\
\xi^{A}= {} & Y^A\lb u, x^A \rb-\frac{1}{r}\eth^A f + \frac{1}{2 r^2}\lb c^{AB}\eth_B f+\lambda_1\eth^A f \rb \nonumber \\
& +\frac{1}{r^3} \bigg[\frac{1}{3}D^{AB}\eth_B f -\frac{1}{16}c^{BC}c_{BC}\eth^A f - \frac{\lambda_1}{3}c^{AB}\eth_B f
\nonumber \\
& +\frac{\lambda_2}{2}\eth^A f +\frac{\lambda_1^2}{12}\lb \omega_0-3\rb\eth^A f\bigg]+O\lb r^{-4}\rb\,.
\end{align}
\end{subequations}

As in Paper I, we use the notation $\delta_\xi$ to denote the transformation of the Bondi metric functions and the functions in the expansion of the scalar field under an infinitesimal transformation parameterized by $\xi^\mu$ in Eq.~\eqref{eq:xi-vector}.
When not using the solutions to the equations of motion, the transformations of $\lambda_1$ and $\lambda_2$ have the same form as in BD theory:
\begin{subequations}\label{eq:lambda_transf_BMS}
\be\label{eq:lambda1_transf}
\dxi\lambda_1=\lb \dxi\lambda_1\rb_{\omega_0}, \qquad 
\dxi\lambda_2 = \lb \dxi\lambda_2\rb_{\omega_0} ,
\ee
where
\begin{align}
    \lb \dxi\lambda_1\rb_{\omega_0} = {} & \frac{1}{2}\lambda_1\psi+Y^B\eth_B\lambda_1+f\partial_u\lambda_1 , \\
     \lb \dxi\lambda_2\rb_{\omega_0} = {} & \lambda_2\psi +f\partial_u\lambda_2 + Y^D\eth_D\lambda_2    \nonumber \\ 
     & -\frac{1}{2}\lambda_1 \ETH^2 f-\eth_Df\eth^D \lambda_1 .
\end{align}
\end{subequations}
Note though that if one were to apply the equations of motion to the expressions for $\delta_\xi \lambda_2$, the term proportional to $\partial_u\lambda_2$ would differ in BD and DEF theories.

The procedure to compute the transformation of the Bondi metric functions was described in Section III of Paper I.
If we were to write the result in terms of the mass aspect $M$ rather than $\mathcal M$, the results in BD and DEF theories would have the same form.
However, because the definition of $\mathcal M$ differs in BD and DEF theory by a $\mathcal B_0$-dependent term, then when $\delta_\xi L_A$ is written in terms of $\mathcal M$, a term proportional to $M \eth_A f$ will cause the angular momentum aspect to transform with an explicit $\mathcal B_0$ dependence.
We find, therefore, that the four metric functions transform in the following ways in DEF theory:
\bw
\begin{subequations}
\begin{equation}
    \dxi c_{AB} = \lb\dxi c_{AB}\rb_{\omega_0} , \qquad
    \delta_{\xi}D_{AB} = \lb\dxi D_{AB}\rb_{\omega_0} , \qquad
    \dxi \m M =  \lb\dxi \m M\rb_{\omega_0} , \qquad
    \delta_{\xi} L_A = (\delta_\xi L_A)_{\omega_0}- \frac{\mathcal{B}_{0}}{8\alpha_0{}^2}  \eth_{A}f \partial_u(\lambda_{1}{}^2)
\end{equation}
where
\begin{align} \label{eq:Met_Func_BMSc}
\lb\dxi c_{AB}\rb_{\omega_0} = {} & f N_{AB} + \m L_{Y} c_{AB} -  \frac{1}{2} c_{AB} \psi - 2 \eth_{B}\eth_{A}f + q_{AB} \ETH^2 f , \\
\lb\dxi D_{AB}\rb_{\omega_0} ={} & \mathcal{L}_{Y}D_{AB}+\lambda_1 \lb \eth_A \eth_B f -\frac{1}{2} q_{AB}\ETH^2 f\rb-\frac{1}{2}f\partial_u\lb \lambda_1 c_{AB}\rb ,\\
\lb\dxi \m M\rb_{\omega_0} = {} & \frac{3}{2} \mathcal{M} \psi -  \frac{1}{4} \lambda_{1}{} \psi + Y^{A} \eth_{A}\mathcal{M} + \frac{1}{2} \eth_{A}\partial_u\lambda_{1}{} \eth^{A}f + \frac{1}{4} \eth_{A}\psi \eth^{A}\lambda_{1}{} + \frac{1}{4} N_{AB} \eth^{A}\eth^{B}f + \frac{1}{2} \eth_{A}f \eth_{B}N^{AB} \nonumber\\
&+ \frac{1}{8} c^{AB} \eth_{B}\eth_{A}\psi+ f \partial_u\mathcal{M} + \frac{1}{4} \ETH^2 f \partial_u\lambda_{1}{} ,\\
\label{eq:Met_Func_BMSL}
\lb\delta_{\xi} L_A\rb_{\omega_0} = {} & f\partial_u L_A+\mathcal{L}_Y L_A+L_A \psi+\frac{1}{96}c^{CD}c_{CD}\eth_A \psi+\frac{1}{6}D_{AB}\eth^B\psi-\mathcal{M}\eth_A f+\frac{1}{12} \lb \ETH^2 f \eth^B c_{AB} -c_{AB}\eth^B \ETH^2 f\rb\nonumber\\
&-\frac{1}{8}\eth_A\lb c^{BC}\eth_B\eth_C f\rb + \frac{1}{4}\lb\eth_C\eth^B c_{AB}-\eth_A\eth^B c_{BC} \rb\eth^C f-\frac{1}{6}c_{AB}\eth^B f-\frac{1}{6}\eth_B\eth_Af\eth_C c^{BC}-\frac{5}{48}c^{BC}N_{BC}\eth_A f \nonumber\\
&+\frac{1}{6}c^{BC}N_{AB}\eth_C f- \frac{1}{12}\left[c_{AB}\eth^B f+\lambda_1 \lb \omega_0+4\rb\eth_A f\right]\partial_u \lambda_1+\frac{1}{24}\left[6\lambda_2+\lb \omega_0-1\rb \lambda_1^2\right]\eth_A \psi\nonumber\\
&-\frac{1}{12}\lb 2\lambda_1+3\partial_u \lambda_2 \rb\eth_A f-\frac{5}{24}\eth_A \lb \lambda_1 \ETH^2 f\rb+\frac{\lambda_1}{12}N_{AB}\eth^B f-\frac{1}{12}\lb \eth_B\eth_A f \eth^B \lambda_1+3\eth_B\eth_A\lambda_1\eth^B f\rb .
\end{align}
\end{subequations}
\ew
We have corrected a typo in the transformation of the angular momentum aspect given in Eq.~(3.10d) of Paper I. The term involving $\eth_C\eth^B c_{AB}$ on the second line was incorrectly written as $\eth^B\eth_C c_{AB}$ in Paper I.
Note that in addition to the explicit $\mathcal B_0$-dependent term written in the transformation of $L_A$ in Eq.~\eqref{eq:Met_Func_BMSc}, if the equations of motion were applied in the term proportional to $\eth_A f\partial_u\lambda_2$ in the final line in Eq.~\eqref{eq:Met_Func_BMSL}, then there would be an additional difference from this term which, although of the same form as the term in Eq.~\eqref{eq:Met_Func_BMSc}, does not cancel.
Thus, there is a difference in the transformation of the angular momentum aspect arising from terms quadratic in the scalar field $\lambda_1$.

\bibliography{memory}

\begin{thebibliography}{146}%
\makeatletter
\providecommand \@ifxundefined [1]{%
 \@ifx{#1\undefined}
}%
\providecommand \@ifnum [1]{%
 \ifnum #1\expandafter \@firstoftwo
 \else \expandafter \@secondoftwo
 \fi
}%
\providecommand \@ifx [1]{%
 \ifx #1\expandafter \@firstoftwo
 \else \expandafter \@secondoftwo
 \fi
}%
\providecommand \natexlab [1]{#1}%
\providecommand \enquote  [1]{``#1''}%
\providecommand \bibnamefont  [1]{#1}%
\providecommand \bibfnamefont [1]{#1}%
\providecommand \citenamefont [1]{#1}%
\providecommand \href@noop [0]{\@secondoftwo}%
\providecommand \href [0]{\begingroup \@sanitize@url \@href}%
\providecommand \@href[1]{\@@startlink{#1}\@@href}%
\providecommand \@@href[1]{\endgroup#1\@@endlink}%
\providecommand \@sanitize@url [0]{\catcode `\\12\catcode `\$12\catcode
  `\&12\catcode `\#12\catcode `\^12\catcode `\_12\catcode `\%12\relax}%
\providecommand \@@startlink[1]{}%
\providecommand \@@endlink[0]{}%
\providecommand \url  [0]{\begingroup\@sanitize@url \@url }%
\providecommand \@url [1]{\endgroup\@href {#1}{\urlprefix }}%
\providecommand \urlprefix  [0]{URL }%
\providecommand \Eprint [0]{\href }%
\providecommand \doibase [0]{http://dx.doi.org/}%
\providecommand \selectlanguage [0]{\@gobble}%
\providecommand \bibinfo  [0]{\@secondoftwo}%
\providecommand \bibfield  [0]{\@secondoftwo}%
\providecommand \translation [1]{[#1]}%
\providecommand \BibitemOpen [0]{}%
\providecommand \bibitemStop [0]{}%
\providecommand \bibitemNoStop [0]{.\EOS\space}%
\providecommand \EOS [0]{\spacefactor3000\relax}%
\providecommand \BibitemShut  [1]{\csname bibitem#1\endcsname}%
\let\auto@bib@innerbib\@empty
\bibitem [{\citenamefont {Abbott}\ \emph
  {et~al.}(2019{\natexlab{a}})\citenamefont {Abbott} \emph
  {et~al.}}]{LIGOScientific:2019fpa}%
  \BibitemOpen
  \bibfield  {author} {\bibinfo {author} {\bibfnamefont {B.~P.}\ \bibnamefont
  {Abbott}} \emph {et~al.} (\bibinfo {collaboration} {LIGO Scientific,
  Virgo}),\ }\bibfield  {title} {\enquote {\bibinfo {title} {{Tests of General
  Relativity with the Binary Black Hole Signals from the LIGO-Virgo Catalog
  GWTC-1}},}\ }\href {\doibase 10.1103/PhysRevD.100.104036} {\bibfield
  {journal} {\bibinfo  {journal} {Phys. Rev.}\ }\textbf {\bibinfo {volume}
  {D100}},\ \bibinfo {pages} {104036} (\bibinfo {year} {2019}{\natexlab{a}})},\
  \Eprint {http://arxiv.org/abs/1903.04467} {arXiv:1903.04467 [gr-qc]}
  \BibitemShut {NoStop}%
\bibitem [{\citenamefont {Abbott}\ \emph
  {et~al.}(2021{\natexlab{a}})\citenamefont {Abbott} \emph
  {et~al.}}]{LIGOScientific:2020tif}%
  \BibitemOpen
  \bibfield  {author} {\bibinfo {author} {\bibfnamefont {R.}~\bibnamefont
  {Abbott}} \emph {et~al.} (\bibinfo {collaboration} {LIGO Scientific,
  Virgo}),\ }\bibfield  {title} {\enquote {\bibinfo {title} {{Tests of general
  relativity with binary black holes from the second LIGO-Virgo
  gravitational-wave transient catalog}},}\ }\href {\doibase
  10.1103/PhysRevD.103.122002} {\bibfield  {journal} {\bibinfo  {journal}
  {Phys. Rev. D}\ }\textbf {\bibinfo {volume} {103}},\ \bibinfo {pages}
  {122002} (\bibinfo {year} {2021}{\natexlab{a}})},\ \Eprint
  {http://arxiv.org/abs/2010.14529} {arXiv:2010.14529 [gr-qc]} \BibitemShut
  {NoStop}%
\bibitem [{\citenamefont {Abbott}\ \emph
  {et~al.}(2021{\natexlab{b}})\citenamefont {Abbott} \emph
  {et~al.}}]{LIGOScientific:2021sio}%
  \BibitemOpen
  \bibfield  {author} {\bibinfo {author} {\bibfnamefont {R.}~\bibnamefont
  {Abbott}} \emph {et~al.} (\bibinfo {collaboration} {LIGO Scientific, VIRGO,
  KAGRA}),\ }\bibfield  {title} {\enquote {\bibinfo {title} {{Tests of General
  Relativity with GWTC-3}},}\ }\href@noop {} {\bibfield  {journal} {\bibinfo
  {journal} {{}}\ } (\bibinfo {year} {2021}{\natexlab{b}})},\ \Eprint
  {http://arxiv.org/abs/2112.06861} {arXiv:2112.06861 [gr-qc]} \BibitemShut
  {NoStop}%
\bibitem [{\citenamefont {Isi}\ \emph {et~al.}(2019)\citenamefont {Isi},
  \citenamefont {Giesler}, \citenamefont {Farr}, \citenamefont {Scheel},\ and\
  \citenamefont {Teukolsky}}]{Isi:2019aib}%
  \BibitemOpen
  \bibfield  {author} {\bibinfo {author} {\bibfnamefont {Maximiliano}\
  \bibnamefont {Isi}}, \bibinfo {author} {\bibfnamefont {Matthew}\ \bibnamefont
  {Giesler}}, \bibinfo {author} {\bibfnamefont {Will~M.}\ \bibnamefont {Farr}},
  \bibinfo {author} {\bibfnamefont {Mark~A.}\ \bibnamefont {Scheel}}, \ and\
  \bibinfo {author} {\bibfnamefont {Saul~A.}\ \bibnamefont {Teukolsky}},\
  }\bibfield  {title} {\enquote {\bibinfo {title} {{Testing the no-hair theorem
  with GW150914}},}\ }\href {\doibase 10.1103/PhysRevLett.123.111102}
  {\bibfield  {journal} {\bibinfo  {journal} {Phys. Rev. Lett.}\ }\textbf
  {\bibinfo {volume} {123}},\ \bibinfo {pages} {111102} (\bibinfo {year}
  {2019})},\ \Eprint {http://arxiv.org/abs/1905.00869} {arXiv:1905.00869
  [gr-qc]} \BibitemShut {NoStop}%
\bibitem [{\citenamefont {Bhagwat}\ \emph {et~al.}(2020)\citenamefont
  {Bhagwat}, \citenamefont {Forteza}, \citenamefont {Pani},\ and\ \citenamefont
  {Ferrari}}]{Bhagwat:2019dtm}%
  \BibitemOpen
  \bibfield  {author} {\bibinfo {author} {\bibfnamefont {Swetha}\ \bibnamefont
  {Bhagwat}}, \bibinfo {author} {\bibfnamefont {Xisco~Jimenez}\ \bibnamefont
  {Forteza}}, \bibinfo {author} {\bibfnamefont {Paolo}\ \bibnamefont {Pani}}, \
  and\ \bibinfo {author} {\bibfnamefont {Valeria}\ \bibnamefont {Ferrari}},\
  }\bibfield  {title} {\enquote {\bibinfo {title} {{Ringdown overtones, black
  hole spectroscopy, and no-hair theorem tests}},}\ }\href {\doibase
  10.1103/PhysRevD.101.044033} {\bibfield  {journal} {\bibinfo  {journal}
  {Phys. Rev. D}\ }\textbf {\bibinfo {volume} {101}},\ \bibinfo {pages}
  {044033} (\bibinfo {year} {2020})},\ \Eprint
  {http://arxiv.org/abs/1910.08708} {arXiv:1910.08708 [gr-qc]} \BibitemShut
  {NoStop}%
\bibitem [{\citenamefont {Isi}\ \emph {et~al.}(2021)\citenamefont {Isi},
  \citenamefont {Farr}, \citenamefont {Giesler}, \citenamefont {Scheel},\ and\
  \citenamefont {Teukolsky}}]{Isi:2020tac}%
  \BibitemOpen
  \bibfield  {author} {\bibinfo {author} {\bibfnamefont {Maximiliano}\
  \bibnamefont {Isi}}, \bibinfo {author} {\bibfnamefont {Will~M.}\ \bibnamefont
  {Farr}}, \bibinfo {author} {\bibfnamefont {Matthew}\ \bibnamefont {Giesler}},
  \bibinfo {author} {\bibfnamefont {Mark~A.}\ \bibnamefont {Scheel}}, \ and\
  \bibinfo {author} {\bibfnamefont {Saul~A.}\ \bibnamefont {Teukolsky}},\
  }\bibfield  {title} {\enquote {\bibinfo {title} {{Testing the Black-Hole Area
  Law with GW150914}},}\ }\href {\doibase 10.1103/PhysRevLett.127.011103}
  {\bibfield  {journal} {\bibinfo  {journal} {Phys. Rev. Lett.}\ }\textbf
  {\bibinfo {volume} {127}},\ \bibinfo {pages} {011103} (\bibinfo {year}
  {2021})},\ \Eprint {http://arxiv.org/abs/2012.04486} {arXiv:2012.04486
  [gr-qc]} \BibitemShut {NoStop}%
\bibitem [{\citenamefont {Calder\'on~Bustillo}\ \emph
  {et~al.}(2021)\citenamefont {Calder\'on~Bustillo}, \citenamefont {Lasky},\
  and\ \citenamefont {Thrane}}]{CalderonBustillo:2020rmh}%
  \BibitemOpen
  \bibfield  {author} {\bibinfo {author} {\bibfnamefont {Juan}\ \bibnamefont
  {Calder\'on~Bustillo}}, \bibinfo {author} {\bibfnamefont {Paul~D.}\
  \bibnamefont {Lasky}}, \ and\ \bibinfo {author} {\bibfnamefont {Eric}\
  \bibnamefont {Thrane}},\ }\bibfield  {title} {\enquote {\bibinfo {title}
  {{Black-hole spectroscopy, the no-hair theorem, and GW150914: Kerr versus
  Occam}},}\ }\href {\doibase 10.1103/PhysRevD.103.024041} {\bibfield
  {journal} {\bibinfo  {journal} {Phys. Rev. D}\ }\textbf {\bibinfo {volume}
  {103}},\ \bibinfo {pages} {024041} (\bibinfo {year} {2021})},\ \Eprint
  {http://arxiv.org/abs/2010.01857} {arXiv:2010.01857 [gr-qc]} \BibitemShut
  {NoStop}%
\bibitem [{\citenamefont {Abbott}\ \emph
  {et~al.}(2019{\natexlab{b}})\citenamefont {Abbott} \emph
  {et~al.}}]{LIGOScientific:2018jsj}%
  \BibitemOpen
  \bibfield  {author} {\bibinfo {author} {\bibfnamefont {B.~P.}\ \bibnamefont
  {Abbott}} \emph {et~al.} (\bibinfo {collaboration} {LIGO Scientific,
  Virgo}),\ }\bibfield  {title} {\enquote {\bibinfo {title} {{Binary Black Hole
  Population Properties Inferred from the First and Second Observing Runs of
  Advanced LIGO and Advanced Virgo}},}\ }\href {\doibase
  10.3847/2041-8213/ab3800} {\bibfield  {journal} {\bibinfo  {journal}
  {Astrophys. J. Lett.}\ }\textbf {\bibinfo {volume} {882}},\ \bibinfo {pages}
  {L24} (\bibinfo {year} {2019}{\natexlab{b}})},\ \Eprint
  {http://arxiv.org/abs/1811.12940} {arXiv:1811.12940 [astro-ph.HE]}
  \BibitemShut {NoStop}%
\bibitem [{\citenamefont {Abbott}\ \emph
  {et~al.}(2021{\natexlab{c}})\citenamefont {Abbott} \emph
  {et~al.}}]{LIGOScientific:2020kqk}%
  \BibitemOpen
  \bibfield  {author} {\bibinfo {author} {\bibfnamefont {R.}~\bibnamefont
  {Abbott}} \emph {et~al.} (\bibinfo {collaboration} {LIGO Scientific,
  Virgo}),\ }\bibfield  {title} {\enquote {\bibinfo {title} {{Population
  Properties of Compact Objects from the Second LIGO-Virgo Gravitational-Wave
  Transient Catalog}},}\ }\href {\doibase 10.3847/2041-8213/abe949} {\bibfield
  {journal} {\bibinfo  {journal} {Astrophys. J. Lett.}\ }\textbf {\bibinfo
  {volume} {913}},\ \bibinfo {pages} {L7} (\bibinfo {year}
  {2021}{\natexlab{c}})},\ \Eprint {http://arxiv.org/abs/2010.14533}
  {arXiv:2010.14533 [astro-ph.HE]} \BibitemShut {NoStop}%
\bibitem [{\citenamefont {Abbott}\ \emph
  {et~al.}(2023{\natexlab{a}})\citenamefont {Abbott} \emph
  {et~al.}}]{KAGRA:2021duu}%
  \BibitemOpen
  \bibfield  {author} {\bibinfo {author} {\bibfnamefont {R.}~\bibnamefont
  {Abbott}} \emph {et~al.} (\bibinfo {collaboration} {KAGRA, VIRGO, LIGO
  Scientific}),\ }\bibfield  {title} {\enquote {\bibinfo {title} {{Population
  of Merging Compact Binaries Inferred Using Gravitational Waves through
  GWTC-3}},}\ }\href {\doibase 10.1103/PhysRevX.13.011048} {\bibfield
  {journal} {\bibinfo  {journal} {Phys. Rev. X}\ }\textbf {\bibinfo {volume}
  {13}},\ \bibinfo {pages} {011048} (\bibinfo {year} {2023}{\natexlab{a}})},\
  \Eprint {http://arxiv.org/abs/2111.03634} {arXiv:2111.03634 [astro-ph.HE]}
  \BibitemShut {NoStop}%
\bibitem [{\citenamefont {Abbott}\ \emph {et~al.}(2017)\citenamefont {Abbott}
  \emph {et~al.}}]{LIGOScientific:2017adf}%
  \BibitemOpen
  \bibfield  {author} {\bibinfo {author} {\bibfnamefont {B.~P.}\ \bibnamefont
  {Abbott}} \emph {et~al.} (\bibinfo {collaboration} {LIGO Scientific, Virgo,
  1M2H, Dark Energy Camera GW-E, DES, DLT40, Las Cumbres Observatory, VINROUGE,
  MASTER}),\ }\bibfield  {title} {\enquote {\bibinfo {title} {{A
  gravitational-wave standard siren measurement of the Hubble constant}},}\
  }\href {\doibase 10.1038/nature24471} {\bibfield  {journal} {\bibinfo
  {journal} {Nature}\ }\textbf {\bibinfo {volume} {551}},\ \bibinfo {pages}
  {85--88} (\bibinfo {year} {2017})},\ \Eprint
  {http://arxiv.org/abs/1710.05835} {arXiv:1710.05835 [astro-ph.CO]}
  \BibitemShut {NoStop}%
\bibitem [{\citenamefont {Abbott}\ \emph
  {et~al.}(2019{\natexlab{c}})\citenamefont {Abbott} \emph
  {et~al.}}]{LIGOScientific:2018mvr}%
  \BibitemOpen
  \bibfield  {author} {\bibinfo {author} {\bibfnamefont {B.P.}\ \bibnamefont
  {Abbott}} \emph {et~al.} (\bibinfo {collaboration} {LIGO Scientific,
  Virgo}),\ }\bibfield  {title} {\enquote {\bibinfo {title} {{GWTC-1: A
  Gravitational-Wave Transient Catalog of Compact Binary Mergers Observed by
  LIGO and Virgo during the First and Second Observing Runs}},}\ }\href
  {\doibase 10.1103/PhysRevX.9.031040} {\bibfield  {journal} {\bibinfo
  {journal} {Phys. Rev. X}\ }\textbf {\bibinfo {volume} {9}},\ \bibinfo {pages}
  {031040} (\bibinfo {year} {2019}{\natexlab{c}})},\ \Eprint
  {http://arxiv.org/abs/1811.12907} {arXiv:1811.12907 [astro-ph.HE]}
  \BibitemShut {NoStop}%
\bibitem [{\citenamefont {Abbott}\ \emph
  {et~al.}(2021{\natexlab{d}})\citenamefont {Abbott} \emph
  {et~al.}}]{LIGOScientific:2020ibl}%
  \BibitemOpen
  \bibfield  {author} {\bibinfo {author} {\bibfnamefont {R.}~\bibnamefont
  {Abbott}} \emph {et~al.} (\bibinfo {collaboration} {LIGO Scientific,
  Virgo}),\ }\bibfield  {title} {\enquote {\bibinfo {title} {{GWTC-2: Compact
  Binary Coalescences Observed by LIGO and Virgo During the First Half of the
  Third Observing Run}},}\ }\href {\doibase 10.1103/PhysRevX.11.021053}
  {\bibfield  {journal} {\bibinfo  {journal} {Phys. Rev. X}\ }\textbf {\bibinfo
  {volume} {11}},\ \bibinfo {pages} {021053} (\bibinfo {year}
  {2021}{\natexlab{d}})},\ \Eprint {http://arxiv.org/abs/2010.14527}
  {arXiv:2010.14527 [gr-qc]} \BibitemShut {NoStop}%
\bibitem [{\citenamefont {Abbott}\ \emph
  {et~al.}(2023{\natexlab{b}})\citenamefont {Abbott} \emph
  {et~al.}}]{KAGRA:2021vkt}%
  \BibitemOpen
  \bibfield  {author} {\bibinfo {author} {\bibfnamefont {R.}~\bibnamefont
  {Abbott}} \emph {et~al.} (\bibinfo {collaboration} {KAGRA, VIRGO, LIGO
  Scientific}),\ }\bibfield  {title} {\enquote {\bibinfo {title} {{GWTC-3:
  Compact Binary Coalescences Observed by LIGO and Virgo during the Second Part
  of the Third Observing Run}},}\ }\href {\doibase 10.1103/PhysRevX.13.041039}
  {\bibfield  {journal} {\bibinfo  {journal} {Phys. Rev. X}\ }\textbf {\bibinfo
  {volume} {13}},\ \bibinfo {pages} {041039} (\bibinfo {year}
  {2023}{\natexlab{b}})},\ \Eprint {http://arxiv.org/abs/2111.03606}
  {arXiv:2111.03606 [gr-qc]} \BibitemShut {NoStop}%
\bibitem [{\citenamefont {Abbott}\ \emph {et~al.}(2016)\citenamefont {Abbott}
  \emph {et~al.}}]{KAGRA:2013rdx}%
  \BibitemOpen
  \bibfield  {author} {\bibinfo {author} {\bibfnamefont {B.~P.}\ \bibnamefont
  {Abbott}} \emph {et~al.} (\bibinfo {collaboration} {KAGRA, LIGO Scientific,
  Virgo}),\ }\bibfield  {title} {\enquote {\bibinfo {title} {{Prospects for
  observing and localizing gravitational-wave transients with Advanced LIGO,
  Advanced Virgo and KAGRA}},}\ }\href {\doibase 10.1007/s41114-020-00026-9}
  {\bibfield  {journal} {\bibinfo  {journal} {Living Rev. Rel.}\ }\textbf
  {\bibinfo {volume} {19}},\ \bibinfo {pages} {1} (\bibinfo {year} {2016})},\
  \Eprint {http://arxiv.org/abs/1304.0670} {arXiv:1304.0670 [gr-qc]}
  \BibitemShut {NoStop}%
\bibitem [{\citenamefont {Amaro-Seoane}\ \emph {et~al.}(2017)\citenamefont
  {Amaro-Seoane} \emph {et~al.}}]{LISA:2017pwj}%
  \BibitemOpen
  \bibfield  {author} {\bibinfo {author} {\bibfnamefont {Pau}\ \bibnamefont
  {Amaro-Seoane}} \emph {et~al.} (\bibinfo {collaboration} {LISA}),\ }\bibfield
   {title} {\enquote {\bibinfo {title} {{Laser Interferometer Space
  Antenna}},}\ }\href@noop {} {\  (\bibinfo {year} {2017})},\ \Eprint
  {http://arxiv.org/abs/1702.00786} {arXiv:1702.00786 [astro-ph.IM]}
  \BibitemShut {NoStop}%
\bibitem [{\citenamefont {Punturo}\ \emph {et~al.}(2010)\citenamefont {Punturo}
  \emph {et~al.}}]{Punturo:2010zz}%
  \BibitemOpen
  \bibfield  {author} {\bibinfo {author} {\bibfnamefont {M.}~\bibnamefont
  {Punturo}} \emph {et~al.},\ }\bibfield  {title} {\enquote {\bibinfo {title}
  {{The Einstein Telescope: A third-generation gravitational wave
  observatory}},}\ }\href {\doibase 10.1088/0264-9381/27/19/194002} {\bibfield
  {journal} {\bibinfo  {journal} {Class. Quant. Grav.}\ }\textbf {\bibinfo
  {volume} {27}},\ \bibinfo {pages} {194002} (\bibinfo {year}
  {2010})}\BibitemShut {NoStop}%
\bibitem [{\citenamefont {Yagi}(2013)}]{Yagi:2013du}%
  \BibitemOpen
  \bibfield  {author} {\bibinfo {author} {\bibfnamefont {Kent}\ \bibnamefont
  {Yagi}},\ }\bibfield  {title} {\enquote {\bibinfo {title} {{Scientific
  Potential of DECIGO Pathfinder and Testing GR with Space-Borne Gravitational
  Wave Interferometers}},}\ }\href {\doibase 10.1142/S0218271813410137}
  {\bibfield  {journal} {\bibinfo  {journal} {Int. J. Mod. Phys. D}\ }\textbf
  {\bibinfo {volume} {22}},\ \bibinfo {pages} {1341013} (\bibinfo {year}
  {2013})},\ \Eprint {http://arxiv.org/abs/1302.2388} {arXiv:1302.2388 [gr-qc]}
  \BibitemShut {NoStop}%
\bibitem [{\citenamefont {Luo}\ \emph {et~al.}(2016)\citenamefont {Luo} \emph
  {et~al.}}]{TianQin:2015yph}%
  \BibitemOpen
  \bibfield  {author} {\bibinfo {author} {\bibfnamefont {Jun}\ \bibnamefont
  {Luo}} \emph {et~al.} (\bibinfo {collaboration} {TianQin}),\ }\bibfield
  {title} {\enquote {\bibinfo {title} {{TianQin: a space-borne gravitational
  wave detector}},}\ }\href {\doibase 10.1088/0264-9381/33/3/035010} {\bibfield
   {journal} {\bibinfo  {journal} {Class. Quant. Grav.}\ }\textbf {\bibinfo
  {volume} {33}},\ \bibinfo {pages} {035010} (\bibinfo {year} {2016})},\
  \Eprint {http://arxiv.org/abs/1512.02076} {arXiv:1512.02076 [astro-ph.IM]}
  \BibitemShut {NoStop}%
\bibitem [{\citenamefont {Gong}\ \emph {et~al.}(2021)\citenamefont {Gong},
  \citenamefont {Luo},\ and\ \citenamefont {Wang}}]{Gong:2021gvw}%
  \BibitemOpen
  \bibfield  {author} {\bibinfo {author} {\bibfnamefont {Yungui}\ \bibnamefont
  {Gong}}, \bibinfo {author} {\bibfnamefont {Jun}\ \bibnamefont {Luo}}, \ and\
  \bibinfo {author} {\bibfnamefont {Bin}\ \bibnamefont {Wang}},\ }\bibfield
  {title} {\enquote {\bibinfo {title} {{Concepts and status of Chinese space
  gravitational wave detection projects}},}\ }\href {\doibase
  10.1038/s41550-021-01480-3} {\bibfield  {journal} {\bibinfo  {journal}
  {Nature Astron.}\ }\textbf {\bibinfo {volume} {5}},\ \bibinfo {pages}
  {881--889} (\bibinfo {year} {2021})},\ \Eprint
  {http://arxiv.org/abs/2109.07442} {arXiv:2109.07442 [astro-ph.IM]}
  \BibitemShut {NoStop}%
\bibitem [{\citenamefont {Reitze}\ \emph {et~al.}(2019)\citenamefont {Reitze}
  \emph {et~al.}}]{Reitze:2019iox}%
  \BibitemOpen
  \bibfield  {author} {\bibinfo {author} {\bibfnamefont {David}\ \bibnamefont
  {Reitze}} \emph {et~al.},\ }\bibfield  {title} {\enquote {\bibinfo {title}
  {{Cosmic Explorer: The U.S. Contribution to Gravitational-Wave Astronomy
  beyond LIGO}},}\ }\href@noop {} {\bibfield  {journal} {\bibinfo  {journal}
  {Bull. Am. Astron. Soc.}\ }\textbf {\bibinfo {volume} {51}},\ \bibinfo
  {pages} {035} (\bibinfo {year} {2019})},\ \Eprint
  {http://arxiv.org/abs/1907.04833} {arXiv:1907.04833 [astro-ph.IM]}
  \BibitemShut {NoStop}%
\bibitem [{\citenamefont {Zel'dovich}\ and\ \citenamefont
  {Polnarev}(1974)}]{Zeldovich:1974gvh}%
  \BibitemOpen
  \bibfield  {author} {\bibinfo {author} {\bibfnamefont {Y.~B.}\ \bibnamefont
  {Zel'dovich}}\ and\ \bibinfo {author} {\bibfnamefont {A.~G.}\ \bibnamefont
  {Polnarev}},\ }\bibfield  {title} {\enquote {\bibinfo {title} {{Radiation of
  gravitational waves by a cluster of superdense stars}},}\ }\href@noop {}
  {\bibfield  {journal} {\bibinfo  {journal} {Sov. Astron.}\ }\textbf {\bibinfo
  {volume} {18}},\ \bibinfo {pages} {17} (\bibinfo {year} {1974})}\BibitemShut
  {NoStop}%
\bibitem [{\citenamefont {Smarr}(1977)}]{Smarr1977}%
  \BibitemOpen
  \bibfield  {author} {\bibinfo {author} {\bibfnamefont {Larry}\ \bibnamefont
  {Smarr}},\ }\bibfield  {title} {\enquote {\bibinfo {title} {Gravitational
  radiation from distant encounters and from head-on collisions of black holes:
  The zero-frequency limit},}\ }\href {\doibase 10.1103/PhysRevD.15.2069}
  {\bibfield  {journal} {\bibinfo  {journal} {Phys. Rev. D}\ }\textbf {\bibinfo
  {volume} {15}},\ \bibinfo {pages} {2069--2077} (\bibinfo {year}
  {1977})}\BibitemShut {NoStop}%
\bibitem [{\citenamefont {{Turner}}(1978)}]{1978Natur.274..565T}%
  \BibitemOpen
  \bibfield  {author} {\bibinfo {author} {\bibfnamefont {M.~S.}\ \bibnamefont
  {{Turner}}},\ }\bibfield  {title} {\enquote {\bibinfo {title} {{Gravitational
  radiation from supernova neutrino bursts}},}\ }\href {\doibase
  10.1038/274565a0} {\bibfield  {journal} {\bibinfo  {journal} {\nat}\ }\textbf
  {\bibinfo {volume} {274}},\ \bibinfo {pages} {565--566} (\bibinfo {year}
  {1978})}\BibitemShut {NoStop}%
\bibitem [{\citenamefont {{Epstein}}(1978)}]{1978ApJ...223.1037E}%
  \BibitemOpen
  \bibfield  {author} {\bibinfo {author} {\bibfnamefont {R.}~\bibnamefont
  {{Epstein}}},\ }\bibfield  {title} {\enquote {\bibinfo {title} {{The
  generation of gravitational radiation by escaping supernova neutrinos.}}}\
  }\href {\doibase 10.1086/156337} {\bibfield  {journal} {\bibinfo  {journal}
  {\apj}\ }\textbf {\bibinfo {volume} {223}},\ \bibinfo {pages} {1037--1045}
  (\bibinfo {year} {1978})}\BibitemShut {NoStop}%
\bibitem [{\citenamefont {{Bontz}}\ and\ \citenamefont
  {{Price}}(1979)}]{1979ApJ}%
  \BibitemOpen
  \bibfield  {author} {\bibinfo {author} {\bibfnamefont {R.~J.}\ \bibnamefont
  {{Bontz}}}\ and\ \bibinfo {author} {\bibfnamefont {R.~H.}\ \bibnamefont
  {{Price}}},\ }\bibfield  {title} {\enquote {\bibinfo {title} {{The spectrum
  of radiation at low frequencies.}}}\ }\href {\doibase 10.1086/156880}
  {\bibfield  {journal} {\bibinfo  {journal} {\apj}\ }\textbf {\bibinfo
  {volume} {228}},\ \bibinfo {pages} {560--575} (\bibinfo {year}
  {1979})}\BibitemShut {NoStop}%
\bibitem [{\citenamefont {Newman}\ and\ \citenamefont
  {Penrose}(1966)}]{Newman:1966ub}%
  \BibitemOpen
  \bibfield  {author} {\bibinfo {author} {\bibfnamefont {E.T.}\ \bibnamefont
  {Newman}}\ and\ \bibinfo {author} {\bibfnamefont {R.}~\bibnamefont
  {Penrose}},\ }\bibfield  {title} {\enquote {\bibinfo {title} {{Note on the
  Bondi-Metzner-Sachs group}},}\ }\href {\doibase 10.1063/1.1931221} {\bibfield
   {journal} {\bibinfo  {journal} {J. Math. Phys.}\ }\textbf {\bibinfo {volume}
  {7}},\ \bibinfo {pages} {863--870} (\bibinfo {year} {1966})}\BibitemShut
  {NoStop}%
\bibitem [{\citenamefont {Christodoulou}(1991)}]{Christodoulou:1991cr}%
  \BibitemOpen
  \bibfield  {author} {\bibinfo {author} {\bibfnamefont {D.}~\bibnamefont
  {Christodoulou}},\ }\bibfield  {title} {\enquote {\bibinfo {title}
  {{Nonlinear nature of gravitation and gravitational wave experiments}},}\
  }\href {\doibase 10.1103/PhysRevLett.67.1486} {\bibfield  {journal} {\bibinfo
   {journal} {Phys. Rev. Lett.}\ }\textbf {\bibinfo {volume} {67}},\ \bibinfo
  {pages} {1486--1489} (\bibinfo {year} {1991})}\BibitemShut {NoStop}%
\bibitem [{\citenamefont {Blanchet}\ and\ \citenamefont
  {Damour}(1992)}]{Blanchet:1992br}%
  \BibitemOpen
  \bibfield  {author} {\bibinfo {author} {\bibfnamefont {Luc}\ \bibnamefont
  {Blanchet}}\ and\ \bibinfo {author} {\bibfnamefont {Thibault}\ \bibnamefont
  {Damour}},\ }\bibfield  {title} {\enquote {\bibinfo {title} {{Hereditary
  effects in gravitational radiation}},}\ }\href {\doibase
  10.1103/PhysRevD.46.4304} {\bibfield  {journal} {\bibinfo  {journal} {Phys.
  Rev. D}\ }\textbf {\bibinfo {volume} {46}},\ \bibinfo {pages} {4304--4319}
  (\bibinfo {year} {1992})}\BibitemShut {NoStop}%
\bibitem [{\citenamefont {Blanchet}(2024)}]{Blanchet:2013haa}%
  \BibitemOpen
  \bibfield  {author} {\bibinfo {author} {\bibfnamefont {Luc}\ \bibnamefont
  {Blanchet}},\ }\bibfield  {title} {\enquote {\bibinfo {title} {{Gravitational
  Radiation from Post-Newtonian Sources and Inspiralling Compact Binaries}},}\
  }\href {\doibase 10.1007/s41114-024-00050-z} {\bibfield  {journal} {\bibinfo
  {journal} {Living Rev. Relativ.}\ }\textbf {\bibinfo {volume} {27}},\
  \bibinfo {pages} {4} (\bibinfo {year} {2024})},\ \Eprint
  {http://arxiv.org/abs/1310.1528} {arXiv:1310.1528 [gr-qc]} \BibitemShut
  {NoStop}%
\bibitem [{\citenamefont {Braginsky}\ and\ \citenamefont
  {Grishchuk}(1985)}]{Braginsky:1986ia}%
  \BibitemOpen
  \bibfield  {author} {\bibinfo {author} {\bibfnamefont {V.B.}\ \bibnamefont
  {Braginsky}}\ and\ \bibinfo {author} {\bibfnamefont {L.P.}\ \bibnamefont
  {Grishchuk}},\ }\bibfield  {title} {\enquote {\bibinfo {title} {{Kinematic
  Resonance and Memory Effect in Free Mass Gravitational Antennas}},}\
  }\href@noop {} {\bibfield  {journal} {\bibinfo  {journal} {Sov. Phys. JETP}\
  }\textbf {\bibinfo {volume} {62}},\ \bibinfo {pages} {427--430} (\bibinfo
  {year} {1985})}\BibitemShut {NoStop}%
\bibitem [{\citenamefont {{Braginsky}}\ and\ \citenamefont
  {{Thorne}}(1987)}]{1987Natur.327..123B}%
  \BibitemOpen
  \bibfield  {author} {\bibinfo {author} {\bibfnamefont {V.~B.}\ \bibnamefont
  {{Braginsky}}}\ and\ \bibinfo {author} {\bibfnamefont {Kip~S.}\ \bibnamefont
  {{Thorne}}},\ }\bibfield  {title} {\enquote {\bibinfo {title}
  {{Gravitational-wave bursts with memory and experimental prospects}},}\
  }\href {\doibase 10.1038/327123a0} {\bibfield  {journal} {\bibinfo  {journal}
  {\nat}\ }\textbf {\bibinfo {volume} {327}},\ \bibinfo {pages} {123--125}
  (\bibinfo {year} {1987})}\BibitemShut {NoStop}%
\bibitem [{\citenamefont {Kennefick}(1994)}]{Kennefick:1994nw}%
  \BibitemOpen
  \bibfield  {author} {\bibinfo {author} {\bibfnamefont {D.}~\bibnamefont
  {Kennefick}},\ }\bibfield  {title} {\enquote {\bibinfo {title} {{Prospects
  for detecting the Christodoulou memory of gravitational waves from a
  coalescing compact binary and using it to measure neutron star radii}},}\
  }\href {\doibase 10.1103/PhysRevD.50.3587} {\bibfield  {journal} {\bibinfo
  {journal} {Phys. Rev. D}\ }\textbf {\bibinfo {volume} {50}},\ \bibinfo
  {pages} {3587--3595} (\bibinfo {year} {1994})}\BibitemShut {NoStop}%
\bibitem [{\citenamefont {Favata}(2009{\natexlab{a}})}]{Favata:2009ii}%
  \BibitemOpen
  \bibfield  {author} {\bibinfo {author} {\bibfnamefont {Marc}\ \bibnamefont
  {Favata}},\ }\bibfield  {title} {\enquote {\bibinfo {title} {{Nonlinear
  gravitational-wave memory from binary black hole mergers}},}\ }\href
  {\doibase 10.1088/0004-637X/696/2/L159} {\bibfield  {journal} {\bibinfo
  {journal} {Astrophys. J. Lett.}\ }\textbf {\bibinfo {volume} {696}},\
  \bibinfo {pages} {L159--L162} (\bibinfo {year} {2009}{\natexlab{a}})},\
  \Eprint {http://arxiv.org/abs/0902.3660} {arXiv:0902.3660 [astro-ph.SR]}
  \BibitemShut {NoStop}%
\bibitem [{\citenamefont {Pollney}\ and\ \citenamefont
  {Reisswig}(2011)}]{Pollney:2010hs}%
  \BibitemOpen
  \bibfield  {author} {\bibinfo {author} {\bibfnamefont {Denis}\ \bibnamefont
  {Pollney}}\ and\ \bibinfo {author} {\bibfnamefont {Christian}\ \bibnamefont
  {Reisswig}},\ }\bibfield  {title} {\enquote {\bibinfo {title} {{Gravitational
  memory in binary black hole mergers}},}\ }\href {\doibase
  10.1088/2041-8205/732/1/L13} {\bibfield  {journal} {\bibinfo  {journal}
  {Astrophys. J. Lett.}\ }\textbf {\bibinfo {volume} {732}},\ \bibinfo {pages}
  {L13} (\bibinfo {year} {2011})},\ \Eprint {http://arxiv.org/abs/1004.4209}
  {arXiv:1004.4209 [gr-qc]} \BibitemShut {NoStop}%
\bibitem [{\citenamefont {Lasky}\ \emph {et~al.}(2016)\citenamefont {Lasky},
  \citenamefont {Thrane}, \citenamefont {Levin}, \citenamefont {Blackman},\
  and\ \citenamefont {Chen}}]{Lasky:2016knh}%
  \BibitemOpen
  \bibfield  {author} {\bibinfo {author} {\bibfnamefont {Paul~D.}\ \bibnamefont
  {Lasky}}, \bibinfo {author} {\bibfnamefont {Eric}\ \bibnamefont {Thrane}},
  \bibinfo {author} {\bibfnamefont {Yuri}\ \bibnamefont {Levin}}, \bibinfo
  {author} {\bibfnamefont {Jonathan}\ \bibnamefont {Blackman}}, \ and\ \bibinfo
  {author} {\bibfnamefont {Yanbei}\ \bibnamefont {Chen}},\ }\bibfield  {title}
  {\enquote {\bibinfo {title} {{Detecting gravitational-wave memory with LIGO:
  implications of GW150914}},}\ }\href {\doibase
  10.1103/PhysRevLett.117.061102} {\bibfield  {journal} {\bibinfo  {journal}
  {Phys. Rev. Lett.}\ }\textbf {\bibinfo {volume} {117}},\ \bibinfo {pages}
  {061102} (\bibinfo {year} {2016})},\ \Eprint
  {http://arxiv.org/abs/1605.01415} {arXiv:1605.01415 [astro-ph.HE]}
  \BibitemShut {NoStop}%
\bibitem [{\citenamefont {Hübner}\ \emph {et~al.}(2020)\citenamefont
  {Hübner}, \citenamefont {Talbot}, \citenamefont {Lasky},\ and\ \citenamefont
  {Thrane}}]{Hubner:2019sly}%
  \BibitemOpen
  \bibfield  {author} {\bibinfo {author} {\bibfnamefont {Moritz}\ \bibnamefont
  {Hübner}}, \bibinfo {author} {\bibfnamefont {Colm}\ \bibnamefont {Talbot}},
  \bibinfo {author} {\bibfnamefont {Paul~D.}\ \bibnamefont {Lasky}}, \ and\
  \bibinfo {author} {\bibfnamefont {Eric}\ \bibnamefont {Thrane}},\ }\bibfield
  {title} {\enquote {\bibinfo {title} {{Measuring gravitational-wave memory in
  the first LIGO/Virgo gravitational-wave transient catalog}},}\ }\href
  {\doibase 10.1103/PhysRevD.101.023011} {\bibfield  {journal} {\bibinfo
  {journal} {Phys. Rev. D}\ }\textbf {\bibinfo {volume} {101}},\ \bibinfo
  {pages} {023011} (\bibinfo {year} {2020})},\ \Eprint
  {http://arxiv.org/abs/1911.12496} {arXiv:1911.12496 [astro-ph.HE]}
  \BibitemShut {NoStop}%
\bibitem [{\citenamefont {Boersma}\ \emph {et~al.}(2020)\citenamefont
  {Boersma}, \citenamefont {Nichols},\ and\ \citenamefont
  {Schmidt}}]{Boersma:2020gxx}%
  \BibitemOpen
  \bibfield  {author} {\bibinfo {author} {\bibfnamefont {Oliver~M.}\
  \bibnamefont {Boersma}}, \bibinfo {author} {\bibfnamefont {David~A.}\
  \bibnamefont {Nichols}}, \ and\ \bibinfo {author} {\bibfnamefont {Patricia}\
  \bibnamefont {Schmidt}},\ }\bibfield  {title} {\enquote {\bibinfo {title}
  {{Forecasts for detecting the gravitational-wave memory effect with Advanced
  LIGO and Virgo}},}\ }\href {\doibase 10.1103/PhysRevD.101.083026} {\bibfield
  {journal} {\bibinfo  {journal} {Phys. Rev. D}\ }\textbf {\bibinfo {volume}
  {101}},\ \bibinfo {pages} {083026} (\bibinfo {year} {2020})},\ \Eprint
  {http://arxiv.org/abs/2002.01821} {arXiv:2002.01821 [astro-ph.HE]}
  \BibitemShut {NoStop}%
\bibitem [{\citenamefont {Grant}\ and\ \citenamefont
  {Nichols}(2023)}]{Grant:2022bla}%
  \BibitemOpen
  \bibfield  {author} {\bibinfo {author} {\bibfnamefont {Alexander~M.}\
  \bibnamefont {Grant}}\ and\ \bibinfo {author} {\bibfnamefont {David~A.}\
  \bibnamefont {Nichols}},\ }\bibfield  {title} {\enquote {\bibinfo {title}
  {{Outlook for detecting the gravitational-wave displacement and spin memory
  effects with current and future gravitational-wave detectors}},}\ }\href
  {\doibase 10.1103/PhysRevD.107.064056} {\bibfield  {journal} {\bibinfo
  {journal} {Phys. Rev. D}\ }\textbf {\bibinfo {volume} {107}},\ \bibinfo
  {pages} {064056} (\bibinfo {year} {2023})},\ \bibinfo {note} {[Erratum:
  Phys.Rev.D 108, 029901 (2023)]},\ \Eprint {http://arxiv.org/abs/2210.16266}
  {arXiv:2210.16266 [gr-qc]} \BibitemShut {NoStop}%
\bibitem [{\citenamefont {H{\"u}bner}\ \emph {et~al.}(2021)\citenamefont
  {H{\"u}bner}, \citenamefont {Lasky},\ and\ \citenamefont
  {Thrane}}]{Hubner:2021amk}%
  \BibitemOpen
  \bibfield  {author} {\bibinfo {author} {\bibfnamefont {Moritz}\ \bibnamefont
  {H{\"u}bner}}, \bibinfo {author} {\bibfnamefont {Paul}\ \bibnamefont
  {Lasky}}, \ and\ \bibinfo {author} {\bibfnamefont {Eric}\ \bibnamefont
  {Thrane}},\ }\bibfield  {title} {\enquote {\bibinfo {title} {{Memory remains
  undetected: Updates from the second LIGO/Virgo gravitational-wave transient
  catalog}},}\ }\href {\doibase 10.1103/PhysRevD.104.023004} {\bibfield
  {journal} {\bibinfo  {journal} {Phys. Rev. D}\ }\textbf {\bibinfo {volume}
  {104}},\ \bibinfo {pages} {023004} (\bibinfo {year} {2021})},\ \Eprint
  {http://arxiv.org/abs/2105.02879} {arXiv:2105.02879 [gr-qc]} \BibitemShut
  {NoStop}%
\bibitem [{\citenamefont {Cheung}\ \emph {et~al.}(2024)\citenamefont {Cheung},
  \citenamefont {Lasky},\ and\ \citenamefont {Thrane}}]{Cheung:2024zow}%
  \BibitemOpen
  \bibfield  {author} {\bibinfo {author} {\bibfnamefont {Shun~Yin}\
  \bibnamefont {Cheung}}, \bibinfo {author} {\bibfnamefont {Paul~D.}\
  \bibnamefont {Lasky}}, \ and\ \bibinfo {author} {\bibfnamefont {Eric}\
  \bibnamefont {Thrane}},\ }\bibfield  {title} {\enquote {\bibinfo {title}
  {{Does spacetime have memories? Searching for gravitational-wave memory in
  the third LIGO-Virgo-KAGRA gravitational-wave transient catalogue}},}\ }\href
  {\doibase 10.1088/1361-6382/ad3ffe} {\bibfield  {journal} {\bibinfo
  {journal} {Class. Quant. Grav.}\ }\textbf {\bibinfo {volume} {41}},\ \bibinfo
  {pages} {115010} (\bibinfo {year} {2024})},\ \Eprint
  {http://arxiv.org/abs/2404.11919} {arXiv:2404.11919 [gr-qc]} \BibitemShut
  {NoStop}%
\bibitem [{\citenamefont {Ebersold}\ and\ \citenamefont
  {Tiwari}(2020)}]{Ebersold:2020zah}%
  \BibitemOpen
  \bibfield  {author} {\bibinfo {author} {\bibfnamefont {Michael}\ \bibnamefont
  {Ebersold}}\ and\ \bibinfo {author} {\bibfnamefont {Shubhanshu}\ \bibnamefont
  {Tiwari}},\ }\bibfield  {title} {\enquote {\bibinfo {title} {{Search for
  nonlinear memory from subsolar mass compact binary mergers}},}\ }\href
  {\doibase 10.1103/PhysRevD.101.104041} {\bibfield  {journal} {\bibinfo
  {journal} {Phys. Rev. D}\ }\textbf {\bibinfo {volume} {101}},\ \bibinfo
  {pages} {104041} (\bibinfo {year} {2020})},\ \Eprint
  {http://arxiv.org/abs/2005.03306} {arXiv:2005.03306 [gr-qc]} \BibitemShut
  {NoStop}%
\bibitem [{\citenamefont {Wang}\ \emph {et~al.}(2015)\citenamefont {Wang} \emph
  {et~al.}}]{Wang:2014zls}%
  \BibitemOpen
  \bibfield  {author} {\bibinfo {author} {\bibfnamefont {J.~B.}\ \bibnamefont
  {Wang}} \emph {et~al.},\ }\bibfield  {title} {\enquote {\bibinfo {title}
  {{Searching for gravitational wave memory bursts with the Parkes Pulsar
  Timing Array}},}\ }\href {\doibase 10.1093/mnras/stu2137} {\bibfield
  {journal} {\bibinfo  {journal} {Mon. Not. Roy. Astron. Soc.}\ }\textbf
  {\bibinfo {volume} {446}},\ \bibinfo {pages} {1657--1671} (\bibinfo {year}
  {2015})},\ \Eprint {http://arxiv.org/abs/1410.3323} {arXiv:1410.3323
  [astro-ph.GA]} \BibitemShut {NoStop}%
\bibitem [{\citenamefont {Agazie}\ \emph {et~al.}(2024)\citenamefont {Agazie}
  \emph {et~al.}}]{NANOGrav:2023vfo}%
  \BibitemOpen
  \bibfield  {author} {\bibinfo {author} {\bibfnamefont {Gabriella}\
  \bibnamefont {Agazie}} \emph {et~al.} (\bibinfo {collaboration} {NANOGrav}),\
  }\bibfield  {title} {\enquote {\bibinfo {title} {{The NANOGrav 12.5 yr Data
  Set: Search for Gravitational Wave Memory}},}\ }\href {\doibase
  10.3847/1538-4357/ad0726} {\bibfield  {journal} {\bibinfo  {journal}
  {Astrophys. J.}\ }\textbf {\bibinfo {volume} {963}},\ \bibinfo {pages} {61}
  (\bibinfo {year} {2024})},\ \Eprint {http://arxiv.org/abs/2307.13797}
  {arXiv:2307.13797 [gr-qc]} \BibitemShut {NoStop}%
\bibitem [{\citenamefont {Johnson}\ \emph {et~al.}(2019)\citenamefont
  {Johnson}, \citenamefont {Kapadia}, \citenamefont {Osborne}, \citenamefont
  {Hixon},\ and\ \citenamefont {Kennefick}}]{Johnson:2018xly}%
  \BibitemOpen
  \bibfield  {author} {\bibinfo {author} {\bibfnamefont {Aaron~D.}\
  \bibnamefont {Johnson}}, \bibinfo {author} {\bibfnamefont {Shasvath~J.}\
  \bibnamefont {Kapadia}}, \bibinfo {author} {\bibfnamefont {Andrew}\
  \bibnamefont {Osborne}}, \bibinfo {author} {\bibfnamefont {Alex}\
  \bibnamefont {Hixon}}, \ and\ \bibinfo {author} {\bibfnamefont {Daniel}\
  \bibnamefont {Kennefick}},\ }\bibfield  {title} {\enquote {\bibinfo {title}
  {{Prospects of detecting the nonlinear gravitational wave memory}},}\ }\href
  {\doibase 10.1103/PhysRevD.99.044045} {\bibfield  {journal} {\bibinfo
  {journal} {Phys. Rev. D}\ }\textbf {\bibinfo {volume} {99}},\ \bibinfo
  {pages} {044045} (\bibinfo {year} {2019})},\ \Eprint
  {http://arxiv.org/abs/1810.09563} {arXiv:1810.09563 [gr-qc]} \BibitemShut
  {NoStop}%
\bibitem [{\citenamefont {Goncharov}\ \emph {et~al.}(2024)\citenamefont
  {Goncharov}, \citenamefont {Donnay},\ and\ \citenamefont
  {Harms}}]{Goncharov:2023woe}%
  \BibitemOpen
  \bibfield  {author} {\bibinfo {author} {\bibfnamefont {Boris}\ \bibnamefont
  {Goncharov}}, \bibinfo {author} {\bibfnamefont {Laura}\ \bibnamefont
  {Donnay}}, \ and\ \bibinfo {author} {\bibfnamefont {Jan}\ \bibnamefont
  {Harms}},\ }\bibfield  {title} {\enquote {\bibinfo {title} {{Inferring
  Fundamental Spacetime Symmetries with Gravitational-Wave Memory: From LISA to
  the Einstein Telescope}},}\ }\href {\doibase 10.1103/PhysRevLett.132.241401}
  {\bibfield  {journal} {\bibinfo  {journal} {Phys. Rev. Lett.}\ }\textbf
  {\bibinfo {volume} {132}},\ \bibinfo {pages} {241401} (\bibinfo {year}
  {2024})},\ \Eprint {http://arxiv.org/abs/2310.10718} {arXiv:2310.10718
  [gr-qc]} \BibitemShut {NoStop}%
\bibitem [{\citenamefont {Islo}\ \emph {et~al.}(2019)\citenamefont {Islo},
  \citenamefont {Simon}, \citenamefont {Burke-Spolaor},\ and\ \citenamefont
  {Siemens}}]{Islo:2019qht}%
  \BibitemOpen
  \bibfield  {author} {\bibinfo {author} {\bibfnamefont {Kristina}\
  \bibnamefont {Islo}}, \bibinfo {author} {\bibfnamefont {Joseph}\ \bibnamefont
  {Simon}}, \bibinfo {author} {\bibfnamefont {Sarah}\ \bibnamefont
  {Burke-Spolaor}}, \ and\ \bibinfo {author} {\bibfnamefont {Xavier}\
  \bibnamefont {Siemens}},\ }\bibfield  {title} {\enquote {\bibinfo {title}
  {{Prospects for Memory Detection with Low-Frequency Gravitational Wave
  Detectors}},}\ }\href@noop {} {\bibfield  {journal} {\bibinfo  {journal}
  {{}}\ } (\bibinfo {year} {2019})},\ \Eprint {http://arxiv.org/abs/1906.11936}
  {arXiv:1906.11936 [astro-ph.HE]} \BibitemShut {NoStop}%
\bibitem [{\citenamefont {Gasparotto}\ \emph {et~al.}(2023)\citenamefont
  {Gasparotto}, \citenamefont {Vicente}, \citenamefont {Blas}, \citenamefont
  {Jenkins},\ and\ \citenamefont {Barausse}}]{Gasparotto:2023fcg}%
  \BibitemOpen
  \bibfield  {author} {\bibinfo {author} {\bibfnamefont {Silvia}\ \bibnamefont
  {Gasparotto}}, \bibinfo {author} {\bibfnamefont {Rodrigo}\ \bibnamefont
  {Vicente}}, \bibinfo {author} {\bibfnamefont {Diego}\ \bibnamefont {Blas}},
  \bibinfo {author} {\bibfnamefont {Alexander~C.}\ \bibnamefont {Jenkins}}, \
  and\ \bibinfo {author} {\bibfnamefont {Enrico}\ \bibnamefont {Barausse}},\
  }\bibfield  {title} {\enquote {\bibinfo {title} {{Can gravitational-wave
  memory help constrain binary black-hole parameters? A LISA case study}},}\
  }\href {\doibase 10.1103/PhysRevD.107.124033} {\bibfield  {journal} {\bibinfo
   {journal} {Phys. Rev. D}\ }\textbf {\bibinfo {volume} {107}},\ \bibinfo
  {pages} {124033} (\bibinfo {year} {2023})},\ \Eprint
  {http://arxiv.org/abs/2301.13228} {arXiv:2301.13228 [gr-qc]} \BibitemShut
  {NoStop}%
\bibitem [{\citenamefont {Inchausp{\'e}}\ \emph {et~al.}(2025)\citenamefont
  {Inchausp{\'e}}, \citenamefont {Gasparotto}, \citenamefont {Blas},
  \citenamefont {Heisenberg}, \citenamefont {Zosso},\ and\ \citenamefont
  {Tiwari}}]{Inchauspe:2024ibs}%
  \BibitemOpen
  \bibfield  {author} {\bibinfo {author} {\bibfnamefont {Henri}\ \bibnamefont
  {Inchausp{\'e}}}, \bibinfo {author} {\bibfnamefont {Silvia}\ \bibnamefont
  {Gasparotto}}, \bibinfo {author} {\bibfnamefont {Diego}\ \bibnamefont
  {Blas}}, \bibinfo {author} {\bibfnamefont {Lavinia}\ \bibnamefont
  {Heisenberg}}, \bibinfo {author} {\bibfnamefont {Jann}\ \bibnamefont
  {Zosso}}, \ and\ \bibinfo {author} {\bibfnamefont {Shubhanshu}\ \bibnamefont
  {Tiwari}},\ }\bibfield  {title} {\enquote {\bibinfo {title} {{Measuring
  gravitational wave memory with LISA}},}\ }\href {\doibase
  10.1103/PhysRevD.111.044044} {\bibfield  {journal} {\bibinfo  {journal}
  {Phys. Rev. D}\ }\textbf {\bibinfo {volume} {111}},\ \bibinfo {pages}
  {044044} (\bibinfo {year} {2025})},\ \Eprint
  {http://arxiv.org/abs/2406.09228} {arXiv:2406.09228 [gr-qc]} \BibitemShut
  {NoStop}%
\bibitem [{\citenamefont {Tiwari}\ \emph {et~al.}(2021)\citenamefont {Tiwari},
  \citenamefont {Ebersold},\ and\ \citenamefont {Hamilton}}]{Tiwari:2021gfl}%
  \BibitemOpen
  \bibfield  {author} {\bibinfo {author} {\bibfnamefont {Shubhanshu}\
  \bibnamefont {Tiwari}}, \bibinfo {author} {\bibfnamefont {Michael}\
  \bibnamefont {Ebersold}}, \ and\ \bibinfo {author} {\bibfnamefont
  {Eleanor~Z.}\ \bibnamefont {Hamilton}},\ }\bibfield  {title} {\enquote
  {\bibinfo {title} {{Leveraging gravitational-wave memory to distinguish
  neutron star-black hole binaries from black hole binaries}},}\ }\href
  {\doibase 10.1103/PhysRevD.104.123024} {\bibfield  {journal} {\bibinfo
  {journal} {Phys. Rev. D}\ }\textbf {\bibinfo {volume} {104}},\ \bibinfo
  {pages} {123024} (\bibinfo {year} {2021})},\ \Eprint
  {http://arxiv.org/abs/2110.11171} {arXiv:2110.11171 [gr-qc]} \BibitemShut
  {NoStop}%
\bibitem [{\citenamefont {Strominger}\ and\ \citenamefont
  {Zhiboedov}(2016)}]{Strominger:2014pwa}%
  \BibitemOpen
  \bibfield  {author} {\bibinfo {author} {\bibfnamefont {Andrew}\ \bibnamefont
  {Strominger}}\ and\ \bibinfo {author} {\bibfnamefont {Alexander}\
  \bibnamefont {Zhiboedov}},\ }\bibfield  {title} {\enquote {\bibinfo {title}
  {{Gravitational Memory, BMS Supertranslations and Soft Theorems}},}\ }\href
  {\doibase 10.1007/JHEP01(2016)086} {\bibfield  {journal} {\bibinfo  {journal}
  {JHEP}\ }\textbf {\bibinfo {volume} {01}},\ \bibinfo {pages} {086} (\bibinfo
  {year} {2016})},\ \Eprint {http://arxiv.org/abs/1411.5745} {arXiv:1411.5745
  [hep-th]} \BibitemShut {NoStop}%
\bibitem [{\citenamefont {Strominger}(2017)}]{Strominger:2017zoo}%
  \BibitemOpen
  \bibfield  {author} {\bibinfo {author} {\bibfnamefont {Andrew}\ \bibnamefont
  {Strominger}},\ }\bibfield  {title} {\enquote {\bibinfo {title} {{Lectures on
  the Infrared Structure of Gravity and Gauge Theory}},}\ }\href@noop {}
  {\bibfield  {journal} {\bibinfo  {journal} {{}}\ } (\bibinfo {year}
  {2017})},\ \Eprint {http://arxiv.org/abs/1703.05448} {arXiv:1703.05448
  [hep-th]} \BibitemShut {NoStop}%
\bibitem [{\citenamefont {Weinberg}(1965)}]{Weinberg:1965nx}%
  \BibitemOpen
  \bibfield  {author} {\bibinfo {author} {\bibfnamefont {Steven}\ \bibnamefont
  {Weinberg}},\ }\bibfield  {title} {\enquote {\bibinfo {title} {{Infrared
  photons and gravitons}},}\ }\href {\doibase 10.1103/PhysRev.140.B516}
  {\bibfield  {journal} {\bibinfo  {journal} {Phys. Rev.}\ }\textbf {\bibinfo
  {volume} {140}},\ \bibinfo {pages} {B516--B524} (\bibinfo {year}
  {1965})}\BibitemShut {NoStop}%
\bibitem [{\citenamefont {Nichols}(2017)}]{Nichols:2017rqr}%
  \BibitemOpen
  \bibfield  {author} {\bibinfo {author} {\bibfnamefont {David~A.}\
  \bibnamefont {Nichols}},\ }\bibfield  {title} {\enquote {\bibinfo {title}
  {{Spin memory effect for compact binaries in the post-Newtonian
  approximation}},}\ }\href {\doibase 10.1103/PhysRevD.95.084048} {\bibfield
  {journal} {\bibinfo  {journal} {Phys. Rev.}\ }\textbf {\bibinfo {volume}
  {D95}},\ \bibinfo {pages} {084048} (\bibinfo {year} {2017})},\ \Eprint
  {http://arxiv.org/abs/1702.03300} {arXiv:1702.03300 [gr-qc]} \BibitemShut
  {NoStop}%
\bibitem [{\citenamefont {Flanagan}\ \emph {et~al.}(2019)\citenamefont
  {Flanagan}, \citenamefont {Grant}, \citenamefont {Harte},\ and\ \citenamefont
  {Nichols}}]{Flanagan:2018yzh}%
  \BibitemOpen
  \bibfield  {author} {\bibinfo {author} {\bibfnamefont {{\'E}anna~{\'E}.}\
  \bibnamefont {Flanagan}}, \bibinfo {author} {\bibfnamefont {Alexander~M.}\
  \bibnamefont {Grant}}, \bibinfo {author} {\bibfnamefont {Abraham~I.}\
  \bibnamefont {Harte}}, \ and\ \bibinfo {author} {\bibfnamefont {David~A.}\
  \bibnamefont {Nichols}},\ }\bibfield  {title} {\enquote {\bibinfo {title}
  {{Persistent gravitational wave observables: general framework}},}\ }\href
  {\doibase 10.1103/PhysRevD.99.084044} {\bibfield  {journal} {\bibinfo
  {journal} {Phys. Rev. D}\ }\textbf {\bibinfo {volume} {99}},\ \bibinfo
  {pages} {084044} (\bibinfo {year} {2019})},\ \Eprint
  {http://arxiv.org/abs/1901.00021} {arXiv:1901.00021 [gr-qc]} \BibitemShut
  {NoStop}%
\bibitem [{\citenamefont {Grant}(2024)}]{Grant:2023ged}%
  \BibitemOpen
  \bibfield  {author} {\bibinfo {author} {\bibfnamefont {Alexander~M.}\
  \bibnamefont {Grant}},\ }\bibfield  {title} {\enquote {\bibinfo {title}
  {{Persistent gravitational wave observables: nonlinearities in (non-)geodesic
  deviation}},}\ }\href {\doibase 10.1088/1361-6382/ad48f5} {\bibfield
  {journal} {\bibinfo  {journal} {Class. Quant. Grav.}\ }\textbf {\bibinfo
  {volume} {41}},\ \bibinfo {pages} {175004} (\bibinfo {year} {2024})},\
  \Eprint {http://arxiv.org/abs/2401.00047} {arXiv:2401.00047 [gr-qc]}
  \BibitemShut {NoStop}%
\bibitem [{\citenamefont {Grant}\ and\ \citenamefont
  {Nichols}(2022)}]{Grant:2021hga}%
  \BibitemOpen
  \bibfield  {author} {\bibinfo {author} {\bibfnamefont {Alexander~M.}\
  \bibnamefont {Grant}}\ and\ \bibinfo {author} {\bibfnamefont {David~A.}\
  \bibnamefont {Nichols}},\ }\bibfield  {title} {\enquote {\bibinfo {title}
  {{Persistent gravitational wave observables: Curve deviation in
  asymptotically flat spacetimes}},}\ }\href {\doibase
  10.1103/PhysRevD.105.024056} {\bibfield  {journal} {\bibinfo  {journal}
  {Phys. Rev. D}\ }\textbf {\bibinfo {volume} {105}},\ \bibinfo {pages}
  {024056} (\bibinfo {year} {2022})},\ \bibinfo {note} {[Erratum: Phys.Rev.D
  107, 109902 (2023)]},\ \Eprint {http://arxiv.org/abs/2109.03832}
  {arXiv:2109.03832 [gr-qc]} \BibitemShut {NoStop}%
\bibitem [{\citenamefont {Pasterski}\ \emph {et~al.}(2016)\citenamefont
  {Pasterski}, \citenamefont {Strominger},\ and\ \citenamefont
  {Zhiboedov}}]{Pasterski:2015tva}%
  \BibitemOpen
  \bibfield  {author} {\bibinfo {author} {\bibfnamefont {Sabrina}\ \bibnamefont
  {Pasterski}}, \bibinfo {author} {\bibfnamefont {Andrew}\ \bibnamefont
  {Strominger}}, \ and\ \bibinfo {author} {\bibfnamefont {Alexander}\
  \bibnamefont {Zhiboedov}},\ }\bibfield  {title} {\enquote {\bibinfo {title}
  {{New Gravitational Memories}},}\ }\href {\doibase 10.1007/JHEP12(2016)053}
  {\bibfield  {journal} {\bibinfo  {journal} {JHEP}\ }\textbf {\bibinfo
  {volume} {12}},\ \bibinfo {pages} {053} (\bibinfo {year} {2016})},\ \Eprint
  {http://arxiv.org/abs/1502.06120} {arXiv:1502.06120 [hep-th]} \BibitemShut
  {NoStop}%
\bibitem [{\citenamefont {Flanagan}\ and\ \citenamefont
  {Nichols}(2017)}]{Flanagan:2015pxa}%
  \BibitemOpen
  \bibfield  {author} {\bibinfo {author} {\bibfnamefont {Eanna~E.}\
  \bibnamefont {Flanagan}}\ and\ \bibinfo {author} {\bibfnamefont {David~A.}\
  \bibnamefont {Nichols}},\ }\bibfield  {title} {\enquote {\bibinfo {title}
  {{Conserved charges of the extended Bondi-Metzner-Sachs algebra}},}\ }\href
  {\doibase 10.1103/PhysRevD.95.044002} {\bibfield  {journal} {\bibinfo
  {journal} {Phys. Rev. D}\ }\textbf {\bibinfo {volume} {95}},\ \bibinfo
  {pages} {044002} (\bibinfo {year} {2017})},\ \Eprint
  {http://arxiv.org/abs/1510.03386} {arXiv:1510.03386 [hep-th]} \BibitemShut
  {NoStop}%
\bibitem [{\citenamefont {Nichols}(2018)}]{Nichols:2018qac}%
  \BibitemOpen
  \bibfield  {author} {\bibinfo {author} {\bibfnamefont {David~A.}\
  \bibnamefont {Nichols}},\ }\bibfield  {title} {\enquote {\bibinfo {title}
  {{Center-of-mass angular momentum and memory effect in asymptotically flat
  spacetimes}},}\ }\href {\doibase 10.1103/PhysRevD.98.064032} {\bibfield
  {journal} {\bibinfo  {journal} {Phys. Rev. D}\ }\textbf {\bibinfo {volume}
  {98}},\ \bibinfo {pages} {064032} (\bibinfo {year} {2018})},\ \Eprint
  {http://arxiv.org/abs/1807.08767} {arXiv:1807.08767 [gr-qc]} \BibitemShut
  {NoStop}%
\bibitem [{\citenamefont {Compère}\ \emph {et~al.}(2018)\citenamefont
  {Compère}, \citenamefont {Fiorucci},\ and\ \citenamefont
  {Ruzziconi}}]{Compere:2018ylh}%
  \BibitemOpen
  \bibfield  {author} {\bibinfo {author} {\bibfnamefont {Geoffrey}\
  \bibnamefont {Compère}}, \bibinfo {author} {\bibfnamefont {Adrien}\
  \bibnamefont {Fiorucci}}, \ and\ \bibinfo {author} {\bibfnamefont {Romain}\
  \bibnamefont {Ruzziconi}},\ }\bibfield  {title} {\enquote {\bibinfo {title}
  {{Superboost transitions, refraction memory and super-Lorentz charge
  algebra}},}\ }\href {\doibase 10.1007/JHEP11(2018)200} {\bibfield  {journal}
  {\bibinfo  {journal} {J. High Energy Phys.}\ }\textbf {\bibinfo {volume}
  {11}},\ \bibinfo {pages} {200} (\bibinfo {year} {2018})},\ \bibinfo {note}
  {[Erratum: J. High Energy Phys. 04, 172 (2020)]},\ \Eprint
  {http://arxiv.org/abs/1810.00377} {arXiv:1810.00377 [hep-th]} \BibitemShut
  {NoStop}%
\bibitem [{\citenamefont {Grant}\ and\ \citenamefont
  {Mitman}(2024)}]{Grant:2023jhd}%
  \BibitemOpen
  \bibfield  {author} {\bibinfo {author} {\bibfnamefont {Alexander~M.}\
  \bibnamefont {Grant}}\ and\ \bibinfo {author} {\bibfnamefont {Keefe}\
  \bibnamefont {Mitman}},\ }\bibfield  {title} {\enquote {\bibinfo {title}
  {{Higher memory effects in numerical simulations of binary black hole
  mergers}},}\ }\href {\doibase 10.1088/1361-6382/ad5d46} {\bibfield  {journal}
  {\bibinfo  {journal} {Classical Quantum Gravity}\ }\textbf {\bibinfo {volume}
  {41}},\ \bibinfo {pages} {175003} (\bibinfo {year} {2024})},\ \Eprint
  {http://arxiv.org/abs/2312.02295} {arXiv:2312.02295 [gr-qc]} \BibitemShut
  {NoStop}%
\bibitem [{\citenamefont {Siddhant}\ \emph {et~al.}(2024)\citenamefont
  {Siddhant}, \citenamefont {Grant},\ and\ \citenamefont
  {Nichols}}]{Siddhant:2024nft}%
  \BibitemOpen
  \bibfield  {author} {\bibinfo {author} {\bibfnamefont {Siddhant}\
  \bibnamefont {Siddhant}}, \bibinfo {author} {\bibfnamefont {Alexander~M.}\
  \bibnamefont {Grant}}, \ and\ \bibinfo {author} {\bibfnamefont {David~A.}\
  \bibnamefont {Nichols}},\ }\bibfield  {title} {\enquote {\bibinfo {title}
  {{Higher memory effects and the post-Newtonian calculation of their
  gravitational-wave signals}},}\ }\href {\doibase 10.1088/1361-6382/ad7663}
  {\bibfield  {journal} {\bibinfo  {journal} {Classical Quantum Gravity}\
  }\textbf {\bibinfo {volume} {41}},\ \bibinfo {pages} {205014} (\bibinfo
  {year} {2024})},\ \Eprint {http://arxiv.org/abs/2403.13907} {arXiv:2403.13907
  [gr-qc]} \BibitemShut {NoStop}%
\bibitem [{\citenamefont {Bondi}\ \emph {et~al.}(1962)\citenamefont {Bondi},
  \citenamefont {van~der Burg},\ and\ \citenamefont {Metzner}}]{Bondi:1962px}%
  \BibitemOpen
  \bibfield  {author} {\bibinfo {author} {\bibfnamefont {H.}~\bibnamefont
  {Bondi}}, \bibinfo {author} {\bibfnamefont {M.G.J.}\ \bibnamefont {van~der
  Burg}}, \ and\ \bibinfo {author} {\bibfnamefont {A.W.K.}\ \bibnamefont
  {Metzner}},\ }\bibfield  {title} {\enquote {\bibinfo {title} {{Gravitational
  waves in general relativity. 7. Waves from axisymmetric isolated systems}},}\
  }\href {\doibase 10.1098/rspa.1962.0161} {\bibfield  {journal} {\bibinfo
  {journal} {Proc. Roy. Soc. Lond. A}\ }\textbf {\bibinfo {volume} {A269}},\
  \bibinfo {pages} {21--52} (\bibinfo {year} {1962})}\BibitemShut {NoStop}%
\bibitem [{\citenamefont {Sachs}(1962)}]{Sachs:1962wk}%
  \BibitemOpen
  \bibfield  {author} {\bibinfo {author} {\bibfnamefont {R.K.}\ \bibnamefont
  {Sachs}},\ }\bibfield  {title} {\enquote {\bibinfo {title} {{Gravitational
  waves in general relativity. 8. Waves in asymptotically flat space-times}},}\
  }\href {\doibase 10.1098/rspa.1962.0206} {\bibfield  {journal} {\bibinfo
  {journal} {Proc. Roy. Soc. Lond. A}\ }\textbf {\bibinfo {volume} {A270}},\
  \bibinfo {pages} {103--126} (\bibinfo {year} {1962})}\BibitemShut {NoStop}%
\bibitem [{\citenamefont {Barnich}\ and\ \citenamefont
  {Troessaert}(2010{\natexlab{a}})}]{Barnich:2009se}%
  \BibitemOpen
  \bibfield  {author} {\bibinfo {author} {\bibfnamefont {Glenn}\ \bibnamefont
  {Barnich}}\ and\ \bibinfo {author} {\bibfnamefont {Cedric}\ \bibnamefont
  {Troessaert}},\ }\bibfield  {title} {\enquote {\bibinfo {title} {{Symmetries
  of asymptotically flat 4 dimensional spacetimes at null infinity
  revisited}},}\ }\href {\doibase 10.1103/PhysRevLett.105.111103} {\bibfield
  {journal} {\bibinfo  {journal} {Phys. Rev. Lett.}\ }\textbf {\bibinfo
  {volume} {105}},\ \bibinfo {pages} {111103} (\bibinfo {year}
  {2010}{\natexlab{a}})},\ \Eprint {http://arxiv.org/abs/0909.2617}
  {arXiv:0909.2617 [gr-qc]} \BibitemShut {NoStop}%
\bibitem [{\citenamefont {Barnich}\ and\ \citenamefont
  {Troessaert}(2010{\natexlab{b}})}]{Barnich:2010eb}%
  \BibitemOpen
  \bibfield  {author} {\bibinfo {author} {\bibfnamefont {Glenn}\ \bibnamefont
  {Barnich}}\ and\ \bibinfo {author} {\bibfnamefont {Cedric}\ \bibnamefont
  {Troessaert}},\ }\bibfield  {title} {\enquote {\bibinfo {title} {{Aspects of
  the BMS/CFT correspondence}},}\ }\href {\doibase 10.1007/JHEP05(2010)062}
  {\bibfield  {journal} {\bibinfo  {journal} {JHEP}\ }\textbf {\bibinfo
  {volume} {05}},\ \bibinfo {pages} {062} (\bibinfo {year}
  {2010}{\natexlab{b}})},\ \Eprint {http://arxiv.org/abs/1001.1541}
  {arXiv:1001.1541 [hep-th]} \BibitemShut {NoStop}%
\bibitem [{\citenamefont {Barnich}\ and\ \citenamefont
  {Troessaert}(2011)}]{Barnich:2011mi}%
  \BibitemOpen
  \bibfield  {author} {\bibinfo {author} {\bibfnamefont {Glenn}\ \bibnamefont
  {Barnich}}\ and\ \bibinfo {author} {\bibfnamefont {Cedric}\ \bibnamefont
  {Troessaert}},\ }\bibfield  {title} {\enquote {\bibinfo {title} {{BMS charge
  algebra}},}\ }\href {\doibase 10.1007/JHEP12(2011)105} {\bibfield  {journal}
  {\bibinfo  {journal} {JHEP}\ }\textbf {\bibinfo {volume} {12}},\ \bibinfo
  {pages} {105} (\bibinfo {year} {2011})},\ \Eprint
  {http://arxiv.org/abs/1106.0213} {arXiv:1106.0213 [hep-th]} \BibitemShut
  {NoStop}%
\bibitem [{\citenamefont {Campiglia}\ and\ \citenamefont
  {Laddha}(2014)}]{Campiglia:2014yka}%
  \BibitemOpen
  \bibfield  {author} {\bibinfo {author} {\bibfnamefont {Miguel}\ \bibnamefont
  {Campiglia}}\ and\ \bibinfo {author} {\bibfnamefont {Alok}\ \bibnamefont
  {Laddha}},\ }\bibfield  {title} {\enquote {\bibinfo {title} {{Asymptotic
  symmetries and subleading soft graviton theorem}},}\ }\href {\doibase
  10.1103/PhysRevD.90.124028} {\bibfield  {journal} {\bibinfo  {journal} {Phys.
  Rev. D}\ }\textbf {\bibinfo {volume} {90}},\ \bibinfo {pages} {124028}
  (\bibinfo {year} {2014})},\ \Eprint {http://arxiv.org/abs/1408.2228}
  {arXiv:1408.2228 [hep-th]} \BibitemShut {NoStop}%
\bibitem [{\citenamefont {Campiglia}\ and\ \citenamefont
  {Laddha}(2015)}]{Campiglia:2015yka}%
  \BibitemOpen
  \bibfield  {author} {\bibinfo {author} {\bibfnamefont {Miguel}\ \bibnamefont
  {Campiglia}}\ and\ \bibinfo {author} {\bibfnamefont {Alok}\ \bibnamefont
  {Laddha}},\ }\bibfield  {title} {\enquote {\bibinfo {title} {{New symmetries
  for the Gravitational S-matrix}},}\ }\href {\doibase 10.1007/JHEP04(2015)076}
  {\bibfield  {journal} {\bibinfo  {journal} {JHEP}\ }\textbf {\bibinfo
  {volume} {04}},\ \bibinfo {pages} {076} (\bibinfo {year} {2015})},\ \Eprint
  {http://arxiv.org/abs/1502.02318} {arXiv:1502.02318 [hep-th]} \BibitemShut
  {NoStop}%
\bibitem [{\citenamefont {Strominger}(2021)}]{Strominger:2021mtt}%
  \BibitemOpen
  \bibfield  {author} {\bibinfo {author} {\bibfnamefont {Andrew}\ \bibnamefont
  {Strominger}},\ }\bibfield  {title} {\enquote {\bibinfo {title}
  {{$w_{1+\infty}$ Algebra and the Celestial Sphere: Infinite Towers of Soft
  Graviton, Photon, and Gluon Symmetries}},}\ }\href {\doibase
  10.1103/PhysRevLett.127.221601} {\bibfield  {journal} {\bibinfo  {journal}
  {Phys. Rev. Lett.}\ }\textbf {\bibinfo {volume} {127}},\ \bibinfo {pages}
  {221601} (\bibinfo {year} {2021})},\ \Eprint
  {http://arxiv.org/abs/2105.14346} {arXiv:2105.14346 [hep-th]} \BibitemShut
  {NoStop}%
\bibitem [{\citenamefont {Freidel}\ \emph {et~al.}(2022)\citenamefont
  {Freidel}, \citenamefont {Pranzetti},\ and\ \citenamefont
  {Raclariu}}]{Freidel:2021ytz}%
  \BibitemOpen
  \bibfield  {author} {\bibinfo {author} {\bibfnamefont {Laurent}\ \bibnamefont
  {Freidel}}, \bibinfo {author} {\bibfnamefont {Daniele}\ \bibnamefont
  {Pranzetti}}, \ and\ \bibinfo {author} {\bibfnamefont {Ana-Maria}\
  \bibnamefont {Raclariu}},\ }\bibfield  {title} {\enquote {\bibinfo {title}
  {{Higher spin dynamics in gravity and w1+\ensuremath{\infty} celestial
  symmetries}},}\ }\href {\doibase 10.1103/PhysRevD.106.086013} {\bibfield
  {journal} {\bibinfo  {journal} {Phys. Rev. D}\ }\textbf {\bibinfo {volume}
  {106}},\ \bibinfo {pages} {086013} (\bibinfo {year} {2022})},\ \Eprint
  {http://arxiv.org/abs/2112.15573} {arXiv:2112.15573 [hep-th]} \BibitemShut
  {NoStop}%
\bibitem [{\citenamefont {Freidel}\ \emph {et~al.}(2024)\citenamefont
  {Freidel}, \citenamefont {Pranzetti},\ and\ \citenamefont
  {Raclariu}}]{Freidel:2022skz}%
  \BibitemOpen
  \bibfield  {author} {\bibinfo {author} {\bibfnamefont {Laurent}\ \bibnamefont
  {Freidel}}, \bibinfo {author} {\bibfnamefont {Daniele}\ \bibnamefont
  {Pranzetti}}, \ and\ \bibinfo {author} {\bibfnamefont {Ana-Maria}\
  \bibnamefont {Raclariu}},\ }\bibfield  {title} {\enquote {\bibinfo {title}
  {{A discrete basis for celestial holography}},}\ }\href {\doibase
  10.1007/JHEP02(2024)176} {\bibfield  {journal} {\bibinfo  {journal} {J. High
  Eenergy Phys.}\ }\textbf {\bibinfo {volume} {02}},\ \bibinfo {pages} {176}
  (\bibinfo {year} {2024})},\ \Eprint {http://arxiv.org/abs/2212.12469}
  {arXiv:2212.12469 [hep-th]} \BibitemShut {NoStop}%
\bibitem [{\citenamefont {Comp\`ere}\ \emph {et~al.}(2022)\citenamefont
  {Comp\`ere}, \citenamefont {Oliveri},\ and\ \citenamefont
  {Seraj}}]{Compere:2022zdz}%
  \BibitemOpen
  \bibfield  {author} {\bibinfo {author} {\bibfnamefont {Geoffrey}\
  \bibnamefont {Comp\`ere}}, \bibinfo {author} {\bibfnamefont {Roberto}\
  \bibnamefont {Oliveri}}, \ and\ \bibinfo {author} {\bibfnamefont {Ali}\
  \bibnamefont {Seraj}},\ }\bibfield  {title} {\enquote {\bibinfo {title}
  {{Metric reconstruction from celestial multipoles}},}\ }\href {\doibase
  10.1007/JHEP11(2022)001} {\bibfield  {journal} {\bibinfo  {journal} {J. High
  Energy Phys.}\ }\textbf {\bibinfo {volume} {11}},\ \bibinfo {pages} {001}
  (\bibinfo {year} {2022})},\ \Eprint {http://arxiv.org/abs/2206.12597}
  {arXiv:2206.12597 [hep-th]} \BibitemShut {NoStop}%
\bibitem [{\citenamefont {Geiller}(2025)}]{Geiller:2024bgf}%
  \BibitemOpen
  \bibfield  {author} {\bibinfo {author} {\bibfnamefont {Marc}\ \bibnamefont
  {Geiller}},\ }\bibfield  {title} {\enquote {\bibinfo {title} {{Celestial
  $w_{1+\infty}$ charges and the subleading structure of asymptotically-flat
  spacetimes}},}\ }\href {\doibase 10.21468/SciPostPhys.18.1.023} {\bibfield
  {journal} {\bibinfo  {journal} {SciPost Phys.}\ }\textbf {\bibinfo {volume}
  {18}},\ \bibinfo {pages} {023} (\bibinfo {year} {2025})},\ \Eprint
  {http://arxiv.org/abs/2403.05195} {arXiv:2403.05195 [hep-th]} \BibitemShut
  {NoStop}%
\bibitem [{\citenamefont {Favata}(2009{\natexlab{b}})}]{Favata:2008yd}%
  \BibitemOpen
  \bibfield  {author} {\bibinfo {author} {\bibfnamefont {Marc}\ \bibnamefont
  {Favata}},\ }\bibfield  {title} {\enquote {\bibinfo {title} {{Post-Newtonian
  corrections to the gravitational-wave memory for quasi-circular, inspiralling
  compact binaries}},}\ }\href {\doibase 10.1103/PhysRevD.80.024002} {\bibfield
   {journal} {\bibinfo  {journal} {Phys. Rev. D}\ }\textbf {\bibinfo {volume}
  {80}},\ \bibinfo {pages} {024002} (\bibinfo {year} {2009}{\natexlab{b}})},\
  \Eprint {http://arxiv.org/abs/0812.0069} {arXiv:0812.0069 [gr-qc]}
  \BibitemShut {NoStop}%
\bibitem [{\citenamefont {Favata}(2011)}]{Favata:2011qi}%
  \BibitemOpen
  \bibfield  {author} {\bibinfo {author} {\bibfnamefont {Marc}\ \bibnamefont
  {Favata}},\ }\bibfield  {title} {\enquote {\bibinfo {title} {{The
  Gravitational-wave memory from eccentric binaries}},}\ }\href {\doibase
  10.1103/PhysRevD.84.124013} {\bibfield  {journal} {\bibinfo  {journal} {Phys.
  Rev. D}\ }\textbf {\bibinfo {volume} {84}},\ \bibinfo {pages} {124013}
  (\bibinfo {year} {2011})},\ \Eprint {http://arxiv.org/abs/1108.3121}
  {arXiv:1108.3121 [gr-qc]} \BibitemShut {NoStop}%
\bibitem [{\citenamefont {Ebersold}\ \emph {et~al.}(2019)\citenamefont
  {Ebersold}, \citenamefont {Boetzel}, \citenamefont {Faye}, \citenamefont
  {Mishra}, \citenamefont {Iyer},\ and\ \citenamefont
  {Jetzer}}]{Ebersold:2019kdc}%
  \BibitemOpen
  \bibfield  {author} {\bibinfo {author} {\bibfnamefont {Michael}\ \bibnamefont
  {Ebersold}}, \bibinfo {author} {\bibfnamefont {Yannick}\ \bibnamefont
  {Boetzel}}, \bibinfo {author} {\bibfnamefont {Guillaume}\ \bibnamefont
  {Faye}}, \bibinfo {author} {\bibfnamefont {Chandra~Kant}\ \bibnamefont
  {Mishra}}, \bibinfo {author} {\bibfnamefont {Bala~R.}\ \bibnamefont {Iyer}},
  \ and\ \bibinfo {author} {\bibfnamefont {Philippe}\ \bibnamefont {Jetzer}},\
  }\bibfield  {title} {\enquote {\bibinfo {title} {{Gravitational-wave
  amplitudes for compact binaries in eccentric orbits at the third
  post-Newtonian order: Memory contributions}},}\ }\href {\doibase
  10.1103/PhysRevD.100.084043} {\bibfield  {journal} {\bibinfo  {journal}
  {Phys. Rev. D}\ }\textbf {\bibinfo {volume} {100}},\ \bibinfo {pages}
  {084043} (\bibinfo {year} {2019})},\ \Eprint
  {http://arxiv.org/abs/1906.06263} {arXiv:1906.06263 [gr-qc]} \BibitemShut
  {NoStop}%
\bibitem [{\citenamefont {Elhashash}\ and\ \citenamefont
  {Nichols}(2025)}]{Elhashash:2024thm}%
  \BibitemOpen
  \bibfield  {author} {\bibinfo {author} {\bibfnamefont {Arwa}\ \bibnamefont
  {Elhashash}}\ and\ \bibinfo {author} {\bibfnamefont {David~A.}\ \bibnamefont
  {Nichols}},\ }\bibfield  {title} {\enquote {\bibinfo {title} {{Waveform
  models for the gravitational-wave memory effect: Extreme mass-ratio limit and
  final memory offset}},}\ }\href {\doibase 10.1103/PhysRevD.111.044052}
  {\bibfield  {journal} {\bibinfo  {journal} {Phys. Rev. D}\ }\textbf {\bibinfo
  {volume} {111}},\ \bibinfo {pages} {044052} (\bibinfo {year} {2025})},\
  \Eprint {http://arxiv.org/abs/2407.19017} {arXiv:2407.19017 [gr-qc]}
  \BibitemShut {NoStop}%
\bibitem [{\citenamefont {Cunningham}\ \emph {et~al.}(2024)\citenamefont
  {Cunningham}, \citenamefont {Kavanagh}, \citenamefont {Pound}, \citenamefont
  {Trestini}, \citenamefont {Warburton},\ and\ \citenamefont
  {Neef}}]{Cunningham:2024dog}%
  \BibitemOpen
  \bibfield  {author} {\bibinfo {author} {\bibfnamefont {Kevin}\ \bibnamefont
  {Cunningham}}, \bibinfo {author} {\bibfnamefont {Chris}\ \bibnamefont
  {Kavanagh}}, \bibinfo {author} {\bibfnamefont {Adam}\ \bibnamefont {Pound}},
  \bibinfo {author} {\bibfnamefont {David}\ \bibnamefont {Trestini}}, \bibinfo
  {author} {\bibfnamefont {Niels}\ \bibnamefont {Warburton}}, \ and\ \bibinfo
  {author} {\bibfnamefont {Jakob}\ \bibnamefont {Neef}},\ }\bibfield  {title}
  {\enquote {\bibinfo {title} {{Gravitational memory: new results from
  post-Newtonian and self-force theory}},}\ }\href@noop {} {\bibfield
  {journal} {\bibinfo  {journal} {{}}\ } (\bibinfo {year} {2024})},\ \Eprint
  {http://arxiv.org/abs/2410.23950} {arXiv:2410.23950 [gr-qc]} \BibitemShut
  {NoStop}%
\bibitem [{\citenamefont {Trestini}\ and\ \citenamefont
  {Blanchet}(2023)}]{Trestini:2023wwg}%
  \BibitemOpen
  \bibfield  {author} {\bibinfo {author} {\bibfnamefont {David}\ \bibnamefont
  {Trestini}}\ and\ \bibinfo {author} {\bibfnamefont {Luc}\ \bibnamefont
  {Blanchet}},\ }\bibfield  {title} {\enquote {\bibinfo {title}
  {{Gravitational-wave tails of memory}},}\ }\href {\doibase
  10.1103/PhysRevD.107.104048} {\bibfield  {journal} {\bibinfo  {journal}
  {Phys. Rev. D}\ }\textbf {\bibinfo {volume} {107}},\ \bibinfo {pages}
  {104048} (\bibinfo {year} {2023})},\ \Eprint
  {http://arxiv.org/abs/2301.09395} {arXiv:2301.09395 [gr-qc]} \BibitemShut
  {NoStop}%
\bibitem [{\citenamefont {Blanchet}\ \emph
  {et~al.}(2023{\natexlab{a}})\citenamefont {Blanchet}, \citenamefont {Faye},
  \citenamefont {Henry}, \citenamefont {Larrouturou},\ and\ \citenamefont
  {Trestini}}]{Blanchet:2023bwj}%
  \BibitemOpen
  \bibfield  {author} {\bibinfo {author} {\bibfnamefont {Luc}\ \bibnamefont
  {Blanchet}}, \bibinfo {author} {\bibfnamefont {Guillaume}\ \bibnamefont
  {Faye}}, \bibinfo {author} {\bibfnamefont {Quentin}\ \bibnamefont {Henry}},
  \bibinfo {author} {\bibfnamefont {Fran\c{c}ois}\ \bibnamefont {Larrouturou}},
  \ and\ \bibinfo {author} {\bibfnamefont {David}\ \bibnamefont {Trestini}},\
  }\bibfield  {title} {\enquote {\bibinfo {title} {{Gravitational-Wave Phasing
  of Quasicircular Compact Binary Systems to the Fourth-and-a-Half
  Post-Newtonian Order}},}\ }\href {\doibase 10.1103/PhysRevLett.131.121402}
  {\bibfield  {journal} {\bibinfo  {journal} {Phys. Rev. Lett.}\ }\textbf
  {\bibinfo {volume} {131}},\ \bibinfo {pages} {121402} (\bibinfo {year}
  {2023}{\natexlab{a}})},\ \Eprint {http://arxiv.org/abs/2304.11185}
  {arXiv:2304.11185 [gr-qc]} \BibitemShut {NoStop}%
\bibitem [{\citenamefont {Blanchet}\ \emph
  {et~al.}(2023{\natexlab{b}})\citenamefont {Blanchet}, \citenamefont {Faye},
  \citenamefont {Henry}, \citenamefont {Larrouturou},\ and\ \citenamefont
  {Trestini}}]{Blanchet:2023sbv}%
  \BibitemOpen
  \bibfield  {author} {\bibinfo {author} {\bibfnamefont {Luc}\ \bibnamefont
  {Blanchet}}, \bibinfo {author} {\bibfnamefont {Guillaume}\ \bibnamefont
  {Faye}}, \bibinfo {author} {\bibfnamefont {Quentin}\ \bibnamefont {Henry}},
  \bibinfo {author} {\bibfnamefont {Fran\c{c}ois}\ \bibnamefont {Larrouturou}},
  \ and\ \bibinfo {author} {\bibfnamefont {David}\ \bibnamefont {Trestini}},\
  }\bibfield  {title} {\enquote {\bibinfo {title} {{Gravitational-wave flux and
  quadrupole modes from quasicircular nonspinning compact binaries to the
  fourth post-Newtonian order}},}\ }\href {\doibase
  10.1103/PhysRevD.108.064041} {\bibfield  {journal} {\bibinfo  {journal}
  {Phys. Rev. D}\ }\textbf {\bibinfo {volume} {108}},\ \bibinfo {pages}
  {064041} (\bibinfo {year} {2023}{\natexlab{b}})},\ \Eprint
  {http://arxiv.org/abs/2304.11186} {arXiv:2304.11186 [gr-qc]} \BibitemShut
  {NoStop}%
\bibitem [{\citenamefont {Mitman}\ \emph {et~al.}(2021)\citenamefont {Mitman}
  \emph {et~al.}}]{Mitman:2020bjf}%
  \BibitemOpen
  \bibfield  {author} {\bibinfo {author} {\bibfnamefont {Keefe}\ \bibnamefont
  {Mitman}} \emph {et~al.},\ }\bibfield  {title} {\enquote {\bibinfo {title}
  {{Adding gravitational memory to waveform catalogs using BMS balance
  laws}},}\ }\href {\doibase 10.1103/PhysRevD.103.024031} {\bibfield  {journal}
  {\bibinfo  {journal} {Phys. Rev. D}\ }\textbf {\bibinfo {volume} {103}},\
  \bibinfo {pages} {024031} (\bibinfo {year} {2021})},\ \Eprint
  {http://arxiv.org/abs/2011.01309} {arXiv:2011.01309 [gr-qc]} \BibitemShut
  {NoStop}%
\bibitem [{\citenamefont {Mitman}\ \emph {et~al.}(2020)\citenamefont {Mitman},
  \citenamefont {Moxon}, \citenamefont {Scheel}, \citenamefont {Teukolsky},
  \citenamefont {Boyle}, \citenamefont {Deppe}, \citenamefont {Kidder},\ and\
  \citenamefont {Throwe}}]{Mitman:2020pbt}%
  \BibitemOpen
  \bibfield  {author} {\bibinfo {author} {\bibfnamefont {Keefe}\ \bibnamefont
  {Mitman}}, \bibinfo {author} {\bibfnamefont {Jordan}\ \bibnamefont {Moxon}},
  \bibinfo {author} {\bibfnamefont {Mark~A.}\ \bibnamefont {Scheel}}, \bibinfo
  {author} {\bibfnamefont {Saul~A.}\ \bibnamefont {Teukolsky}}, \bibinfo
  {author} {\bibfnamefont {Michael}\ \bibnamefont {Boyle}}, \bibinfo {author}
  {\bibfnamefont {Nils}\ \bibnamefont {Deppe}}, \bibinfo {author}
  {\bibfnamefont {Lawrence~E.}\ \bibnamefont {Kidder}}, \ and\ \bibinfo
  {author} {\bibfnamefont {William}\ \bibnamefont {Throwe}},\ }\bibfield
  {title} {\enquote {\bibinfo {title} {{Computation of displacement and spin
  gravitational memory in numerical relativity}},}\ }\href {\doibase
  10.1103/PhysRevD.102.104007} {\bibfield  {journal} {\bibinfo  {journal}
  {Phys. Rev. D}\ }\textbf {\bibinfo {volume} {102}},\ \bibinfo {pages}
  {104007} (\bibinfo {year} {2020})},\ \Eprint
  {http://arxiv.org/abs/2007.11562} {arXiv:2007.11562 [gr-qc]} \BibitemShut
  {NoStop}%
\bibitem [{\citenamefont {Khera}\ \emph {et~al.}(2021)\citenamefont {Khera},
  \citenamefont {Krishnan}, \citenamefont {Ashtekar},\ and\ \citenamefont
  {De~Lorenzo}}]{Khera:2020mcz}%
  \BibitemOpen
  \bibfield  {author} {\bibinfo {author} {\bibfnamefont {Neev}\ \bibnamefont
  {Khera}}, \bibinfo {author} {\bibfnamefont {Badri}\ \bibnamefont {Krishnan}},
  \bibinfo {author} {\bibfnamefont {Abhay}\ \bibnamefont {Ashtekar}}, \ and\
  \bibinfo {author} {\bibfnamefont {Tommaso}\ \bibnamefont {De~Lorenzo}},\
  }\bibfield  {title} {\enquote {\bibinfo {title} {{Inferring the gravitational
  wave memory for binary coalescence events}},}\ }\href {\doibase
  10.1103/PhysRevD.103.044012} {\bibfield  {journal} {\bibinfo  {journal}
  {Phys. Rev. D}\ }\textbf {\bibinfo {volume} {103}},\ \bibinfo {pages}
  {044012} (\bibinfo {year} {2021})},\ \Eprint
  {http://arxiv.org/abs/2009.06351} {arXiv:2009.06351 [gr-qc]} \BibitemShut
  {NoStop}%
\bibitem [{\citenamefont {Boyle}\ \emph {et~al.}(2019)\citenamefont {Boyle}
  \emph {et~al.}}]{Boyle:2019kee}%
  \BibitemOpen
  \bibfield  {author} {\bibinfo {author} {\bibfnamefont {Michael}\ \bibnamefont
  {Boyle}} \emph {et~al.},\ }\bibfield  {title} {\enquote {\bibinfo {title}
  {{The SXS Collaboration catalog of binary black hole simulations}},}\ }\href
  {\doibase 10.1088/1361-6382/ab34e2} {\bibfield  {journal} {\bibinfo
  {journal} {Class. Quant. Grav.}\ }\textbf {\bibinfo {volume} {36}},\ \bibinfo
  {pages} {195006} (\bibinfo {year} {2019})},\ \Eprint
  {http://arxiv.org/abs/1904.04831} {arXiv:1904.04831 [gr-qc]} \BibitemShut
  {NoStop}%
\bibitem [{\citenamefont {Winicour}(2009)}]{Winicour:2008vpn}%
  \BibitemOpen
  \bibfield  {author} {\bibinfo {author} {\bibfnamefont {Jeffrey}\ \bibnamefont
  {Winicour}},\ }\bibfield  {title} {\enquote {\bibinfo {title}
  {{Characteristic Evolution and Matching}},}\ }\href {\doibase
  10.12942/lrr-2009-3} {\bibfield  {journal} {\bibinfo  {journal} {Living Rev.
  Rel.}\ }\textbf {\bibinfo {volume} {12}},\ \bibinfo {pages} {3} (\bibinfo
  {year} {2009})},\ \Eprint {http://arxiv.org/abs/0810.1903} {arXiv:0810.1903
  [gr-qc]} \BibitemShut {NoStop}%
\bibitem [{\citenamefont {Mitman}\ \emph {et~al.}(2024)\citenamefont {Mitman}
  \emph {et~al.}}]{Mitman:2024uss}%
  \BibitemOpen
  \bibfield  {author} {\bibinfo {author} {\bibfnamefont {Keefe}\ \bibnamefont
  {Mitman}} \emph {et~al.},\ }\bibfield  {title} {\enquote {\bibinfo {title}
  {{A review of gravitational memory and BMS frame fixing in numerical
  relativity}},}\ }\href {\doibase 10.1088/1361-6382/ad83c2} {\bibfield
  {journal} {\bibinfo  {journal} {Class. Quant. Grav.}\ }\textbf {\bibinfo
  {volume} {41}},\ \bibinfo {pages} {223001} (\bibinfo {year} {2024})},\
  \Eprint {http://arxiv.org/abs/2405.08868} {arXiv:2405.08868 [gr-qc]}
  \BibitemShut {NoStop}%
\bibitem [{\citenamefont {Moxon}\ \emph {et~al.}(2020)\citenamefont {Moxon},
  \citenamefont {Scheel},\ and\ \citenamefont {Teukolsky}}]{Moxon:2020gha}%
  \BibitemOpen
  \bibfield  {author} {\bibinfo {author} {\bibfnamefont {Jordan}\ \bibnamefont
  {Moxon}}, \bibinfo {author} {\bibfnamefont {Mark~A.}\ \bibnamefont {Scheel}},
  \ and\ \bibinfo {author} {\bibfnamefont {Saul~A.}\ \bibnamefont
  {Teukolsky}},\ }\bibfield  {title} {\enquote {\bibinfo {title} {{Improved
  Cauchy-characteristic evolution system for high-precision numerical
  relativity waveforms}},}\ }\href {\doibase 10.1103/PhysRevD.102.044052}
  {\bibfield  {journal} {\bibinfo  {journal} {Phys. Rev. D}\ }\textbf {\bibinfo
  {volume} {102}},\ \bibinfo {pages} {044052} (\bibinfo {year} {2020})},\
  \Eprint {http://arxiv.org/abs/2007.01339} {arXiv:2007.01339 [gr-qc]}
  \BibitemShut {NoStop}%
\bibitem [{\citenamefont {Moxon}\ \emph {et~al.}(2023)\citenamefont {Moxon},
  \citenamefont {Scheel}, \citenamefont {Teukolsky}, \citenamefont {Deppe},
  \citenamefont {Fischer}, \citenamefont {H\'ebert}, \citenamefont {Kidder},\
  and\ \citenamefont {Throwe}}]{Moxon:2021gbv}%
  \BibitemOpen
  \bibfield  {author} {\bibinfo {author} {\bibfnamefont {Jordan}\ \bibnamefont
  {Moxon}}, \bibinfo {author} {\bibfnamefont {Mark~A.}\ \bibnamefont {Scheel}},
  \bibinfo {author} {\bibfnamefont {Saul~A.}\ \bibnamefont {Teukolsky}},
  \bibinfo {author} {\bibfnamefont {Nils}\ \bibnamefont {Deppe}}, \bibinfo
  {author} {\bibfnamefont {Nils}\ \bibnamefont {Fischer}}, \bibinfo {author}
  {\bibfnamefont {Francois}\ \bibnamefont {H\'ebert}}, \bibinfo {author}
  {\bibfnamefont {Lawrence~E.}\ \bibnamefont {Kidder}}, \ and\ \bibinfo
  {author} {\bibfnamefont {William}\ \bibnamefont {Throwe}},\ }\bibfield
  {title} {\enquote {\bibinfo {title} {{SpECTRE Cauchy-characteristic evolution
  system for rapid, precise waveform extraction}},}\ }\href {\doibase
  10.1103/PhysRevD.107.064013} {\bibfield  {journal} {\bibinfo  {journal}
  {Phys. Rev. D}\ }\textbf {\bibinfo {volume} {107}},\ \bibinfo {pages}
  {064013} (\bibinfo {year} {2023})},\ \Eprint
  {http://arxiv.org/abs/2110.08635} {arXiv:2110.08635 [gr-qc]} \BibitemShut
  {NoStop}%
\bibitem [{\citenamefont {Ma}\ \emph {et~al.}(2024)\citenamefont {Ma} \emph
  {et~al.}}]{Ma:2023qjn}%
  \BibitemOpen
  \bibfield  {author} {\bibinfo {author} {\bibfnamefont {Sizheng}\ \bibnamefont
  {Ma}} \emph {et~al.},\ }\bibfield  {title} {\enquote {\bibinfo {title}
  {{Fully relativistic three-dimensional Cauchy-characteristic matching for
  physical degrees of freedom}},}\ }\href {\doibase
  10.1103/PhysRevD.109.124027} {\bibfield  {journal} {\bibinfo  {journal}
  {Phys. Rev. D}\ }\textbf {\bibinfo {volume} {109}},\ \bibinfo {pages}
  {124027} (\bibinfo {year} {2024})},\ \Eprint
  {http://arxiv.org/abs/2308.10361} {arXiv:2308.10361 [gr-qc]} \BibitemShut
  {NoStop}%
\bibitem [{\citenamefont {Deppe}\ \emph {et~al.}(2024)\citenamefont {Deppe},
  \citenamefont {Throwe}, \citenamefont {Kidder}, \citenamefont {Vu},
  \citenamefont {Nelli}, \citenamefont {Armaza}, \citenamefont {Bonilla},
  \citenamefont {H\'ebert}, \citenamefont {Kim}, \citenamefont {Kumar},
  \citenamefont {Lovelace}, \citenamefont {Macedo}, \citenamefont {Moxon},
  \citenamefont {O'Shea}, \citenamefont {Pfeiffer}, \citenamefont {Scheel},
  \citenamefont {Teukolsky}, \citenamefont {Wittek} \emph
  {et~al.}}]{spectrecode}%
  \BibitemOpen
  \bibfield  {author} {\bibinfo {author} {\bibfnamefont {Nils}\ \bibnamefont
  {Deppe}}, \bibinfo {author} {\bibfnamefont {William}\ \bibnamefont {Throwe}},
  \bibinfo {author} {\bibfnamefont {Lawrence~E.}\ \bibnamefont {Kidder}},
  \bibinfo {author} {\bibfnamefont {Nils~L.}\ \bibnamefont {Vu}}, \bibinfo
  {author} {\bibfnamefont {Kyle~C.}\ \bibnamefont {Nelli}}, \bibinfo {author}
  {\bibfnamefont {Crist\'obal}\ \bibnamefont {Armaza}}, \bibinfo {author}
  {\bibfnamefont {Marceline~S.}\ \bibnamefont {Bonilla}}, \bibinfo {author}
  {\bibfnamefont {Fran\c{c}ois}\ \bibnamefont {H\'ebert}}, \bibinfo {author}
  {\bibfnamefont {Yoonsoo}\ \bibnamefont {Kim}}, \bibinfo {author}
  {\bibfnamefont {Prayush}\ \bibnamefont {Kumar}}, \bibinfo {author}
  {\bibfnamefont {Geoffrey}\ \bibnamefont {Lovelace}}, \bibinfo {author}
  {\bibfnamefont {Alexandra}\ \bibnamefont {Macedo}}, \bibinfo {author}
  {\bibfnamefont {Jordan}\ \bibnamefont {Moxon}}, \bibinfo {author}
  {\bibfnamefont {Eamonn}\ \bibnamefont {O'Shea}}, \bibinfo {author}
  {\bibfnamefont {Harald~P.}\ \bibnamefont {Pfeiffer}}, \bibinfo {author}
  {\bibfnamefont {Mark~A.}\ \bibnamefont {Scheel}}, \bibinfo {author}
  {\bibfnamefont {Saul~A.}\ \bibnamefont {Teukolsky}}, \bibinfo {author}
  {\bibfnamefont {Nikolas~A.}\ \bibnamefont {Wittek}},  \emph {et~al.},\ }\href
  {\doibase 10.5281/zenodo.13858965} {\enquote {\bibinfo {title}
  {\texttt{SpECTRE v2024.09.29}},}\ }\bibinfo {howpublished}
  {\href{https://doi.org/10.5281/zenodo.13858965}{10.5281/zenodo.13858965}}
  (\bibinfo {year} {2024})\BibitemShut {NoStop}%
\bibitem [{\citenamefont {Yunes}\ and\ \citenamefont
  {Siemens}(2013)}]{Yunes:2013dva}%
  \BibitemOpen
  \bibfield  {author} {\bibinfo {author} {\bibfnamefont {Nicol\'as}\
  \bibnamefont {Yunes}}\ and\ \bibinfo {author} {\bibfnamefont {Xavier}\
  \bibnamefont {Siemens}},\ }\bibfield  {title} {\enquote {\bibinfo {title}
  {{Gravitational-Wave Tests of General Relativity with Ground-Based Detectors
  and Pulsar Timing-Arrays}},}\ }\href {\doibase 10.12942/lrr-2013-9}
  {\bibfield  {journal} {\bibinfo  {journal} {Living Rev. Rel.}\ }\textbf
  {\bibinfo {volume} {16}},\ \bibinfo {pages} {9} (\bibinfo {year} {2013})},\
  \Eprint {http://arxiv.org/abs/1304.3473} {arXiv:1304.3473 [gr-qc]}
  \BibitemShut {NoStop}%
\bibitem [{\citenamefont {Yunes}\ \emph {et~al.}(2024)\citenamefont {Yunes},
  \citenamefont {Siemens},\ and\ \citenamefont {Yagi}}]{Yunes:2024lzm}%
  \BibitemOpen
  \bibfield  {author} {\bibinfo {author} {\bibfnamefont {Nicolas}\ \bibnamefont
  {Yunes}}, \bibinfo {author} {\bibfnamefont {Xavier}\ \bibnamefont {Siemens}},
  \ and\ \bibinfo {author} {\bibfnamefont {Kent}\ \bibnamefont {Yagi}},\
  }\bibfield  {title} {\enquote {\bibinfo {title} {{Gravitational-Wave Tests of
  General Relativity with Ground-Based Detectors and Pulsar-Timing Arrays}},}\
  }\href@noop {} {\  (\bibinfo {year} {2024})},\ \Eprint
  {http://arxiv.org/abs/2408.05240} {arXiv:2408.05240 [gr-qc]} \BibitemShut
  {NoStop}%
\bibitem [{\citenamefont {Tahura}\ \emph
  {et~al.}(2021{\natexlab{a}})\citenamefont {Tahura}, \citenamefont {Nichols},
  \citenamefont {Saffer}, \citenamefont {Stein},\ and\ \citenamefont
  {Yagi}}]{Tahura:2020vsa}%
  \BibitemOpen
  \bibfield  {author} {\bibinfo {author} {\bibfnamefont {Shammi}\ \bibnamefont
  {Tahura}}, \bibinfo {author} {\bibfnamefont {David~A.}\ \bibnamefont
  {Nichols}}, \bibinfo {author} {\bibfnamefont {Alexander}\ \bibnamefont
  {Saffer}}, \bibinfo {author} {\bibfnamefont {Leo~C.}\ \bibnamefont {Stein}},
  \ and\ \bibinfo {author} {\bibfnamefont {Kent}\ \bibnamefont {Yagi}},\
  }\bibfield  {title} {\enquote {\bibinfo {title} {{Brans-Dicke theory in
  Bondi-Sachs form: Asymptotically flat solutions, asymptotic symmetries and
  gravitational-wave memory effects}},}\ }\href {\doibase
  10.1103/PhysRevD.103.104026} {\bibfield  {journal} {\bibinfo  {journal}
  {Phys. Rev. D}\ }\textbf {\bibinfo {volume} {103}},\ \bibinfo {pages}
  {104026} (\bibinfo {year} {2021}{\natexlab{a}})},\ \Eprint
  {http://arxiv.org/abs/2007.13799} {arXiv:2007.13799 [gr-qc]} \BibitemShut
  {NoStop}%
\bibitem [{\citenamefont {Tahura}\ \emph
  {et~al.}(2021{\natexlab{b}})\citenamefont {Tahura}, \citenamefont {Nichols},\
  and\ \citenamefont {Yagi}}]{Tahura:2021hbk}%
  \BibitemOpen
  \bibfield  {author} {\bibinfo {author} {\bibfnamefont {Shammi}\ \bibnamefont
  {Tahura}}, \bibinfo {author} {\bibfnamefont {David~A.}\ \bibnamefont
  {Nichols}}, \ and\ \bibinfo {author} {\bibfnamefont {Kent}\ \bibnamefont
  {Yagi}},\ }\bibfield  {title} {\enquote {\bibinfo {title}
  {{Gravitational-wave memory effects in Brans-Dicke theory: Waveforms and
  effects in the post-Newtonian approximation}},}\ }\href {\doibase
  10.1103/PhysRevD.104.104010} {\bibfield  {journal} {\bibinfo  {journal}
  {Phys. Rev. D}\ }\textbf {\bibinfo {volume} {104}},\ \bibinfo {pages}
  {104010} (\bibinfo {year} {2021}{\natexlab{b}})},\ \Eprint
  {http://arxiv.org/abs/2107.02208} {arXiv:2107.02208 [gr-qc]} \BibitemShut
  {NoStop}%
\bibitem [{\citenamefont {Hou}\ and\ \citenamefont
  {Zhu}(2021{\natexlab{a}})}]{Hou:2020tnd}%
  \BibitemOpen
  \bibfield  {author} {\bibinfo {author} {\bibfnamefont {Shaoqi}\ \bibnamefont
  {Hou}}\ and\ \bibinfo {author} {\bibfnamefont {Zong-Hong}\ \bibnamefont
  {Zhu}},\ }\bibfield  {title} {\enquote {\bibinfo {title} {{Gravitational
  memory effects and Bondi-Metzner-Sachs symmetries in scalar-tensor
  theories}},}\ }\href {\doibase 10.1007/JHEP01(2021)083} {\bibfield  {journal}
  {\bibinfo  {journal} {JHEP}\ }\textbf {\bibinfo {volume} {01}},\ \bibinfo
  {pages} {083} (\bibinfo {year} {2021}{\natexlab{a}})},\ \Eprint
  {http://arxiv.org/abs/2005.01310} {arXiv:2005.01310 [gr-qc]} \BibitemShut
  {NoStop}%
\bibitem [{\citenamefont {Hou}\ and\ \citenamefont
  {Zhu}(2021{\natexlab{b}})}]{Hou:2020wbo}%
  \BibitemOpen
  \bibfield  {author} {\bibinfo {author} {\bibfnamefont {Shaoqi}\ \bibnamefont
  {Hou}}\ and\ \bibinfo {author} {\bibfnamefont {Zong-Hong}\ \bibnamefont
  {Zhu}},\ }\bibfield  {title} {\enquote {\bibinfo {title} {{''Conserved
  charges'' of the Bondi-Metzner-Sachs algebra in the Brans-Dicke theory}},}\
  }\href {\doibase 10.1088/1674-1137/abd087} {\bibfield  {journal} {\bibinfo
  {journal} {Chin. Phys. C}\ }\textbf {\bibinfo {volume} {45}},\ \bibinfo
  {pages} {023122} (\bibinfo {year} {2021}{\natexlab{b}})},\ \Eprint
  {http://arxiv.org/abs/2008.05154} {arXiv:2008.05154 [gr-qc]} \BibitemShut
  {NoStop}%
\bibitem [{\citenamefont {Hou}\ \emph {et~al.}(2022{\natexlab{a}})\citenamefont
  {Hou}, \citenamefont {Zhu},\ and\ \citenamefont {Zhu}}]{Hou:2021oxe}%
  \BibitemOpen
  \bibfield  {author} {\bibinfo {author} {\bibfnamefont {Shaoqi}\ \bibnamefont
  {Hou}}, \bibinfo {author} {\bibfnamefont {Tao}\ \bibnamefont {Zhu}}, \ and\
  \bibinfo {author} {\bibfnamefont {Zong-Hong}\ \bibnamefont {Zhu}},\
  }\bibfield  {title} {\enquote {\bibinfo {title} {{Asymptotic analysis of
  Chern-Simons modified gravity and its memory effects}},}\ }\href {\doibase
  10.1103/PhysRevD.105.024025} {\bibfield  {journal} {\bibinfo  {journal}
  {Phys. Rev. D}\ }\textbf {\bibinfo {volume} {105}},\ \bibinfo {pages}
  {024025} (\bibinfo {year} {2022}{\natexlab{a}})},\ \Eprint
  {http://arxiv.org/abs/2109.04238} {arXiv:2109.04238 [gr-qc]} \BibitemShut
  {NoStop}%
\bibitem [{\citenamefont {Hou}\ \emph {et~al.}(2022{\natexlab{b}})\citenamefont
  {Hou}, \citenamefont {Zhu},\ and\ \citenamefont {Zhu}}]{Hou:2021bxz}%
  \BibitemOpen
  \bibfield  {author} {\bibinfo {author} {\bibfnamefont {Shaoqi}\ \bibnamefont
  {Hou}}, \bibinfo {author} {\bibfnamefont {Tao}\ \bibnamefont {Zhu}}, \ and\
  \bibinfo {author} {\bibfnamefont {Zong-Hong}\ \bibnamefont {Zhu}},\
  }\bibfield  {title} {\enquote {\bibinfo {title} {{Conserved charges in
  Chern-Simons modified theory and memory effects}},}\ }\href {\doibase
  10.1088/1475-7516/2022/04/032} {\bibfield  {journal} {\bibinfo  {journal}
  {JCAP}\ }\textbf {\bibinfo {volume} {04}},\ \bibinfo {pages} {032} (\bibinfo
  {year} {2022}{\natexlab{b}})},\ \Eprint {http://arxiv.org/abs/2112.13049}
  {arXiv:2112.13049 [gr-qc]} \BibitemShut {NoStop}%
\bibitem [{\citenamefont {Capuano}\ \emph {et~al.}(2023)\citenamefont
  {Capuano}, \citenamefont {Santoni},\ and\ \citenamefont
  {Barausse}}]{Capuano:2023yyh}%
  \BibitemOpen
  \bibfield  {author} {\bibinfo {author} {\bibfnamefont {Lodovico}\
  \bibnamefont {Capuano}}, \bibinfo {author} {\bibfnamefont {Luca}\
  \bibnamefont {Santoni}}, \ and\ \bibinfo {author} {\bibfnamefont {Enrico}\
  \bibnamefont {Barausse}},\ }\bibfield  {title} {\enquote {\bibinfo {title}
  {{Black hole hairs in scalar-tensor gravity and the lack thereof}},}\ }\href
  {\doibase 10.1103/PhysRevD.108.064058} {\bibfield  {journal} {\bibinfo
  {journal} {Phys. Rev. D}\ }\textbf {\bibinfo {volume} {108}},\ \bibinfo
  {pages} {064058} (\bibinfo {year} {2023})},\ \Eprint
  {http://arxiv.org/abs/2304.12750} {arXiv:2304.12750 [gr-qc]} \BibitemShut
  {NoStop}%
\bibitem [{\citenamefont {Heisenberg}\ \emph {et~al.}(2023)\citenamefont
  {Heisenberg}, \citenamefont {Yunes},\ and\ \citenamefont
  {Zosso}}]{Heisenberg:2023prj}%
  \BibitemOpen
  \bibfield  {author} {\bibinfo {author} {\bibfnamefont {Lavinia}\ \bibnamefont
  {Heisenberg}}, \bibinfo {author} {\bibfnamefont {Nicol\'as}\ \bibnamefont
  {Yunes}}, \ and\ \bibinfo {author} {\bibfnamefont {Jann}\ \bibnamefont
  {Zosso}},\ }\bibfield  {title} {\enquote {\bibinfo {title} {{Gravitational
  wave memory beyond general relativity}},}\ }\href {\doibase
  10.1103/PhysRevD.108.024010} {\bibfield  {journal} {\bibinfo  {journal}
  {Phys. Rev. D}\ }\textbf {\bibinfo {volume} {108}},\ \bibinfo {pages}
  {024010} (\bibinfo {year} {2023})},\ \Eprint
  {http://arxiv.org/abs/2303.02021} {arXiv:2303.02021 [gr-qc]} \BibitemShut
  {NoStop}%
\bibitem [{\citenamefont {Heisenberg}\ \emph {et~al.}(2024)\citenamefont
  {Heisenberg}, \citenamefont {Xu},\ and\ \citenamefont
  {Zosso}}]{Heisenberg:2024cjk}%
  \BibitemOpen
  \bibfield  {author} {\bibinfo {author} {\bibfnamefont {Lavinia}\ \bibnamefont
  {Heisenberg}}, \bibinfo {author} {\bibfnamefont {Guangzi}\ \bibnamefont
  {Xu}}, \ and\ \bibinfo {author} {\bibfnamefont {Jann}\ \bibnamefont
  {Zosso}},\ }\bibfield  {title} {\enquote {\bibinfo {title} {{Unifying
  ordinary and null memory}},}\ }\href {\doibase 10.1088/1475-7516/2024/05/119}
  {\bibfield  {journal} {\bibinfo  {journal} {JCAP}\ }\textbf {\bibinfo
  {volume} {05}},\ \bibinfo {pages} {119} (\bibinfo {year} {2024})},\ \Eprint
  {http://arxiv.org/abs/2401.05936} {arXiv:2401.05936 [gr-qc]} \BibitemShut
  {NoStop}%
\bibitem [{\citenamefont {Isaacson}(1968{\natexlab{a}})}]{Isaacson:1968hbi}%
  \BibitemOpen
  \bibfield  {author} {\bibinfo {author} {\bibfnamefont {Richard~A.}\
  \bibnamefont {Isaacson}},\ }\bibfield  {title} {\enquote {\bibinfo {title}
  {{Gravitational Radiation in the Limit of High Frequency. I. The Linear
  Approximation and Geometrical Optics}},}\ }\href {\doibase
  10.1103/PhysRev.166.1263} {\bibfield  {journal} {\bibinfo  {journal} {Phys.
  Rev.}\ }\textbf {\bibinfo {volume} {166}},\ \bibinfo {pages} {1263--1271}
  (\bibinfo {year} {1968}{\natexlab{a}})}\BibitemShut {NoStop}%
\bibitem [{\citenamefont {Isaacson}(1968{\natexlab{b}})}]{Isaacson:1968zza}%
  \BibitemOpen
  \bibfield  {author} {\bibinfo {author} {\bibfnamefont {Richard~A.}\
  \bibnamefont {Isaacson}},\ }\bibfield  {title} {\enquote {\bibinfo {title}
  {{Gravitational Radiation in the Limit of High Frequency. II. Nonlinear Terms
  and the Ef fective Stress Tensor}},}\ }\href {\doibase
  10.1103/PhysRev.166.1272} {\bibfield  {journal} {\bibinfo  {journal} {Phys.
  Rev.}\ }\textbf {\bibinfo {volume} {166}},\ \bibinfo {pages} {1272--1279}
  (\bibinfo {year} {1968}{\natexlab{b}})}\BibitemShut {NoStop}%
\bibitem [{\citenamefont {Stein}\ and\ \citenamefont
  {Yunes}(2011)}]{Stein:2010pn}%
  \BibitemOpen
  \bibfield  {author} {\bibinfo {author} {\bibfnamefont {Leo~C.}\ \bibnamefont
  {Stein}}\ and\ \bibinfo {author} {\bibfnamefont {Nicolas}\ \bibnamefont
  {Yunes}},\ }\bibfield  {title} {\enquote {\bibinfo {title} {{Effective
  Gravitational Wave Stress-energy Tensor in Alternative Theories of
  Gravity}},}\ }\href {\doibase 10.1103/PhysRevD.83.064038} {\bibfield
  {journal} {\bibinfo  {journal} {Phys. Rev. D}\ }\textbf {\bibinfo {volume}
  {83}},\ \bibinfo {pages} {064038} (\bibinfo {year} {2011})},\ \Eprint
  {http://arxiv.org/abs/1012.3144} {arXiv:1012.3144 [gr-qc]} \BibitemShut
  {NoStop}%
\bibitem [{\citenamefont {Hou}\ \emph {et~al.}(2024)\citenamefont {Hou},
  \citenamefont {Wang},\ and\ \citenamefont {Zhu}}]{Hou:2023pfz}%
  \BibitemOpen
  \bibfield  {author} {\bibinfo {author} {\bibfnamefont {Shaoqi}\ \bibnamefont
  {Hou}}, \bibinfo {author} {\bibfnamefont {Anzhong}\ \bibnamefont {Wang}}, \
  and\ \bibinfo {author} {\bibfnamefont {Zong-Hong}\ \bibnamefont {Zhu}},\
  }\bibfield  {title} {\enquote {\bibinfo {title} {{Asymptotic analysis of
  Einstein-\AE{}ther theory and its memory effects: The linearized case}},}\
  }\href {\doibase 10.1103/PhysRevD.109.044025} {\bibfield  {journal} {\bibinfo
   {journal} {Phys. Rev. D}\ }\textbf {\bibinfo {volume} {109}},\ \bibinfo
  {pages} {044025} (\bibinfo {year} {2024})},\ \Eprint
  {http://arxiv.org/abs/2309.01165} {arXiv:2309.01165 [gr-qc]} \BibitemShut
  {NoStop}%
\bibitem [{\citenamefont {Hou}(2024)}]{Hou:2024exz}%
  \BibitemOpen
  \bibfield  {author} {\bibinfo {author} {\bibfnamefont {Shaoqi}\ \bibnamefont
  {Hou}},\ }\bibfield  {title} {\enquote {\bibinfo {title} {{The general
  property of the tensor gravitational memory effect in theories of
  gravity}},}\ }\href@noop {} {\bibfield  {journal} {\bibinfo  {journal} {{}}\
  } (\bibinfo {year} {2024})},\ \Eprint {http://arxiv.org/abs/2411.17318}
  {arXiv:2411.17318 [gr-qc]} \BibitemShut {NoStop}%
\bibitem [{\citenamefont {Seraj}(2021)}]{Seraj:2021qja}%
  \BibitemOpen
  \bibfield  {author} {\bibinfo {author} {\bibfnamefont {Ali}\ \bibnamefont
  {Seraj}},\ }\bibfield  {title} {\enquote {\bibinfo {title} {{Gravitational
  breathing memory and dual symmetries}},}\ }\href {\doibase
  10.1007/JHEP05(2021)283} {\bibfield  {journal} {\bibinfo  {journal} {J. High
  Energy Phys.}\ }\textbf {\bibinfo {volume} {05}},\ \bibinfo {pages} {283}
  (\bibinfo {year} {2021})},\ \Eprint {http://arxiv.org/abs/2103.12185}
  {arXiv:2103.12185 [hep-th]} \BibitemShut {NoStop}%
\bibitem [{\citenamefont {Barausse}\ \emph {et~al.}(2013)\citenamefont
  {Barausse}, \citenamefont {Palenzuela}, \citenamefont {Ponce},\ and\
  \citenamefont {Lehner}}]{Barausse:2012da}%
  \BibitemOpen
  \bibfield  {author} {\bibinfo {author} {\bibfnamefont {Enrico}\ \bibnamefont
  {Barausse}}, \bibinfo {author} {\bibfnamefont {Carlos}\ \bibnamefont
  {Palenzuela}}, \bibinfo {author} {\bibfnamefont {Marcelo}\ \bibnamefont
  {Ponce}}, \ and\ \bibinfo {author} {\bibfnamefont {Luis}\ \bibnamefont
  {Lehner}},\ }\bibfield  {title} {\enquote {\bibinfo {title} {{Neutron-star
  mergers in scalar-tensor theories of gravity}},}\ }\href {\doibase
  10.1103/PhysRevD.87.081506} {\bibfield  {journal} {\bibinfo  {journal} {Phys.
  Rev. D}\ }\textbf {\bibinfo {volume} {87}},\ \bibinfo {pages} {081506}
  (\bibinfo {year} {2013})},\ \Eprint {http://arxiv.org/abs/1212.5053}
  {arXiv:1212.5053 [gr-qc]} \BibitemShut {NoStop}%
\bibitem [{\citenamefont {Shibata}\ \emph {et~al.}(2014)\citenamefont
  {Shibata}, \citenamefont {Taniguchi}, \citenamefont {Okawa},\ and\
  \citenamefont {Buonanno}}]{Shibata:2013pra}%
  \BibitemOpen
  \bibfield  {author} {\bibinfo {author} {\bibfnamefont {Masaru}\ \bibnamefont
  {Shibata}}, \bibinfo {author} {\bibfnamefont {Keisuke}\ \bibnamefont
  {Taniguchi}}, \bibinfo {author} {\bibfnamefont {Hirotada}\ \bibnamefont
  {Okawa}}, \ and\ \bibinfo {author} {\bibfnamefont {Alessandra}\ \bibnamefont
  {Buonanno}},\ }\bibfield  {title} {\enquote {\bibinfo {title} {{Coalescence
  of binary neutron stars in a scalar-tensor theory of gravity}},}\ }\href
  {\doibase 10.1103/PhysRevD.89.084005} {\bibfield  {journal} {\bibinfo
  {journal} {Phys. Rev. D}\ }\textbf {\bibinfo {volume} {89}},\ \bibinfo
  {pages} {084005} (\bibinfo {year} {2014})},\ \Eprint
  {http://arxiv.org/abs/1310.0627} {arXiv:1310.0627 [gr-qc]} \BibitemShut
  {NoStop}%
\bibitem [{\citenamefont {Palenzuela}\ \emph {et~al.}(2014)\citenamefont
  {Palenzuela}, \citenamefont {Barausse}, \citenamefont {Ponce},\ and\
  \citenamefont {Lehner}}]{Palenzuela:2013hsa}%
  \BibitemOpen
  \bibfield  {author} {\bibinfo {author} {\bibfnamefont {Carlos}\ \bibnamefont
  {Palenzuela}}, \bibinfo {author} {\bibfnamefont {Enrico}\ \bibnamefont
  {Barausse}}, \bibinfo {author} {\bibfnamefont {Marcelo}\ \bibnamefont
  {Ponce}}, \ and\ \bibinfo {author} {\bibfnamefont {Luis}\ \bibnamefont
  {Lehner}},\ }\bibfield  {title} {\enquote {\bibinfo {title} {{Dynamical
  scalarization of neutron stars in scalar-tensor gravity theories}},}\ }\href
  {\doibase 10.1103/PhysRevD.89.044024} {\bibfield  {journal} {\bibinfo
  {journal} {Phys. Rev. D}\ }\textbf {\bibinfo {volume} {89}},\ \bibinfo
  {pages} {044024} (\bibinfo {year} {2014})},\ \Eprint
  {http://arxiv.org/abs/1310.4481} {arXiv:1310.4481 [gr-qc]} \BibitemShut
  {NoStop}%
\bibitem [{\citenamefont {Damour}\ and\ \citenamefont
  {Esposito-Farese}(1992)}]{Damour:1992we}%
  \BibitemOpen
  \bibfield  {author} {\bibinfo {author} {\bibfnamefont {Thibault}\
  \bibnamefont {Damour}}\ and\ \bibinfo {author} {\bibfnamefont {Gilles}\
  \bibnamefont {Esposito-Farese}},\ }\bibfield  {title} {\enquote {\bibinfo
  {title} {{Tensor multiscalar theories of gravitation}},}\ }\href {\doibase
  10.1088/0264-9381/9/9/015} {\bibfield  {journal} {\bibinfo  {journal} {Class.
  Quant. Grav.}\ }\textbf {\bibinfo {volume} {9}},\ \bibinfo {pages}
  {2093--2176} (\bibinfo {year} {1992})}\BibitemShut {NoStop}%
\bibitem [{\citenamefont {Ma}\ \emph {et~al.}(2023)\citenamefont {Ma},
  \citenamefont {Varma}, \citenamefont {Stein}, \citenamefont {Foucart},
  \citenamefont {Duez}, \citenamefont {Kidder}, \citenamefont {Pfeiffer},\ and\
  \citenamefont {Scheel}}]{Ma:2023sok}%
  \BibitemOpen
  \bibfield  {author} {\bibinfo {author} {\bibfnamefont {Sizheng}\ \bibnamefont
  {Ma}}, \bibinfo {author} {\bibfnamefont {Vijay}\ \bibnamefont {Varma}},
  \bibinfo {author} {\bibfnamefont {Leo~C.}\ \bibnamefont {Stein}}, \bibinfo
  {author} {\bibfnamefont {Francois}\ \bibnamefont {Foucart}}, \bibinfo
  {author} {\bibfnamefont {Matthew~D.}\ \bibnamefont {Duez}}, \bibinfo {author}
  {\bibfnamefont {Lawrence~E.}\ \bibnamefont {Kidder}}, \bibinfo {author}
  {\bibfnamefont {Harald~P.}\ \bibnamefont {Pfeiffer}}, \ and\ \bibinfo
  {author} {\bibfnamefont {Mark~A.}\ \bibnamefont {Scheel}},\ }\bibfield
  {title} {\enquote {\bibinfo {title} {{Numerical simulations of black
  hole-neutron star mergers in scalar-tensor gravity}},}\ }\href {\doibase
  10.1103/PhysRevD.107.124051} {\bibfield  {journal} {\bibinfo  {journal}
  {Phys. Rev. D}\ }\textbf {\bibinfo {volume} {107}},\ \bibinfo {pages}
  {124051} (\bibinfo {year} {2023})},\ \Eprint
  {http://arxiv.org/abs/2304.11836} {arXiv:2304.11836 [gr-qc]} \BibitemShut
  {NoStop}%
\bibitem [{\citenamefont {Ma}\ \emph {et~al.}(2025)\citenamefont {Ma},
  \citenamefont {Nelli}, \citenamefont {Moxon}, \citenamefont {Scheel},
  \citenamefont {Deppe}, \citenamefont {Kidder}, \citenamefont {Throwe},\ and\
  \citenamefont {Vu}}]{Ma:2024bed}%
  \BibitemOpen
  \bibfield  {author} {\bibinfo {author} {\bibfnamefont {Sizheng}\ \bibnamefont
  {Ma}}, \bibinfo {author} {\bibfnamefont {Kyle~C.}\ \bibnamefont {Nelli}},
  \bibinfo {author} {\bibfnamefont {Jordan}\ \bibnamefont {Moxon}}, \bibinfo
  {author} {\bibfnamefont {Mark~A.}\ \bibnamefont {Scheel}}, \bibinfo {author}
  {\bibfnamefont {Nils}\ \bibnamefont {Deppe}}, \bibinfo {author}
  {\bibfnamefont {Lawrence~E.}\ \bibnamefont {Kidder}}, \bibinfo {author}
  {\bibfnamefont {William}\ \bibnamefont {Throwe}}, \ and\ \bibinfo {author}
  {\bibfnamefont {Nils~L.}\ \bibnamefont {Vu}},\ }\bibfield  {title} {\enquote
  {\bibinfo {title} {{Einstein{\textendash}Klein{\textendash}Gordon system via
  Cauchy-characteristic evolution: computation of memory and ringdown tail}},}\
  }\href {\doibase 10.1088/1361-6382/adaf6f} {\bibfield  {journal} {\bibinfo
  {journal} {Classical Quantum Gravity}\ }\textbf {\bibinfo {volume} {42}},\
  \bibinfo {pages} {055006} (\bibinfo {year} {2025})},\ \Eprint
  {http://arxiv.org/abs/2409.06141} {arXiv:2409.06141 [gr-qc]} \BibitemShut
  {NoStop}%
\bibitem [{\citenamefont {Wagoner}(1970)}]{Wagoner1970}%
  \BibitemOpen
  \bibfield  {author} {\bibinfo {author} {\bibfnamefont {Robert~V.}\
  \bibnamefont {Wagoner}},\ }\bibfield  {title} {\enquote {\bibinfo {title}
  {Scalar-tensor theory and gravitational waves},}\ }\href {\doibase
  10.1103/PhysRevD.1.3209} {\bibfield  {journal} {\bibinfo  {journal} {Phys.
  Rev. D}\ }\textbf {\bibinfo {volume} {1}},\ \bibinfo {pages} {3209--3216}
  (\bibinfo {year} {1970})}\BibitemShut {NoStop}%
\bibitem [{\citenamefont {Bergmann}(1968)}]{Bergmann:1968ve}%
  \BibitemOpen
  \bibfield  {author} {\bibinfo {author} {\bibfnamefont {Peter~G.}\
  \bibnamefont {Bergmann}},\ }\bibfield  {title} {\enquote {\bibinfo {title}
  {{Comments on the scalar tensor theory}},}\ }\href {\doibase
  10.1007/BF00668828} {\bibfield  {journal} {\bibinfo  {journal} {Int. J.
  Theor. Phys.}\ }\textbf {\bibinfo {volume} {1}},\ \bibinfo {pages} {25--36}
  (\bibinfo {year} {1968})}\BibitemShut {NoStop}%
\bibitem [{\citenamefont {Damour}\ and\ \citenamefont
  {Esposito-Farese}(1993)}]{Damour:1993hw}%
  \BibitemOpen
  \bibfield  {author} {\bibinfo {author} {\bibfnamefont {Thibault}\
  \bibnamefont {Damour}}\ and\ \bibinfo {author} {\bibfnamefont {Gilles}\
  \bibnamefont {Esposito-Farese}},\ }\bibfield  {title} {\enquote {\bibinfo
  {title} {{Nonperturbative strong field effects in tensor - scalar theories of
  gravitation}},}\ }\href {\doibase 10.1103/PhysRevLett.70.2220} {\bibfield
  {journal} {\bibinfo  {journal} {Phys.Rev.Lett.}\ }\textbf {\bibinfo {volume}
  {70}},\ \bibinfo {pages} {2220--2223} (\bibinfo {year} {1993})}\BibitemShut
  {NoStop}%
\bibitem [{\citenamefont {Misner}\ \emph {et~al.}(1973)\citenamefont {Misner},
  \citenamefont {Thorne},\ and\ \citenamefont {Wheeler}}]{Misner:1974qy}%
  \BibitemOpen
  \bibfield  {author} {\bibinfo {author} {\bibfnamefont {Charles~W.}\
  \bibnamefont {Misner}}, \bibinfo {author} {\bibfnamefont {K.S.}\ \bibnamefont
  {Thorne}}, \ and\ \bibinfo {author} {\bibfnamefont {J.A.}\ \bibnamefont
  {Wheeler}},\ }\href@noop {} {\emph {\bibinfo {title} {{Gravitation}}}}\
  (\bibinfo  {publisher} {W. H. Freeman},\ \bibinfo {address} {San Francisco},\
  \bibinfo {year} {1973})\BibitemShut {NoStop}%
\bibitem [{\citenamefont {Mart\'in-Garc\'ia}(2025)}]{xAct}%
  \BibitemOpen
  \bibfield  {author} {\bibinfo {author} {\bibfnamefont {Jos\'e~M.}\
  \bibnamefont {Mart\'in-Garc\'ia}},\ }\href@noop {} {\enquote {\bibinfo
  {title} {{xAct: Efficient tensor computer algebra for the Wolfram
  Language}},}\ }\bibinfo {howpublished} {\url{http://www.xact.es/}} (\bibinfo
  {year} {2025})\BibitemShut {NoStop}%
\bibitem [{\citenamefont
  {Mart\'\i{}n-Garc\'\i{}a}(2008)}]{Martin-Garcia:2008ysv}%
  \BibitemOpen
  \bibfield  {author} {\bibinfo {author} {\bibfnamefont {Jos\'e~M.}\
  \bibnamefont {Mart\'\i{}n-Garc\'\i{}a}},\ }\bibfield  {title} {\enquote
  {\bibinfo {title} {{xPerm: fast index canonicalization for tensor computer
  algebra}},}\ }\href {\doibase 10.1016/j.cpc.2008.05.009} {\bibfield
  {journal} {\bibinfo  {journal} {Comput. Phys. Commun.}\ }\textbf {\bibinfo
  {volume} {179}},\ \bibinfo {pages} {597--603} (\bibinfo {year} {2008})},\
  \Eprint {http://arxiv.org/abs/0803.0862} {arXiv:0803.0862 [cs.SC]}
  \BibitemShut {NoStop}%
\bibitem [{\citenamefont {Tahura}(2025)}]{def_jordan}%
  \BibitemOpen
  \bibfield  {author} {\bibinfo {author} {\bibfnamefont {Shammi}\ \bibnamefont
  {Tahura}},\ }\href {https://github.com/stshammi/DEF-Jordan-BondiSachs-Public}
  {\enquote {\bibinfo {title} {{Damour-Esposito-Farese Scalar-Tensor Theory in
  Jordan Frame in Bondi-Sach Framework,
  https://github.com/stshammi/DEF-Jordan-BondiSachs-Public}},}\ } (\bibinfo
  {year} {2025})\BibitemShut {NoStop}%
\bibitem [{\citenamefont {Berti}\ \emph {et~al.}(2015)\citenamefont {Berti}
  \emph {et~al.}}]{Berti:2015itd}%
  \BibitemOpen
  \bibfield  {author} {\bibinfo {author} {\bibfnamefont {Emanuele}\
  \bibnamefont {Berti}} \emph {et~al.},\ }\bibfield  {title} {\enquote
  {\bibinfo {title} {{Testing General Relativity with Present and Future
  Astrophysical Observations}},}\ }\href {\doibase
  10.1088/0264-9381/32/24/243001} {\bibfield  {journal} {\bibinfo  {journal}
  {Class. Quant. Grav.}\ }\textbf {\bibinfo {volume} {32}},\ \bibinfo {pages}
  {243001} (\bibinfo {year} {2015})},\ \Eprint
  {http://arxiv.org/abs/1501.07274} {arXiv:1501.07274 [gr-qc]} \BibitemShut
  {NoStop}%
\bibitem [{\citenamefont {Mädler}\ and\ \citenamefont
  {Winicour}(2016)}]{Madler:2016xju}%
  \BibitemOpen
  \bibfield  {author} {\bibinfo {author} {\bibfnamefont {Thomas}\ \bibnamefont
  {Mädler}}\ and\ \bibinfo {author} {\bibfnamefont {Jeffrey}\ \bibnamefont
  {Winicour}},\ }\bibfield  {title} {\enquote {\bibinfo {title} {{Bondi-Sachs
  Formalism}},}\ }\href {\doibase 10.4249/scholarpedia.33528} {\bibfield
  {journal} {\bibinfo  {journal} {Scholarpedia}\ }\textbf {\bibinfo {volume}
  {11}},\ \bibinfo {pages} {33528} (\bibinfo {year} {2016})},\ \Eprint
  {http://arxiv.org/abs/1609.01731} {arXiv:1609.01731 [gr-qc]} \BibitemShut
  {NoStop}%
\bibitem [{\citenamefont {Carroll}(2019)}]{Carroll:2004st}%
  \BibitemOpen
  \bibfield  {author} {\bibinfo {author} {\bibfnamefont {Sean~M.}\ \bibnamefont
  {Carroll}},\ }\href@noop {} {\emph {\bibinfo {title} {{Spacetime and
  Geometry}}}}\ (\bibinfo  {publisher} {Cambridge University Press},\ \bibinfo
  {year} {2019})\BibitemShut {NoStop}%
\bibitem [{\citenamefont {Brans}\ and\ \citenamefont
  {Dicke}(1961)}]{Brans:1961sx}%
  \BibitemOpen
  \bibfield  {author} {\bibinfo {author} {\bibfnamefont {C.}~\bibnamefont
  {Brans}}\ and\ \bibinfo {author} {\bibfnamefont {R.H.}\ \bibnamefont
  {Dicke}},\ }\bibfield  {title} {\enquote {\bibinfo {title} {{Mach's principle
  and a relativistic theory of gravitation}},}\ }\href {\doibase
  10.1103/PhysRev.124.925} {\bibfield  {journal} {\bibinfo  {journal} {Phys.
  Rev.}\ }\textbf {\bibinfo {volume} {124}},\ \bibinfo {pages} {925--935}
  (\bibinfo {year} {1961})}\BibitemShut {NoStop}%
\bibitem [{\citenamefont {Mirshekari}\ and\ \citenamefont
  {Will}(2013)}]{Mirshekari:2013vb}%
  \BibitemOpen
  \bibfield  {author} {\bibinfo {author} {\bibfnamefont {Saeed}\ \bibnamefont
  {Mirshekari}}\ and\ \bibinfo {author} {\bibfnamefont {Clifford~M.}\
  \bibnamefont {Will}},\ }\bibfield  {title} {\enquote {\bibinfo {title}
  {{Compact binary systems in scalar-tensor gravity: Equations of motion to 2.5
  post-Newtonian order}},}\ }\href {\doibase 10.1103/PhysRevD.87.084070}
  {\bibfield  {journal} {\bibinfo  {journal} {Phys. Rev. D}\ }\textbf {\bibinfo
  {volume} {87}},\ \bibinfo {pages} {084070} (\bibinfo {year} {2013})},\
  \Eprint {http://arxiv.org/abs/1301.4680} {arXiv:1301.4680 [gr-qc]}
  \BibitemShut {NoStop}%
\bibitem [{\citenamefont {{Friedrich}}(1998)}]{1998JGP....24...83F}%
  \BibitemOpen
  \bibfield  {author} {\bibinfo {author} {\bibfnamefont {Helmut}\ \bibnamefont
  {{Friedrich}}},\ }\bibfield  {title} {\enquote {\bibinfo {title}
  {{Gravitational fields near space-like and null infinity}},}\ }\href
  {\doibase 10.1016/S0393-0440(97)82168-7} {\bibfield  {journal} {\bibinfo
  {journal} {Journal of Geometry and Physics}\ }\textbf {\bibinfo {volume}
  {24}},\ \bibinfo {pages} {83--163} (\bibinfo {year} {1998})}\BibitemShut
  {NoStop}%
\bibitem [{\citenamefont {Satishchandran}\ and\ \citenamefont
  {Wald}(2019)}]{Satishchandran:2019pyc}%
  \BibitemOpen
  \bibfield  {author} {\bibinfo {author} {\bibfnamefont {Gautam}\ \bibnamefont
  {Satishchandran}}\ and\ \bibinfo {author} {\bibfnamefont {Robert~M.}\
  \bibnamefont {Wald}},\ }\bibfield  {title} {\enquote {\bibinfo {title}
  {{Asymptotic behavior of massless fields and the memory effect}},}\ }\href
  {\doibase 10.1103/PhysRevD.99.084007} {\bibfield  {journal} {\bibinfo
  {journal} {Phys. Rev. D}\ }\textbf {\bibinfo {volume} {99}},\ \bibinfo
  {pages} {084007} (\bibinfo {year} {2019})},\ \Eprint
  {http://arxiv.org/abs/1901.05942} {arXiv:1901.05942 [gr-qc]} \BibitemShut
  {NoStop}%
\bibitem [{\citenamefont {Thorne}(1992)}]{PhysRevD.45.520}%
  \BibitemOpen
  \bibfield  {author} {\bibinfo {author} {\bibfnamefont {Kip~S.}\ \bibnamefont
  {Thorne}},\ }\bibfield  {title} {\enquote {\bibinfo {title}
  {Gravitational-wave bursts with memory: The christodoulou effect},}\ }\href
  {\doibase 10.1103/PhysRevD.45.520} {\bibfield  {journal} {\bibinfo  {journal}
  {Phys. Rev. D}\ }\textbf {\bibinfo {volume} {45}},\ \bibinfo {pages}
  {520--524} (\bibinfo {year} {1992})}\BibitemShut {NoStop}%
\bibitem [{\citenamefont {Wald}\ and\ \citenamefont
  {Zoupas}(2000)}]{Wald:1999wa}%
  \BibitemOpen
  \bibfield  {author} {\bibinfo {author} {\bibfnamefont {Robert~M.}\
  \bibnamefont {Wald}}\ and\ \bibinfo {author} {\bibfnamefont {Andreas}\
  \bibnamefont {Zoupas}},\ }\bibfield  {title} {\enquote {\bibinfo {title} {{A
  General definition of 'conserved quantities' in general relativity and other
  theories of gravity}},}\ }\href {\doibase 10.1103/PhysRevD.61.084027}
  {\bibfield  {journal} {\bibinfo  {journal} {Phys. Rev. D}\ }\textbf {\bibinfo
  {volume} {61}},\ \bibinfo {pages} {084027} (\bibinfo {year} {2000})},\
  \Eprint {http://arxiv.org/abs/gr-qc/9911095} {arXiv:gr-qc/9911095}
  \BibitemShut {NoStop}%
\bibitem [{\citenamefont {Bieri}\ and\ \citenamefont
  {Garfinkle}(2014)}]{Bieri:2013ada}%
  \BibitemOpen
  \bibfield  {author} {\bibinfo {author} {\bibfnamefont {Lydia}\ \bibnamefont
  {Bieri}}\ and\ \bibinfo {author} {\bibfnamefont {David}\ \bibnamefont
  {Garfinkle}},\ }\bibfield  {title} {\enquote {\bibinfo {title} {{Perturbative
  and gauge invariant treatment of gravitational wave memory}},}\ }\href
  {\doibase 10.1103/PhysRevD.89.084039} {\bibfield  {journal} {\bibinfo
  {journal} {Phys. Rev. D}\ }\textbf {\bibinfo {volume} {89}},\ \bibinfo
  {pages} {084039} (\bibinfo {year} {2014})},\ \Eprint
  {http://arxiv.org/abs/1312.6871} {arXiv:1312.6871 [gr-qc]} \BibitemShut
  {NoStop}%
\bibitem [{\citenamefont {Lang}(2014)}]{Lang:2013fna}%
  \BibitemOpen
  \bibfield  {author} {\bibinfo {author} {\bibfnamefont {Ryan~N.}\ \bibnamefont
  {Lang}},\ }\bibfield  {title} {\enquote {\bibinfo {title} {{Compact binary
  systems in scalar-tensor gravity. II. Tensor gravitational waves to second
  post-Newtonian order}},}\ }\href {\doibase 10.1103/PhysRevD.89.084014}
  {\bibfield  {journal} {\bibinfo  {journal} {Phys. Rev. D}\ }\textbf {\bibinfo
  {volume} {89}},\ \bibinfo {pages} {084014} (\bibinfo {year} {2014})},\
  \Eprint {http://arxiv.org/abs/1310.3320} {arXiv:1310.3320 [gr-qc]}
  \BibitemShut {NoStop}%
\bibitem [{\citenamefont {Lang}(2015)}]{Lang:2014osa}%
  \BibitemOpen
  \bibfield  {author} {\bibinfo {author} {\bibfnamefont {Ryan~N.}\ \bibnamefont
  {Lang}},\ }\bibfield  {title} {\enquote {\bibinfo {title} {{Compact binary
  systems in scalar-tensor gravity. III. Scalar waves and energy flux}},}\
  }\href {\doibase 10.1103/PhysRevD.91.084027} {\bibfield  {journal} {\bibinfo
  {journal} {Phys. Rev. D}\ }\textbf {\bibinfo {volume} {91}},\ \bibinfo
  {pages} {084027} (\bibinfo {year} {2015})},\ \Eprint
  {http://arxiv.org/abs/1411.3073} {arXiv:1411.3073 [gr-qc]} \BibitemShut
  {NoStop}%
\bibitem [{\citenamefont {Sennett}\ \emph {et~al.}(2016)\citenamefont
  {Sennett}, \citenamefont {Marsat},\ and\ \citenamefont
  {Buonanno}}]{Sennett:2016klh}%
  \BibitemOpen
  \bibfield  {author} {\bibinfo {author} {\bibfnamefont {Noah}\ \bibnamefont
  {Sennett}}, \bibinfo {author} {\bibfnamefont {Sylvain}\ \bibnamefont
  {Marsat}}, \ and\ \bibinfo {author} {\bibfnamefont {Alessandra}\ \bibnamefont
  {Buonanno}},\ }\bibfield  {title} {\enquote {\bibinfo {title} {{Gravitational
  waveforms in scalar-tensor gravity at 2PN relative order}},}\ }\href
  {\doibase 10.1103/PhysRevD.94.084003} {\bibfield  {journal} {\bibinfo
  {journal} {Phys. Rev. D}\ }\textbf {\bibinfo {volume} {94}},\ \bibinfo
  {pages} {084003} (\bibinfo {year} {2016})},\ \Eprint
  {http://arxiv.org/abs/1607.01420} {arXiv:1607.01420 [gr-qc]} \BibitemShut
  {NoStop}%
\bibitem [{\citenamefont {Bernard}(2018)}]{Bernard:2018hta}%
  \BibitemOpen
  \bibfield  {author} {\bibinfo {author} {\bibfnamefont {Laura}\ \bibnamefont
  {Bernard}},\ }\bibfield  {title} {\enquote {\bibinfo {title} {{Dynamics of
  compact binary systems in scalar-tensor theories: Equations of motion to the
  third post-Newtonian order}},}\ }\href {\doibase 10.1103/PhysRevD.98.044004}
  {\bibfield  {journal} {\bibinfo  {journal} {Phys. Rev. D}\ }\textbf {\bibinfo
  {volume} {98}},\ \bibinfo {pages} {044004} (\bibinfo {year} {2018})},\
  \Eprint {http://arxiv.org/abs/1802.10201} {arXiv:1802.10201 [gr-qc]}
  \BibitemShut {NoStop}%
\bibitem [{\citenamefont {Bernard}(2019)}]{Bernard:2018ivi}%
  \BibitemOpen
  \bibfield  {author} {\bibinfo {author} {\bibfnamefont {Laura}\ \bibnamefont
  {Bernard}},\ }\bibfield  {title} {\enquote {\bibinfo {title} {{Dynamics of
  compact binary systems in scalar-tensor theories: II. Center-of-mass and
  conserved quantities to 3PN order}},}\ }\href {\doibase
  10.1103/PhysRevD.99.044047} {\bibfield  {journal} {\bibinfo  {journal} {Phys.
  Rev. D}\ }\textbf {\bibinfo {volume} {99}},\ \bibinfo {pages} {044047}
  (\bibinfo {year} {2019})},\ \Eprint {http://arxiv.org/abs/1812.04169}
  {arXiv:1812.04169 [gr-qc]} \BibitemShut {NoStop}%
\bibitem [{\citenamefont {Bernard}\ \emph {et~al.}(2022)\citenamefont
  {Bernard}, \citenamefont {Blanchet},\ and\ \citenamefont
  {Trestini}}]{Bernard:2022noq}%
  \BibitemOpen
  \bibfield  {author} {\bibinfo {author} {\bibfnamefont {Laura}\ \bibnamefont
  {Bernard}}, \bibinfo {author} {\bibfnamefont {Luc}\ \bibnamefont {Blanchet}},
  \ and\ \bibinfo {author} {\bibfnamefont {David}\ \bibnamefont {Trestini}},\
  }\bibfield  {title} {\enquote {\bibinfo {title} {{Gravitational waves in
  scalar-tensor theory to one-and-a-half post-Newtonian order}},}\ }\href
  {\doibase 10.1088/1475-7516/2022/08/008} {\bibfield  {journal} {\bibinfo
  {journal} {JCAP}\ }\textbf {\bibinfo {volume} {08}},\ \bibinfo {pages} {008}
  (\bibinfo {year} {2022})},\ \Eprint {http://arxiv.org/abs/2201.10924}
  {arXiv:2201.10924 [gr-qc]} \BibitemShut {NoStop}%
\bibitem [{\citenamefont {Trestini}(2024{\natexlab{a}})}]{Trestini:2024zpi}%
  \BibitemOpen
  \bibfield  {author} {\bibinfo {author} {\bibfnamefont {David}\ \bibnamefont
  {Trestini}},\ }\bibfield  {title} {\enquote {\bibinfo {title}
  {{Quasi-Keplerian parametrization for eccentric compact binaries in
  scalar-tensor theories at second post-Newtonian order and applications}},}\
  }\href {\doibase 10.1103/PhysRevD.109.104003} {\bibfield  {journal} {\bibinfo
   {journal} {Phys. Rev. D}\ }\textbf {\bibinfo {volume} {109}},\ \bibinfo
  {pages} {104003} (\bibinfo {year} {2024}{\natexlab{a}})},\ \Eprint
  {http://arxiv.org/abs/2401.06844} {arXiv:2401.06844 [gr-qc]} \BibitemShut
  {NoStop}%
\bibitem [{\citenamefont {Trestini}(2024{\natexlab{b}})}]{Trestini:2024mfs}%
  \BibitemOpen
  \bibfield  {author} {\bibinfo {author} {\bibfnamefont {David}\ \bibnamefont
  {Trestini}},\ }\bibfield  {title} {\enquote {\bibinfo {title} {{Gravitational
  waves from quasielliptic compact binaries in scalar-tensor theory to
  one-and-a-half post-Newtonian order}},}\ }\href@noop {} {\  (\bibinfo {year}
  {2024}{\natexlab{b}})},\ \Eprint {http://arxiv.org/abs/2410.12898}
  {arXiv:2410.12898 [gr-qc]} \BibitemShut {NoStop}%
\bibitem [{\citenamefont {Will}(2014)}]{Will:2014kxa}%
  \BibitemOpen
  \bibfield  {author} {\bibinfo {author} {\bibfnamefont {Clifford~M.}\
  \bibnamefont {Will}},\ }\bibfield  {title} {\enquote {\bibinfo {title} {{The
  Confrontation between General Relativity and Experiment}},}\ }\href {\doibase
  10.12942/lrr-2014-4} {\bibfield  {journal} {\bibinfo  {journal} {Living Rev.
  Rel.}\ }\textbf {\bibinfo {volume} {17}},\ \bibinfo {pages} {4} (\bibinfo
  {year} {2014})},\ \Eprint {http://arxiv.org/abs/1403.7377} {arXiv:1403.7377
  [gr-qc]} \BibitemShut {NoStop}%
\bibitem [{\citenamefont {Bertotti}\ \emph {et~al.}(2003)\citenamefont
  {Bertotti}, \citenamefont {Iess},\ and\ \citenamefont
  {Tortora}}]{Bertotti:2003rm}%
  \BibitemOpen
  \bibfield  {author} {\bibinfo {author} {\bibfnamefont {B.}~\bibnamefont
  {Bertotti}}, \bibinfo {author} {\bibfnamefont {L.}~\bibnamefont {Iess}}, \
  and\ \bibinfo {author} {\bibfnamefont {P.}~\bibnamefont {Tortora}},\
  }\bibfield  {title} {\enquote {\bibinfo {title} {{A test of general
  relativity using radio links with the Cassini spacecraft}},}\ }\href
  {\doibase 10.1038/nature01997} {\bibfield  {journal} {\bibinfo  {journal}
  {Nature}\ }\textbf {\bibinfo {volume} {425}},\ \bibinfo {pages} {374--376}
  (\bibinfo {year} {2003})}\BibitemShut {NoStop}%
\bibitem [{LIG(2025)}]{LIGOScientific:2025pvj}%
  \BibitemOpen
  \bibfield  {title} {\enquote {\bibinfo {title} {{GWTC-4.0: Population
  Properties of Merging Compact Binaries}},}\ }\href@noop {} {\bibfield
  {journal} {\bibinfo  {journal} {{}}\ } (\bibinfo {year} {2025})},\ \Eprint
  {http://arxiv.org/abs/2508.18083} {arXiv:2508.18083 [astro-ph.HE]}
  \BibitemShut {NoStop}%
\bibitem [{\citenamefont {Kuntz}\ and\ \citenamefont
  {Barausse}(2024)}]{Kuntz:2024jxo}%
  \BibitemOpen
  \bibfield  {author} {\bibinfo {author} {\bibfnamefont {Adrien}\ \bibnamefont
  {Kuntz}}\ and\ \bibinfo {author} {\bibfnamefont {Enrico}\ \bibnamefont
  {Barausse}},\ }\bibfield  {title} {\enquote {\bibinfo {title} {{Angular
  momentum sensitivities in scalar-tensor theories}},}\ }\href {\doibase
  10.1103/PhysRevD.109.124001} {\bibfield  {journal} {\bibinfo  {journal}
  {Phys. Rev. D}\ }\textbf {\bibinfo {volume} {109}},\ \bibinfo {pages}
  {124001} (\bibinfo {year} {2024})},\ \Eprint
  {http://arxiv.org/abs/2403.07980} {arXiv:2403.07980 [gr-qc]} \BibitemShut
  {NoStop}%
\bibitem [{\citenamefont {Takeda}\ \emph {et~al.}(2024)\citenamefont {Takeda},
  \citenamefont {Tsujikawa},\ and\ \citenamefont {Nishizawa}}]{Takeda:2023wqn}%
  \BibitemOpen
  \bibfield  {author} {\bibinfo {author} {\bibfnamefont {Hiroki}\ \bibnamefont
  {Takeda}}, \bibinfo {author} {\bibfnamefont {Shinji}\ \bibnamefont
  {Tsujikawa}}, \ and\ \bibinfo {author} {\bibfnamefont {Atsushi}\ \bibnamefont
  {Nishizawa}},\ }\bibfield  {title} {\enquote {\bibinfo {title}
  {{Gravitational-wave constraints on scalar-tensor gravity from a neutron star
  and black-hole binary GW200115}},}\ }\href {\doibase
  10.1103/PhysRevD.109.104072} {\bibfield  {journal} {\bibinfo  {journal}
  {Phys. Rev. D}\ }\textbf {\bibinfo {volume} {109}},\ \bibinfo {pages}
  {104072} (\bibinfo {year} {2024})},\ \Eprint
  {http://arxiv.org/abs/2311.09281} {arXiv:2311.09281 [gr-qc]} \BibitemShut
  {NoStop}%
\end{thebibliography}%

\end{document}